%% file: master.tex
\documentclass[runningheads]{llncs}

\usepackage[T1]{fontenc}
\usepackage[utf8]{inputenc}
\usepackage{amsmath,amssymb,amsfonts}
\usepackage{tikz}
\usepackage{caption}
\usepackage{subcaption}
\usepackage{cleveref}
\usepackage{pgfplots}
\usepackage{algorithmic}
\usepackage{graphicx}
\usepackage{textcomp}
\usepackage{xcolor}
\usepackage{csquotes}
\usepackage{xspace}
\usepackage{booktabs}

\newcommand*{\ie}{i.e.\@\xspace}

\usepackage[utf8]{inputenc}
\usepackage[backend=biber,
    bibstyle=lncs,
    sortcites=true,
    maxcitenames=2,
    mincitenames=1,
    maxbibnames=3,
    uniquelist=false,
    defernumbers=true,
]{biblatex}
\makeatletter
\let\blx@rerun@biber\relax
\makeatother

\begin{document}

\title{On the Routing Convergence Delay in the Lightning Network}

\author{Niklas Gögge\inst{1}
\and Elias Rohrer\inst{2}
\and Florian Tschorsch\inst{2}}

\authorrunning{N. Gögge et al.}

\institute{Distributed Security Infrastructures\\
Technical University of Berlin\\
\email{n.goegge@campus.tu-berlin.de}
\and
Distributed Security Infrastructures\\
Technical University of Berlin\\
\email{\{elias.rohrer, florian.tschorsch\}@tu-berlin.de}}

\maketitle

\begin{abstract}
Nodes in the Lightning Network synchronise routing information through a
gossip protocol that makes use of a staggered broadcast mechanism. In this
work, we show that the convergence delay in the network is larger than what
would be expected from the protocol's specification and that payment attempt
failures caused by the delay are more frequent, the larger the delay is. To
this end, we measure the convergence delay incurred in the network and analyse
what its primary causes are. Moreover, we further investigate and confirm our
findings through a time-discrete simulation of the Lightning Network gossip
protocol. We explore the use of alternative gossip protocols as well as
parameter variations of the current protocol and evaluate them by the
resulting bandwidth usage and convergence delay. Our research shows that there
are multiple ways of lowering the convergence delay, ranging from simple
parameter changes to overhauling the entire protocol.
\end{abstract}

\section{Introduction}\label{sec:intro}

Since its inception in 2008, the Bitcoin~\cite{nakamoto2008bitcoin} network
showed an inability to scale to a high volume of
transactions~\cite{sompolinsky2013accelerating}. The Bitcoin Lightning
Network~\cite{poon2016bitcoin} is a second-layer payment channel network (PCN)
that enables a high volume of low-cost off-chain Bitcoin transactions.

In the Lightning Network, nodes route payments by finding a path to the
destination based on a local copy of the public channel graph that each node
maintains. In order to keep their channel graph views in sync, nodes propagate
update messages via a peer-to-peer gossip protocol that utilizes a so-called
\emph{staggered broadcast} mechanism. As a result of the gossip protocol, it can---in
the worst case---take more than $10$ minutes for a message to reach all nodes
in the network.

To avoid issues caused by stale routing information, a convergence delay of
this magnitude goes against the common goal of routing protocols to reach
convergence quickly and reliably. The larger the convergence delay is, the
more likely it is for payment attempts to fail since a source node might be
computing a route based on stale information. Payment attempt failures
stemming from the convergence delay currently account for roughly $1.24\%$ of
all failures according to~\cite{waugh2020empirical}. These failures can not be
eliminated completely given that message propagation cannot be instant.
Moreover, improved routing algorithms such as multi-part payments~(MPPs) do
not improve the rate at which these failures occur. In fact, they may even
increase their occurrences as the probability of such failures only increases
with the number of channels involved in a payment.

In this work, we investigate the convergence delay of routing information and
its effects on payments in the Lightning Network. Our main goal is to present
the state of the convergence delay in the Lightning Network, the issues it
causes, and to layout potential improvement ideas. Our contributions can be
summarized as follows:

\begin{itemize}
\item We analyze the Lightning Network's gossip protocol in its current state
	by looking at and comparing \texttt{c-lightning} and \texttt{LND}, the two
	most popular node implementations. We measure the delay seen in the real
	network through a passive experiment and catalog the seen gossip messages
	(specifically all channel updates) to understand why and when gossip
	messages are broadcast by nodes. The catalog is also useful to understand
	which types of channel updates are potentially disruptive to payment
	routing. (\Cref{sec:analysis})
\item We implemented a simulator capable of simulating the Lightning Network's
	gossip protocol as well as payments in the Lightning Network. We can
	bootstrap our simulation from historical topology data and replay recorded
	gossip messages. We use the simulation to gain further inside into how the
	gossip protocol operates and where its inefficiencies lie.
	(\Cref{sec:eval})
\item We evaluate the use of alternative message propagation mechanisms in the
	Lightning Network. Through simulation, we compare flooding, a structured
	broadcast utilizing the channel graph topology, inventory based gossip, as
	well as efficient set reconciliation using Minisketch~\cite{minisketchreadme}. (\Cref{sec:eval})
\end{itemize}

To our knowledge, there exists no prior related work on the convergence delay in
the Lightning Network. However, there is a long history of convergence delay
research in internet routing through the Border Gateway Protocol (BGP), which we
use to draw inspiration for potential improvement ideas~\cite{BGPdelayed,
BGPsurvey, BGPtables}. We discuss these and other related works in
\Cref{sec:relwork}. In the following, we give a primer on information
propagation and the convergence delay in the Lightning Network.

\section{Information Propagation in the Lightning Network}\label{sec:motivation}

The Bitcoin Lightning Network~\cite{poon2016bitcoin} is a second-layer payment
channel network (PCN) that enables a high volume of low-cost off-chain Bitcoin
transactions. A payment channel describes a type of smart contract that
enables two parties to transact off-chain, with the only bottleneck being the
network latency between the two parties. A PCN enables
payments between nodes that do not have direct channels with each other by
routing payments over intermediary nodes to reach the destination. In order to
ensure that payment forwarding requires no trust towards these intermediaries,
such multi-hop payments are secured through so-called \textit{Hash Time Locked
Contracts} (HTLCs). Candidate routes are discovered by the originators through
a source-routing algorithm operating on a local copy of the network
graph, \ie, the routing information base (RIB). These local information are
regularly kept in sync by gossiping update messages in the network.

The \texttt{channel\_announcement}, \texttt{node\_announcement} and
\texttt{channel\_update} messages are the three main messages of the Lightning
Network's gossip protocol. Channel announcements are used by two nodes to prove
that there is a channel between them. The proof comes in the form of four
signatures tying the nodes to the keys used in the funding transaction. Node
announcements are used to provide additional information about a node such as
reachable network addresses. Channel updates provide routing information for a
channel edge, such as routing fees and lock times. Each channel counterparty is
able to broadcast a channel update for its outgoing channel edge. In order for
a channel to be operational the network has to see three messages, one channel
announcement and two channel updates (one for each edge of the channel).

\subsection{Influences on the Convergence Delay}
While the details of the information dissemination protocols are left to the
implementations, the most common implementations, such as
\texttt{c-lightning}\footnote{\url{https://github.com/ElementsProject/lightning}}
and \texttt{LND}\footnote{\url{https://github.com/lightningnetwork/lnd}},
generally follow the same concepts. As we show later, the concepts presented in
the following and their concrete parameterizations can have a significant
impact on the convergence delay.

\paragraph{Staggered Broadcast.}
The gossip protocol of the Lightning Network uses a staggered broadcast that
acts as a natural rate limiting mechanism to ensure that the network is
resistant to certain types of denial-of-service~(DoS) attacks. In a staggered
broadcast, each node listens for gossip messages for a specified
interval~(stagger interval) before broadcasting all messages to a subset of
peers. While listening, messages concerning the same channels are deduplicated
by the timestamp field provided in the messages. If two channel updates for the
same channel edge are seen, only the most recent update is kept in the
broadcast queue. The value chosen for the stagger interval has a big impact on
the convergence delay, since the higher it is the longer messages take to reach
a majority of nodes. The
specification\footnote{\url{https://github.com/lightning/bolts}} recommends a
$60$ second stagger interval.

\paragraph{Gossip Syncers.}
The \texttt{gossip\_timestamp\_filter} message allows nodes to manage from
which peers they want to receive new gossip. Not sending the filter message is
equivalent to not requesting any gossip. By default, nodes only send filters to
a subset of their peers, which are called \textit{active gossip syncers}, while
all other peers are \textit{passive gossip syncers}. The number of active
syncer connections each node maintains has an impact on the convergence delay
since it determines how well nodes are connected. The more active syncer nodes
choose the faster messages will propagate.

\paragraph{Rate Limiting.}
While the staggered broadcast already offers a form of rate limiting, nodes in
addition apply a second rate limit on a per-edge basis. Only a certain number
of updates from the same edge are allowed for each rate limiting interval. Such
policies exist to prevent nodes from spamming the network with channel updates,
but also to prevent I/O DoS attacks, since nodes write new channel updates to
disk. A third rate limiting applies to redundant channel updates (only
differing in the timestamp of the message), which are also considered as
\textit{keep alive updates}. A node will broadcast keep alive updates to
indicate that its channels are still active and should not be pruned from other
nodes' views of the network. To rate limit keep alive updates, nodes usually
only allow them in a defined frequency, but the details differ from
implementation to implementation.

\paragraph{Comparing Node Implementations}
While the Lightning implementations generally follow the concepts just
discussed, the specific parameters used by these implementations can differ
quite a bit. In the following, we therefore discuss the relevant details of the
two most popular implementations of the Lightning Network protocol,
\texttt{c-lightning} and \texttt{LND}.
\setlength{\tabcolsep}{2ex}
\renewcommand{\arraystretch}{1}
\begin{table}[t]
    \centering
	\begin{tabular}{l p{0.3\textwidth} p{0.3\textwidth} }
		\toprule
		& \texttt{c-lightning} & \texttt{LND}\\
		\midrule
        \textit{Staggered Broadcast}
		& 60 second stagger interval.
		& 90 second stagger interval, batches are broadcast in 5 second intervals.\\ 
		\midrule
        \textit{Gossip Syncers}
		& Five syncers, individual rotations every hour.
		& Three syncers, one being rotated every 20 minutes.\\
		\midrule
        \textit{Rate Limiting}
		& One channel update per day, burst up to 4.
		& One channel update per minute, burst of up to 10.\\ 
		\bottomrule
    \end{tabular}
    \caption{Comparison of \texttt{c-lightning} and \texttt{LND} with regard to  the most influential
		concepts on the convergence delay.}
    \label{tab:compareimpls}
\end{table}

As also shown in \Cref{tab:compareimpls}, the behavior of \texttt{c-lightning} generally sticks to
the specification's guidance, while \texttt{LND} differs from it significantly
with a stagger interval of $90$ seconds. When the timer
expires, all seen messages are split up into batches and broadcast to all
relevant peers in $5$ second intervals. 
The function for calculating the batch
size from the total number of messages $n$ to broadcast is the following: 
\begin{equation*}
	sb(n) = min\left(10, \frac{n \cdot 5s + 90s - 1}{90s}\right) 
\end{equation*}
The number of broadcast batches increases with the number of messages, but is capped at $18$
in order to prevent the overlapping of stagger intervals.
With $5$ seconds between batches and a maximum of $18$ batches, the
last message may potentially be broadcast $17 \cdot 5 = 85$ seconds after the stagger timer
expires. A plot of $sb(n)$ can be seen in \Cref{fig:subbatchfn}. Only if there
are more than $162$ messages seen per $90$-second stagger interval, all
$18$ batches will be filled. If the general rate of messages in the network is lower
than that, less batches will be used lowering the convergence delay.\footnote{The
stagger interval was increased in January 2019 from $30$ to $90$ seconds with
the reasoning to lower bandwidth usage by slowing the propagation of
messages~\cite{staggerPR}. In April 2019, the sub-batch broadcast was introduced
with the reasoning to eliminate bursty resource usage after the stagger timer
expires~\cite{subbatchPR}. We could not find records of detailed discussion on
how the exact parameter values for these changes were chosen.}

The rate limiting policies of these two node implementations do not play
together without friction. If a channel is updated once per minute, a
\texttt{c-lightning} node would disregard all updates after the fourth for up to one
hour, while a \texttt{LND} node would happily accept all updates. The
\texttt{c-lightning} node
will not relay disregarded updates, which can cause the convergence delay for
these updates to increase. However, this is not an observable issue, since the majority
of nodes are running \texttt{LND}. 

\section{Gossip Traffic Analysis}\label{sec:analysis}
In the following, we describe our methodology for measuring and analysing gossip traffic in
the Lightning Network.

\subsection{Measuring the Convergence Delay}\label{section:measureconv}

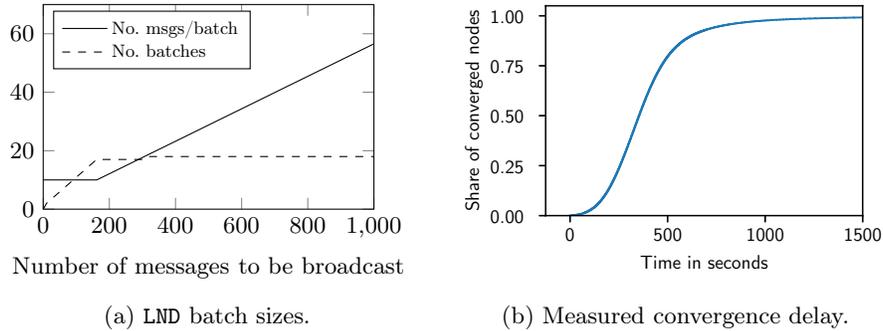
\begin{figure*}[t]
	\centering
	\begin{subfigure}[t]{.49\textwidth}
		\centering
		\begin{tikzpicture}
			\begin{axis}[
				xmin = 0, xmax = 1000,
				ymin = 0, ymax = 70,
				legend cell align = {left},
				legend pos = north west,
				width = \textwidth,
				height = \textwidth * 0.72,
				xlabel={Number of messages to be broadcast},
				xlabel style={font=\small},
				legend style={nodes={scale=0.75, transform shape}}]
				\addplot[samples=100,domain=0:1000]{max(10, (x*5000 + 89999) / 90000)};
				\addplot[dashed,samples=100,domain=0:1000]{ceil(x / max(10, (x*5000 + 89999) / 90000))};
				\legend{
				No.\ msgs/batch,
				No.\ batches
				}
			\end{axis}
		\end{tikzpicture}
		\captionof{figure}{\texttt{LND} batch sizes.}
		\label{fig:subbatchfn}
	\end{subfigure}
	\hspace*{\fill}
	\begin{subfigure}[t]{.49\textwidth}
		\centering
        \resizebox{\textwidth}{!}{\input{figs/analysis/messages.pgf}}
		\captionof{figure}{Measured convergence delay.}
		\label{fig:convdelay_real}
	\end{subfigure}
	\label{fig:convergence_delay}
	\caption{\texttt{LND}'s broadcast batching and measured convergence delay.}
	\vspace{-1.5em}
\end{figure*}

In order to measure the convergence delay in the Lightning Network, we used
the python
\texttt{pyln-proto}\footnote{\url{https://github.com/ElementsProject/lightning/tree/master/contrib/pyln-proto}}
package to connect to and communicate with nodes on the network. The node
addresses were extracted from a topology snapshot collected from an
\texttt{LND} node right before the start of the experiment (Oct.~30,~2021). We
connected to as many nodes as possible and chose all of them as our active
gossip syncers. We recorded all received messages including at which times
$\{t_1,\ldots,t_n\}$ and from which node we got the message. The recorded timestamps can
then be used to estimate the convergence delay in the network by looking at
the difference between the first and last timestamp. This estimation method
assumes that the first timestamps in these lists correspond to the time of
initial broadcast and that all nodes have seen the message after the last
timestamp.

In total, we received $69,942$ unique gossip messages from $1,046$ nodes over a time span
of close to $10$ hours. To estimate the convergence delay, we used all messages
that were received at least from $500$ different nodes. 

\Cref{fig:convdelay_real} shows the share of nodes
that have seen a message in relation to the time since initial broadcast: the
average time it takes for a node to see a message is $359.9$ seconds, with
$95\%$ of nodes seeing messages after $753$ seconds and $100\%$ of nodes
seeing messages after $2,500$ seconds.

\subsection{Dissecting Recorded Gossip}\label{section:recordedgossip}
We then categorized the collected data and examined which
share of gossip messages are node announcements, channel announcements or
channel updates. We also analyzed the contents of all channel updates to
understand when nodes send updates and how they typically update channel
policies.

\begin{figure*}[t]
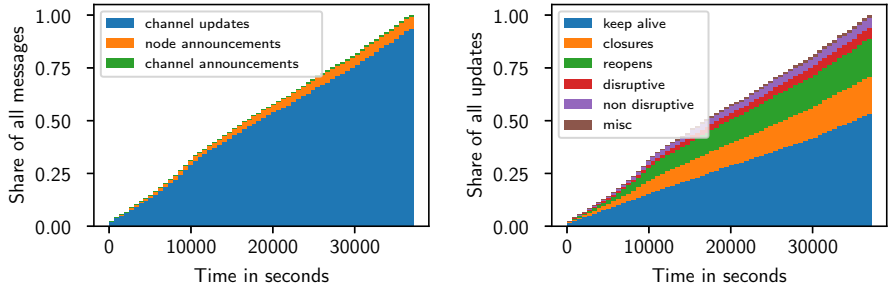

    \centering
    \begin{subfigure}{.49\textwidth}
        \centering
        \resizebox{\textwidth}{!}{\input{figs/analysis/catalog_0_plot.pgf}}
        \captionof{figure}{Observed shares of update messages.}
        \label{fig:catalog}
    \end{subfigure}
    \begin{subfigure}{.49\textwidth}
        \centering
        \resizebox{\textwidth}{!}{\input{figs/analysis/update_catalog_0_plot.pgf}}
        \captionof{figure}{Channel updates cataloged by type.}
        \label{fig:update_catalog}
    \end{subfigure}
    \label{fig:convergence_delay}
    \caption{Categorization of observed gossip messages.}
	\vspace{-1.5em}
\end{figure*}

As seen in \Cref{fig:catalog}, the rate at which new messages arrive is more
or less constant. 
Of all messages we recorded, $5.13\%$ were node
announcements, $0.34\%$ were channel announcements and $94.53\%$ were channel
updates. This distribution matches our expectations, as channel announcements
are directly rate limited by the blockchain, node announcements only need to
be broadcast infrequently to modify network addresses or add new feature
announcements, and channel updates change channel policies, which happens
regularly over the course of a channel's lifespan. 

We categorized channel updates into six different
categories:

\begin{itemize}
	\item \emph{Keep-alive} updates only differ in the timestamp field.
		These updates are meant to tell the network that a channel is still
		active. They made up $45.32\%$ of all recorded messages.
	\item \emph{Channel closure} updates close a channel temporarily or
		permanently. Temporary channel closures can happen if a peer
		goes offline due to network issues, in which case the other peer will
		broadcast such an update to inform the network not to route over the
		offline peer. These updates made up $19.29\%$ of all recorded
		messages.  
	\item \emph{Channel re-open} updates open a channel that was previously
		closed. These updates made up $18.66\%$ of all recorded messages.
	\item \emph{Disruptive} updates change the channel policy in a way that
		could cause payment failures, if the payment source does not know of
		the update. Channel closures are excluded because we categorize
		them separately. Disruptive updates made up $8.57\%$ of all
		recorded messages.  
	\item \emph{Non-disruptive} updates change the channel policy in a way
		that could cause a payment source to over-pay on fees or use a higher
		lock time than needed. These updates made up $7.22\%$ of all recorded
		messages.  
	\item \emph{Misc.}\ updates are all other updates that we saw. For
		example, updates that change the \texttt{htlc\_minimum\_msat} field
		fall into this category. These updates made up $0.99\%$ of all
		recorded messages.
\end{itemize}

\begin{figure*}[t]
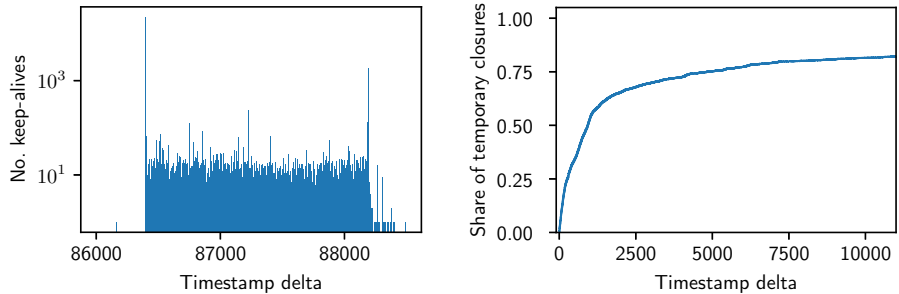

    \centering
    \begin{subfigure}[t]{.481\textwidth}
        \centering
        \resizebox{\textwidth}{!}{\input{figs/analysis/keepalive_times_0_plot.pgf}}
        \captionof{figure}{Time since last keep-alive.}
        \label{fig:keepalivefreq}
    \end{subfigure}
	\hspace*{\fill}
    \begin{subfigure}[t]{.499\textwidth}
        \centering
        \resizebox{\textwidth}{!}{\input{figs/analysis/reopen_0_plot.pgf}}
        \captionof{figure}{Cumulative channel closure durations.}
        \label{fig:reopenfreq}
    \end{subfigure}
    \label{fig:timestampdiffs}
    \caption{Timestamp differences of keep-alive and channel closure updates.}
	\vspace{-1.5em}
\end{figure*}

The observed amount of keep-alive updates is slightly concerning, as they make up
roughly $50\%$ of all seen updates. This amount of keep-alive updates cannot
be explained by nodes broadcasting them at a reasonable rate. In theory, a
keep-alive only has to be sent for channels that did not have an update within
$14$ days. Therefore, transmitting a keep-alive update every $13$ days should
be sufficient to prevent other nodes from pruning the channel.
\Cref{fig:keepalivefreq} shows the difference in the timestamp field between
the keep-alive and the previous update: we observe that for almost all of the
keep-alive updates the differences lie between $86,400$ and $88,200$ seconds,
which corresponds to exactly $1$ day and $1$ day plus $30$
minutes. We found that \texttt{LND} nodes are responsible for these updates, because
they check every $30$ minutes if any of their channels had an update
within the last day, and will broadcast a keep-alive update otherwise. 
However, we were not able to explain the large peaks seen in \Cref{fig:keepalivefreq} at the
interval boundaries. Moreover, we did not observe any keep-alive updates with
a smaller difference, because \texttt{LND} nodes do not relay such updates and
therefore they do not propagate through the network.

Looking at the timestamp differences for all updates in our channel re-open
category (cf.~\Cref{fig:reopenfreq}), we see that most channels edges that get
re-opened were disabled for short periods of time. For example, $60\%$ of
edges were closed for less than $22$ minutes. This is likely caused by network
issues that lead nodes to temporarily disable edges.

\section{Simulation Study}\label{sec:eval}
In the following, we discuss the conducted model-based simulation study on the routing convergence
delay in the Lightning Network.

\subsection{Simulation Model}
The behavior of real-world peer-to-peer networks is influenced by many
different variables. Nodes participating in such networks can be diverse in
geographical location, bandwidth restrictions, software implementation,
software version or configuration, and simulating all different permutations
is simply not feasible. In the context of investigating the gossip protocol of
the Lightning Network, we restrict the scope of our simulation by making the
following assumptions: if two nodes are connected through a channel, they have a
constant TCP connection. The snapshot we use to bootstrap our simulation
contains all nodes and all channels that exist in the network. We ignore any
non-listening nodes that were not announced to the network, as well as private
channels. Our simulation propagates node and channel announcements, but does
not actually add them to the simulated topology. Only channel updates are
applied to the simulated topology. The gossip algorithm is the main influence
on the convergence delay, and we do not simulate other potential influences
such as an overhead caused by cryptographic functions. Payments are atomic and
instant. All nodes in each simulation follow the same gossip protocol.
All nodes have the same bandwidth of $1\,MB/s$ in up- and download.

We chose to implement our discrete-event simulator\footnote{\url{https://github.com/dergoegge/lnconv-paper-sim}} 
in the Go programming language and bootstrap the simulation from historical
topology snapshots that were extracted from an \texttt{LND} node with a fully
synced network graph. These snapshots contain a list of nodes and channels
which we use to build our simulation network. The snapshot we use for all
simulations contains $17,332$ nodes, $77,921$ channels and was taken on Oct.\
30, 2021. In order to simulate a realistic amount of traffic, we replay gossip
messages that we recorded in the real network. This works well as most gossip
messages can be traced back to an origin node in the network as long the
snapshot we use to bootstrap the simulation is not much older than the start
of the recorded period. For messages for which we could not find an origin in
our snapshot we choose a random origin. 
Bandwidth is modeled by each node having an incoming byte counter that gets
incremented with every message that is being downloaded and decremented with
every message that is fully received. The arrival time of a new message is
calculated based on a fixed bandwidth, the number of incoming bytes and a fixed
latency overhead of $100\,ms$.

\subsection{Simulation Results}\label{section:simresults}
In this section, we present the data collected on an \texttt{LND} simulation scenario in
which we replayed the first hour of the gossip we recorded in
\Cref{sec:analysis}, consisting of $7,217$ network messages. We simulate
$100,000$ payment attempts which were uniformly distributed over the hour.
Payment sources and destinations are chosen randomly and the payment amount is
set to $1\,sat$ in order to reduce interference by failures originating from 
anything else than outdated routing information.

\paragraph{Bandwidth.} The simulated network transferred a total of
$40.77\,GB$ to deliver the $7,217$ messages to all nodes. The theoretical
lower bound for bandwidth usage $B_{min}$ is the product of the number of all
nodes, the total number of messages and the average message size, \ie,
\begin{equation*} 
	B_{min} = num\_nodes \cdot num\_messages \cdot avg\_message\_size 
\end{equation*}
Assuming all messages are channel updates with a size of $128\,bytes$, 
$B_{min} = 16.01\,GB$. We therefore found that the network uses
$2.55$ times the theoretically needed bandwidth $B_{min}$.

\paragraph{Redundancy.} $6.29\%$ of messages will be seen only once, $33.28\%$
will be seen twice, $59.93\%$ will be seen three, and $0.5\%$ will be seen four
times. All nodes have $3$ active gossip syncers which explains why most
messages are seen three times or less. A message is only seen $4$ times if it
is received as part of the initial broadcast, which goes out to all connected
peers. On average each message is seen $2.55$ times. Note that this is the same
factor as the one from our bandwidth calculations: every message that is
received more than once is exactly the overhead to a perfect broadcast in which
every message is received only once by each node. 

\begin{figure*}[t]
    \centering
	\begin{subfigure}[t]{.49\textwidth}
        \centering
        \resizebox{\textwidth}{!}{\input{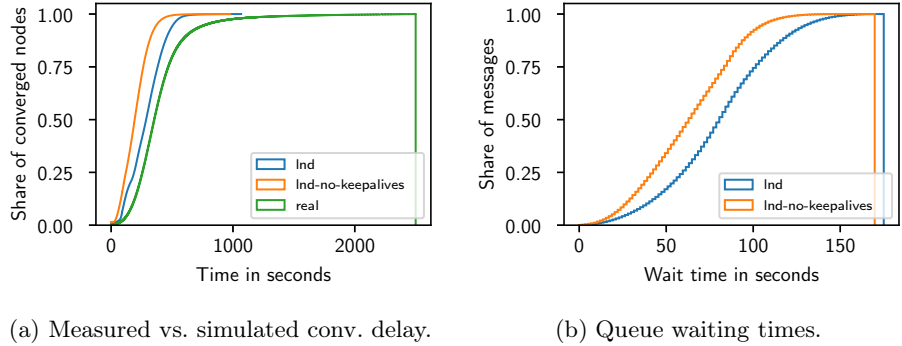}}
        \captionof{figure}{Measured vs.\ simulated conv.\ delay.}
        \label{fig:lndconvdelaycompare}
    \end{subfigure}
    \hspace*{\fill}
	\begin{subfigure}[t]{.49\textwidth}
        \centering
        \resizebox{\textwidth}{!}{\input{figs/sim/waited_buckets_0_plot.pgf}}
        \captionof{figure}{Queue waiting times.}
        \label{fig:lndwaitingtimes}
    \end{subfigure}
    \label{fig:lndresults}
    \caption{Convergence delay and broadcast queue waiting times in a simulated network consisting only of \texttt{LND} nodes.}
	\vspace{-1.5em}
\end{figure*}

\paragraph{Convergence Delay.} We measure the convergence delay by recording how
long it takes a message to be seen for the first time by every node. This
is very similar to the measurements conducted in \Cref{sec:analysis},
but within a simulation we get much more accurate data since we have an
omniscient view. \Cref{fig:lndconvdelaycompare} compares the convergence
delay we recorded in the real network to the one we observed in the simulation. In
our simulation, the average time it took for a node to see a message is
$291.21$ seconds, with $95\%$ of nodes seeing messages after $510$ seconds and
$100\%$ of nodes seeing messages after $1,075$ seconds. The convergence delay
seen in the simulation slightly differs from the delay measured in the real
network with messages in the simulation propagating faster after initially
being broadcast and messages taking longer to reach all nodes in the real
network. From $20\%$ to $80\%$ of nodes having seen the messages it takes
$240$ seconds in the simulation while in took $265$ seconds in the real
network. 

As mentioned previously, roughly $50\%$ of the messages that we recorded are
keep-alive updates. We ran a simulation without the keep-alive updates
(\texttt{lnd-no-keepalives}) and found that the convergence delay was
significantly reduced, with $95\%$ of nodes converging after $374.19$ instead
of $510$ seconds.

\paragraph{Waiting Times.} Looking at the broadcast queue waiting times of
messages we observed that waiting times and hence the convergence delay become
larger the more messages are propagating through the network. This is
explained by the sub-batch trickling approach that \texttt{LND} has chosen which makes
waiting times dynamic to a certain degree. The growth of waiting times is
bounded by the maximum number of sub-batches that \texttt{LND} will send. A plot of the
waiting times can be seen in \Cref{fig:lndwaitingtimes}. The minimum
waiting time is $0$ seconds and the maximum is $175$ seconds. A message will
wait $175$ seconds, if it arrives at the beginning of the $90$ second stagger
interval and gets broadcast in the last sub-batch, $85$ seconds after the stagger
timer ticks.

\paragraph{Failed Payment Attempts.} Out of the $100,000$ tried payment attempts,
$42\%$ were successful and $58\%$ failed. $0.114\%$ of attempts failed because
the payment source did not have a recent update for one of the channel edges
in the payment route.

As we have seen, the staggered broadcast is quite inefficient in its bandwidth
usage with messages being seen $2.55$ times on average by the same node and
$95\%$ of nodes converging after $510$ seconds. The share of unconverged
payment attempts ($0.114\%$) does not seem that problematic but it could be
argued that in absolute numbers the total number of unconverged payment
attempts can still be large. The research by \citeauthor{waugh2020empirical} suggests
that this rate is actually higher at around $1.2\%$~\cite{waugh2020empirical}.
Exploring alternative gossip algorithms seems worthwhile based on these
results.

\subsection{Evaluating Alternative Gossip Strategies}
In this section, we layout ideas for potential alternative gossip algorithms
that the Lightning Network could employ. We use our simulator 
to compare the different algorithms and evaluate the
feasibility of these alternatives being used in the real network based on
bandwidth usage, convergence delays, and their impact on payment attempts. We
compare the following alternative strategies: flooding, a structured broadcast
using a global spanning tree, inventory based gossip, parameter variations of
the current protocol, as well as set reconciliation using
Minisketch~\cite{minisketchreadme}.

We compare all alternative strategies to each other and the simulation data
from \Cref{section:simresults}. We specifically compare bandwidth usage,
convergence times and the number of unconverged payment attempts and simulate
each algorithm using the same snapshot and replaying the same messages
as before~($17,332$ nodes, $77,921$ channels, $7,217$ messages over $1$
hour, $100,000$ payment attempts). The convergence delays and bandwidth usage
for all the different algorithms are listed in \Cref{tab:evaconvband}.

\begin{table}[t]
    \centering
    \begin{tabular}{lrrr}
		\toprule
        Algorithm & Conv.\ Delay & Bandwidth Usage & Payment attempts \\
        \midrule
        \texttt{lnd} & 509.75s & 40.47\,GB & 602 \\
        \texttt{lnd-t1s} & 312.65s & 39.36\,GB & 349 \\
        \texttt{lnd-sb100} & 266.54s & 38.9\,GB & 316 \\
        \texttt{lnd-inv} & 509.46s & 19.26\,GB & 592 \\
        \texttt{lnd-inv-t1s} & 313.45s & 19.41\,GB & 394 \\
        \texttt{lnd-inv-sb100} & 267.93s & 20.23\,GB & 274 \\
        \texttt{c-lightning} & 101.29s & 59.52\,GB & 171 \\
        \texttt{c-lightning-inv} & 103.2s & 26.36\,GB & 161 \\
        \texttt{spanning (BFS)} & 1.11s & 15.7\,GB & 5 \\
        \texttt{flooding-4} & 2.72s & 50.7\,GB & 3 \\
        \texttt{flooding-8} & 1.72s & 94.7\,GB & 1 \\
        \texttt{flooding-16} & 1.16s & 180.92\,GB & 2 \\
        \texttt{flooding-32} & 0.82s & 353.21\,GB & 4 \\
        \texttt{minisketch-4} & 19.25s & 19.15\,GB & 33 \\
        \texttt{minisketch-8} & 20.24s & 19.84\,GB & 43 \\
        \texttt{minisketch-16} & 20.7s & 21.45\,GB & 43 \\
        \texttt{minisketch-32} & 20.54s & 21.46\,GB & 30 \\
		\bottomrule
    \end{tabular}
    \caption{Convergence Delays (95\%), bandwidth usage, and unconverged payment attempts.}
    \label{tab:evaconvband}
	\vspace{-1.5em}
\end{table}

As expected, flooding has the highest bandwidth usage with low convergence
delays and the spanning tree algorithm (global tree constructed using
breadth-first search) has the lowest bandwidth usage and the lowest convergence
delay. With flooding, we see the bandwidth consumption scaling proportionally
with increased connectivity (number of active syncer connections). The
convergence delay is naturally smaller with increased connectivity.

\texttt{LND}'s choice of staggered broadcast parameters results in a roughly
five times increase in the convergence delay compared to \texttt{c-lightning}.
While \texttt{LND}'s approach leads to a larger convergence delay it also
reduces bandwidth usage by about $33\%$. We simulated two variations of
\texttt{LND}'s algorithm, one with a minimum sub-batch size of $100$ instead
of $10$ messages (\texttt{lnd-sb100}), and one with a sub-batch delay of one
instead of five seconds (\texttt{lnd-t1s}). Both of these parameter changes
lead to faster messages broadcast after the stagger timer expires leading to
an decrease in convergence delay of $39\%$ for \texttt{lnd-t1s} and $48\%$ for
\texttt{lnd-sb100}.

Inventory-based protocols announce a shortened version of the full message to
give the receiver the chance to only request the full message once. For gossip
messages in the lightning network, the size of an inventory message can be
$64$ bits~\cite{mailminisketch}. We see that inventory based protocols reduce
bandwidth usage significantly when compared to their regular variants. With
\texttt{lnd-inv} requiring $52.4\%$ less bandwidth than \texttt{lnd} and
\texttt{c-lightning-inv} requiring $55.7\%$ less bandwidth than
\texttt{c-lightning}. The convergence delays however are unaffected by the
decrease in bandwidth usage. Usually it would be expected that latency
increases with an inventory-based gossip protocol but the extra round trip has no impact
here, given that the stagger interval is multiples larger than the round trip
time.

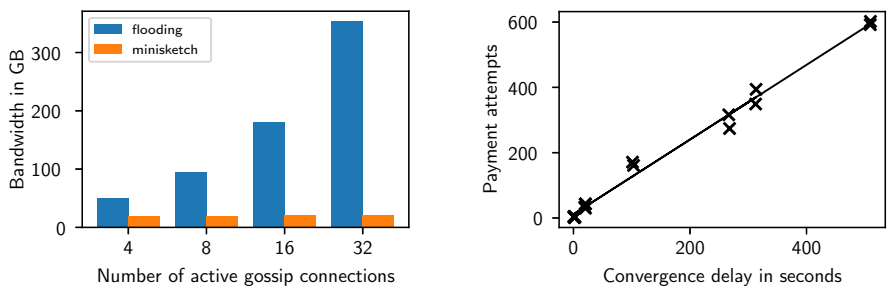
\begin{figure*}[t]
    \centering
	\begin{subfigure}[t]{.49\textwidth}
        \centering
        \resizebox{\textwidth}{!}{\input{figs/eval/bandwidth_conns_0_plot.pgf}}
        \caption{Bandwidth usage vs.\ increased connectivity.}
        \label{fig:evafloodrecon}
    \end{subfigure}
    \hspace*{\fill}
	\begin{subfigure}[t]{.49\textwidth}
        \centering
        \resizebox{.96\textwidth}{!}{\input{figs/eval/convdelay_payments_0_plot.pgf}}
        \caption{Unconverged payment attempts vs.\ convergence delay.}
        \label{fig:evapayments}
    \end{subfigure}
    \label{fig:evamisc}
	\caption{Simulated parameter interdependencies.}
	\vspace{-1.5em}
\end{figure*}

In \Cref{fig:evafloodrecon}, we compare the bandwidth usage of flooding and
set reconciliation, in relation to the connections made by each node. Our set
reconciliation algorithm is based on the Erlay protocol that was proposed for
the transaction relay in the Bitcoin network~\cite{naumenko2019erlay}. In our
protocol, we implemented no fan-out flooding and hence all messages are exchanged via set
reconciliation. We observe that bandwidth usage does not increase proportionally with the
number of connections made for the set reconciliation protocol. Instead, the
bandwidth usage scales with the rate of messages in the network, just like the
Erlay protocol.

We observe that the number of unconverged payment attempts is highly correlated
with the convergence delay. We do not distinguish between failed payment
attempts and attempts that arise due to opportunity costs, as the combined
number of these attempts is sufficient in evaluating different protocols. As
seen in \Cref{fig:evapayments}, based on our limited data set of the different
algorithms, the relationship between the convergence delay and the number of
unconverged payment attempts is linear. The lower the convergence delay, the
fewer unconverged payment attempts can be observed.

\section{Discussion}\label{sec:discussion}
The staggered broadcast protocols rate-limit the propagation of channel
updates by de-duplicating updates for the same channel with in the stagger
interval. This means that a node will only forward one channel update for the
same channel edge in every stagger interval. No potentially important updates
are discarded, since the newest update that was seen will always be forwarded.
This form of rate limiting prevents the network from witnessing rapid changes
in channel policies, while still propagating the newest updates. The
propagation of the newest updates is significantly delayed as we have shown
through the simulations and measured in \Cref{sec:analysis}. We argue that
this form of rate limiting implicitly discourages frequent channel updates at
the cost of delivering the newest updates with large delays. Explicitly
discouraging frequent updates through strict per-channel rate limiting as
discussed in \Cref{sec:analysis} could be well suited for some of our
alternative protocols that aim to deliver messages faster. A strict rate limit
would discard newer updates that violate the rate limit, so honest nodes
should never broadcast messages for the same channel in violation of the
limit.

\texttt{LND}'s choice of parameters for its staggered broadcast is a bit of a mystery,
since there is no public record on how the exact values were chosen. However,
broadcasting messages in sub-batches instead of one large batch after the
stagger timer expires is a good choice to reduce bursty resource usage. We
would however recommend that the \texttt{LND} developers revisit their choice of parameters for
the staggered broadcast, because reducing bandwidth usage by $33\%$ while
increasing the convergence delay by a factor of five does not seem like a
reasonable trade-off (compared to parameters mentioned in the specification).
As we have shown through the simulations, adjusting the parameters can have a
big impact on the convergence delay. Adjusting these parameters would be the
least complex software change to address the large convergence delay, while
maintaining the rate limiting properties of the staggered broadcast.

Introducing an inventory-based gossip protocol reduces the bandwidth usage without
changing the convergence delay at all. In combination with adjusting the
parameters of the staggered broadcast the convergence delay could also be
lowered. An inventory-based gossip protocol could remain a staggered broadcast and
thereby maintain its rate limiting effect without introducing strict rate
limiting. The added software complexity of an inventory-based gossip is fairly
low and there already exists a proposal on the specification~\cite{invPR}. 

Increasing the number of connections that nodes make to gossip (connectivity)
can lead to better reliability in adversarial environments. With low
connectivity an attacker has to control less connections to be able to censor
information from reaching a victim. For some protocols an increase in
connectivity can also lead to a reduction in convergence times because the
spread factor is higher.

Even though the spanning tree protocol seems great based on the results, it is
not a great fit for the real network. As mentioned earlier, the protocol makes
the assumption that all nodes agree on the exact same static spanning tree,
which would not trivially work in the real network. A single tree is also not
going work for security and reliability reasons. If one node in the tree goes
offline, none of the nodes in its sub-tree would receive new messages.
Introducing multiple trees to gain redundancy would increase the bandwidth usage
which makes using a spanning tree less desirable in the first place. A
spanning tree protocol with multiple trees would probably turnout to be
similar in efficiency to a flooding protocol.

A flooding protocol comes with a small convergence delay of one to two seconds
but increases bandwidth usage above that of the current algorithm
(\texttt{lnd}). Bandwidth usage increases linearly with increased
connectivity. If an increase in connectivity is wanted then flooding would not
be suitable. In fact all protocols besides set reconciliation lead to a
proportional increase in bandwidth with increased connectivity.

Compared to the other protocols, set reconciliation has a small convergence
delay and low bandwidth usage. Increasing connectivity is also possible
without increasing bandwidth usage, as the bandwidth usage scales with the
rate of messages seen in the network. Introducing set reconciliation comes
with much greater software complexity than any of the other protocols.
Multiple new message types would need to be introduced and the Minisketch
library adds a dependency.

Decreasing the number of unconverged payment attempts can also be done without
changing the gossip protocol. Nodes could temporarily allow payments that use
old channel policies, after broadcasting a new policy. This would work well
for fee or lock time adjustments. In the end this depends on the channel
owners preferences on whether or not they want updates to immediately take
effect.

\section{Related Work}\label{sec:relwork}

The explosive growth of the internet in its topological complexity as well as
user count has led to a lot of research on the convergence delay for routing
protocols, such as the Border Gateway Protocol~(BGP). Large convergence delays
in BGP can cause routing failures similar to how large convergence delays in
the Lightning Network can cause payment failures. \citeauthor{BGPdelayed}
showed through a $2$-year study that the convergence delay of BGP was much
higher than previously expected. By injecting routing events to simulate
failures and collecting data on these events, the authors were able to figure out the
convergence delay for different types of events. Convergence delays were
primarily caused by different router vendor's implementations of the BGP
specification with regard to the choice of timer values~\cite{BGPdelayed}.
\citeauthor{BGPsurvey} suggested ways on how to lower the BGP convergence
delay which included adjusting timer values of implementations and
centralizing control of networks~\cite{BGPsurvey}. \citeauthor{BGPtables}
investigated slow BGP table transfers which increase the convergence delay.
They found that gaps, in which both sender and receiver are idle, during table
transfers are a common occurrence caused by timer driven implementations, with
different vendors choosing different timer values~\cite{BGPtables}. Similar to
this BGP research, we found that a big part of the convergence delay in the
Lightning Network is driven by the parameter choices for the staggered
broadcast of different implementations. 

\citeauthor{decker2013propagation} measured block propagation times in the Bitcoin network
and verified that the propagation time is the primary cause for forks in the
blockchain. They measured the propagation times by connecting to a large
number of nodes and listening for block announcements. With this setup they
recorded when blocks where seen and from which nodes. From this data they are
able to estimate how long it takes blocks to traverse the network after the
initial broadcast~\cite{decker2013propagation}. Our work is methodically similar, since we
also measure the convergence delay in the Lightning Network by connecting to
many nodes in the network and record arrival times of messages.

\citeauthor{naumenko2019erlay} proposed \textit{Erlay}, a protocol for transaction relay
in the Bitcoin network that makes use of efficient set reconciliation in
combination with flooding. It aims to lower the bandwidth requirements needed
for transaction relay with the trade-off of higher latency. The authors evaluated the
bandwidth and latency trade-off of Erlay and compared it to the current
flood-only protocol~\cite{naumenko2019erlay}. We used the Erlay protocol as inspiration
for simulating a similar protocol in the Lightning Network and specifically
used their prior research when choosing the parameters for our protocol.

\citeauthor{waugh2020empirical} studied availability and reliability properties of the
Lightning Network. They tested the network's ability to route payments of
different amounts and created a taxonomy of permanent and temporary failures
that occurred. They looked at the availability of nodes in the network and
measured how much churn (nodes joining and leaving the network)
exists~\cite{waugh2020empirical}. This work listed payment attempt failure types that
were caused by outdated routing information, by probing the network with real
payments. We only simulated payments to investigate these failure types.

\section{Conclusion}\label{sec:conclusion}
In this work, we analyzed the convergence delay in the Lightning Network,
described the effect it can have on payments, and evaluated alternative gossip
protocols that could reduce the delay.
We found the network to have a significant convergence delay, with $95\%$ of
nodes only having converged after roughly $10$ minutes. A majority of the gossip
traffic consists of redundant channel updates (keep-alive messages), which further
increase the delay given the parameter choices of the \texttt{LND}
implementation. Our simulations show that payment attempt failures due to
unconverged routing information are rare (occuring in $\ll1\%$ of payment
attempts). However, the convergence delay may still be lowered while also
reducing the bandwidth usage, either by switching to alternative gossip
algorithms or adjusting the parameters of the current protocol. By switching to
a set reconciliation based protocol, the connectivity of the network could be
increased with nodes receiving gossip updates from more peers without
suffering from significant increases in bandwidth.

\printbibliography
\end{document}

%% file: figs/analysis/messages.pgf
\begingroup%
\makeatletter%
\begin{pgfpicture}%
\pgfpathrectangle{\pgfpointorigin}{\pgfqpoint{3.298166in}{2.161603in}}%
\pgfusepath{use as bounding box, clip}%
\begin{pgfscope}%
\pgfsetbuttcap%
\pgfsetmiterjoin%
\definecolor{currentfill}{rgb}{1.000000,1.000000,1.000000}%
\pgfsetfillcolor{currentfill}%
\pgfsetlinewidth{0.000000pt}%
\definecolor{currentstroke}{rgb}{1.000000,1.000000,1.000000}%
\pgfsetstrokecolor{currentstroke}%
\pgfsetdash{}{0pt}%
\pgfpathmoveto{\pgfqpoint{0.000000in}{0.000000in}}%
\pgfpathlineto{\pgfqpoint{3.298166in}{0.000000in}}%
\pgfpathlineto{\pgfqpoint{3.298166in}{2.161603in}}%
\pgfpathlineto{\pgfqpoint{0.000000in}{2.161603in}}%
\pgfpathclose%
\pgfusepath{fill}%
\end{pgfscope}%
\begin{pgfscope}%
\pgfsetbuttcap%
\pgfsetmiterjoin%
\definecolor{currentfill}{rgb}{1.000000,1.000000,1.000000}%
\pgfsetfillcolor{currentfill}%
\pgfsetlinewidth{0.000000pt}%
\definecolor{currentstroke}{rgb}{0.000000,0.000000,0.000000}%
\pgfsetstrokecolor{currentstroke}%
\pgfsetstrokeopacity{0.000000}%
\pgfsetdash{}{0pt}%
\pgfpathmoveto{\pgfqpoint{0.696435in}{0.521603in}}%
\pgfpathlineto{\pgfqpoint{3.021435in}{0.521603in}}%
\pgfpathlineto{\pgfqpoint{3.021435in}{2.061603in}}%
\pgfpathlineto{\pgfqpoint{0.696435in}{2.061603in}}%
\pgfpathclose%
\pgfusepath{fill}%
\end{pgfscope}%
\begin{pgfscope}%
\pgfsetbuttcap%
\pgfsetroundjoin%
\definecolor{currentfill}{rgb}{0.000000,0.000000,0.000000}%
\pgfsetfillcolor{currentfill}%
\pgfsetlinewidth{0.803000pt}%
\definecolor{currentstroke}{rgb}{0.000000,0.000000,0.000000}%
\pgfsetstrokecolor{currentstroke}%
\pgfsetdash{}{0pt}%
\pgfsys@defobject{currentmarker}{\pgfqpoint{0.000000in}{-0.048611in}}{\pgfqpoint{0.000000in}{0.000000in}}{%
\pgfpathmoveto{\pgfqpoint{0.000000in}{0.000000in}}%
\pgfpathlineto{\pgfqpoint{0.000000in}{-0.048611in}}%
\pgfusepath{stroke,fill}%
}%
\begin{pgfscope}%
\pgfsys@transformshift{0.875282in}{0.521603in}%
\pgfsys@useobject{currentmarker}{}%
\end{pgfscope}%
\end{pgfscope}%
\begin{pgfscope}%
\definecolor{textcolor}{rgb}{0.000000,0.000000,0.000000}%
\pgfsetstrokecolor{textcolor}%
\pgfsetfillcolor{textcolor}%
\pgftext[x=0.875282in,y=0.424381in,,top]{\color{textcolor}\sffamily\fontsize{10.000000}{12.000000}\selectfont 0}%
\end{pgfscope}%
\begin{pgfscope}%
\pgfsetbuttcap%
\pgfsetroundjoin%
\definecolor{currentfill}{rgb}{0.000000,0.000000,0.000000}%
\pgfsetfillcolor{currentfill}%
\pgfsetlinewidth{0.803000pt}%
\definecolor{currentstroke}{rgb}{0.000000,0.000000,0.000000}%
\pgfsetstrokecolor{currentstroke}%
\pgfsetdash{}{0pt}%
\pgfsys@defobject{currentmarker}{\pgfqpoint{0.000000in}{-0.048611in}}{\pgfqpoint{0.000000in}{0.000000in}}{%
\pgfpathmoveto{\pgfqpoint{0.000000in}{0.000000in}}%
\pgfpathlineto{\pgfqpoint{0.000000in}{-0.048611in}}%
\pgfusepath{stroke,fill}%
}%
\begin{pgfscope}%
\pgfsys@transformshift{1.590666in}{0.521603in}%
\pgfsys@useobject{currentmarker}{}%
\end{pgfscope}%
\end{pgfscope}%
\begin{pgfscope}%
\definecolor{textcolor}{rgb}{0.000000,0.000000,0.000000}%
\pgfsetstrokecolor{textcolor}%
\pgfsetfillcolor{textcolor}%
\pgftext[x=1.590666in,y=0.424381in,,top]{\color{textcolor}\sffamily\fontsize{10.000000}{12.000000}\selectfont 500}%
\end{pgfscope}%
\begin{pgfscope}%
\pgfsetbuttcap%
\pgfsetroundjoin%
\definecolor{currentfill}{rgb}{0.000000,0.000000,0.000000}%
\pgfsetfillcolor{currentfill}%
\pgfsetlinewidth{0.803000pt}%
\definecolor{currentstroke}{rgb}{0.000000,0.000000,0.000000}%
\pgfsetstrokecolor{currentstroke}%
\pgfsetdash{}{0pt}%
\pgfsys@defobject{currentmarker}{\pgfqpoint{0.000000in}{-0.048611in}}{\pgfqpoint{0.000000in}{0.000000in}}{%
\pgfpathmoveto{\pgfqpoint{0.000000in}{0.000000in}}%
\pgfpathlineto{\pgfqpoint{0.000000in}{-0.048611in}}%
\pgfusepath{stroke,fill}%
}%
\begin{pgfscope}%
\pgfsys@transformshift{2.306051in}{0.521603in}%
\pgfsys@useobject{currentmarker}{}%
\end{pgfscope}%
\end{pgfscope}%
\begin{pgfscope}%
\definecolor{textcolor}{rgb}{0.000000,0.000000,0.000000}%
\pgfsetstrokecolor{textcolor}%
\pgfsetfillcolor{textcolor}%
\pgftext[x=2.306051in,y=0.424381in,,top]{\color{textcolor}\sffamily\fontsize{10.000000}{12.000000}\selectfont 1000}%
\end{pgfscope}%
\begin{pgfscope}%
\pgfsetbuttcap%
\pgfsetroundjoin%
\definecolor{currentfill}{rgb}{0.000000,0.000000,0.000000}%
\pgfsetfillcolor{currentfill}%
\pgfsetlinewidth{0.803000pt}%
\definecolor{currentstroke}{rgb}{0.000000,0.000000,0.000000}%
\pgfsetstrokecolor{currentstroke}%
\pgfsetdash{}{0pt}%
\pgfsys@defobject{currentmarker}{\pgfqpoint{0.000000in}{-0.048611in}}{\pgfqpoint{0.000000in}{0.000000in}}{%
\pgfpathmoveto{\pgfqpoint{0.000000in}{0.000000in}}%
\pgfpathlineto{\pgfqpoint{0.000000in}{-0.048611in}}%
\pgfusepath{stroke,fill}%
}%
\begin{pgfscope}%
\pgfsys@transformshift{3.021435in}{0.521603in}%
\pgfsys@useobject{currentmarker}{}%
\end{pgfscope}%
\end{pgfscope}%
\begin{pgfscope}%
\definecolor{textcolor}{rgb}{0.000000,0.000000,0.000000}%
\pgfsetstrokecolor{textcolor}%
\pgfsetfillcolor{textcolor}%
\pgftext[x=3.021435in,y=0.424381in,,top]{\color{textcolor}\sffamily\fontsize{10.000000}{12.000000}\selectfont 1500}%
\end{pgfscope}%
\begin{pgfscope}%
\definecolor{textcolor}{rgb}{0.000000,0.000000,0.000000}%
\pgfsetstrokecolor{textcolor}%
\pgfsetfillcolor{textcolor}%
\pgftext[x=1.858935in,y=0.234413in,,top]{\color{textcolor}\sffamily\fontsize{10.000000}{12.000000}\selectfont Time in seconds}%
\end{pgfscope}%
\begin{pgfscope}%
\pgfsetbuttcap%
\pgfsetroundjoin%
\definecolor{currentfill}{rgb}{0.000000,0.000000,0.000000}%
\pgfsetfillcolor{currentfill}%
\pgfsetlinewidth{0.803000pt}%
\definecolor{currentstroke}{rgb}{0.000000,0.000000,0.000000}%
\pgfsetstrokecolor{currentstroke}%
\pgfsetdash{}{0pt}%
\pgfsys@defobject{currentmarker}{\pgfqpoint{-0.048611in}{0.000000in}}{\pgfqpoint{-0.000000in}{0.000000in}}{%
\pgfpathmoveto{\pgfqpoint{-0.000000in}{0.000000in}}%
\pgfpathlineto{\pgfqpoint{-0.048611in}{0.000000in}}%
\pgfusepath{stroke,fill}%
}%
\begin{pgfscope}%
\pgfsys@transformshift{0.696435in}{0.521603in}%
\pgfsys@useobject{currentmarker}{}%
\end{pgfscope}%
\end{pgfscope}%
\begin{pgfscope}%
\definecolor{textcolor}{rgb}{0.000000,0.000000,0.000000}%
\pgfsetstrokecolor{textcolor}%
\pgfsetfillcolor{textcolor}%
\pgftext[x=0.289968in, y=0.468842in, left, base]{\color{textcolor}\sffamily\fontsize{10.000000}{12.000000}\selectfont 0.00}%
\end{pgfscope}%
\begin{pgfscope}%
\pgfsetbuttcap%
\pgfsetroundjoin%
\definecolor{currentfill}{rgb}{0.000000,0.000000,0.000000}%
\pgfsetfillcolor{currentfill}%
\pgfsetlinewidth{0.803000pt}%
\definecolor{currentstroke}{rgb}{0.000000,0.000000,0.000000}%
\pgfsetstrokecolor{currentstroke}%
\pgfsetdash{}{0pt}%
\pgfsys@defobject{currentmarker}{\pgfqpoint{-0.048611in}{0.000000in}}{\pgfqpoint{-0.000000in}{0.000000in}}{%
\pgfpathmoveto{\pgfqpoint{-0.000000in}{0.000000in}}%
\pgfpathlineto{\pgfqpoint{-0.048611in}{0.000000in}}%
\pgfusepath{stroke,fill}%
}%
\begin{pgfscope}%
\pgfsys@transformshift{0.696435in}{0.888270in}%
\pgfsys@useobject{currentmarker}{}%
\end{pgfscope}%
\end{pgfscope}%
\begin{pgfscope}%
\definecolor{textcolor}{rgb}{0.000000,0.000000,0.000000}%
\pgfsetstrokecolor{textcolor}%
\pgfsetfillcolor{textcolor}%
\pgftext[x=0.289968in, y=0.835508in, left, base]{\color{textcolor}\sffamily\fontsize{10.000000}{12.000000}\selectfont 0.25}%
\end{pgfscope}%
\begin{pgfscope}%
\pgfsetbuttcap%
\pgfsetroundjoin%
\definecolor{currentfill}{rgb}{0.000000,0.000000,0.000000}%
\pgfsetfillcolor{currentfill}%
\pgfsetlinewidth{0.803000pt}%
\definecolor{currentstroke}{rgb}{0.000000,0.000000,0.000000}%
\pgfsetstrokecolor{currentstroke}%
\pgfsetdash{}{0pt}%
\pgfsys@defobject{currentmarker}{\pgfqpoint{-0.048611in}{0.000000in}}{\pgfqpoint{-0.000000in}{0.000000in}}{%
\pgfpathmoveto{\pgfqpoint{-0.000000in}{0.000000in}}%
\pgfpathlineto{\pgfqpoint{-0.048611in}{0.000000in}}%
\pgfusepath{stroke,fill}%
}%
\begin{pgfscope}%
\pgfsys@transformshift{0.696435in}{1.254937in}%
\pgfsys@useobject{currentmarker}{}%
\end{pgfscope}%
\end{pgfscope}%
\begin{pgfscope}%
\definecolor{textcolor}{rgb}{0.000000,0.000000,0.000000}%
\pgfsetstrokecolor{textcolor}%
\pgfsetfillcolor{textcolor}%
\pgftext[x=0.289968in, y=1.202175in, left, base]{\color{textcolor}\sffamily\fontsize{10.000000}{12.000000}\selectfont 0.50}%
\end{pgfscope}%
\begin{pgfscope}%
\pgfsetbuttcap%
\pgfsetroundjoin%
\definecolor{currentfill}{rgb}{0.000000,0.000000,0.000000}%
\pgfsetfillcolor{currentfill}%
\pgfsetlinewidth{0.803000pt}%
\definecolor{currentstroke}{rgb}{0.000000,0.000000,0.000000}%
\pgfsetstrokecolor{currentstroke}%
\pgfsetdash{}{0pt}%
\pgfsys@defobject{currentmarker}{\pgfqpoint{-0.048611in}{0.000000in}}{\pgfqpoint{-0.000000in}{0.000000in}}{%
\pgfpathmoveto{\pgfqpoint{-0.000000in}{0.000000in}}%
\pgfpathlineto{\pgfqpoint{-0.048611in}{0.000000in}}%
\pgfusepath{stroke,fill}%
}%
\begin{pgfscope}%
\pgfsys@transformshift{0.696435in}{1.621603in}%
\pgfsys@useobject{currentmarker}{}%
\end{pgfscope}%
\end{pgfscope}%
\begin{pgfscope}%
\definecolor{textcolor}{rgb}{0.000000,0.000000,0.000000}%
\pgfsetstrokecolor{textcolor}%
\pgfsetfillcolor{textcolor}%
\pgftext[x=0.289968in, y=1.568842in, left, base]{\color{textcolor}\sffamily\fontsize{10.000000}{12.000000}\selectfont 0.75}%
\end{pgfscope}%
\begin{pgfscope}%
\pgfsetbuttcap%
\pgfsetroundjoin%
\definecolor{currentfill}{rgb}{0.000000,0.000000,0.000000}%
\pgfsetfillcolor{currentfill}%
\pgfsetlinewidth{0.803000pt}%
\definecolor{currentstroke}{rgb}{0.000000,0.000000,0.000000}%
\pgfsetstrokecolor{currentstroke}%
\pgfsetdash{}{0pt}%
\pgfsys@defobject{currentmarker}{\pgfqpoint{-0.048611in}{0.000000in}}{\pgfqpoint{-0.000000in}{0.000000in}}{%
\pgfpathmoveto{\pgfqpoint{-0.000000in}{0.000000in}}%
\pgfpathlineto{\pgfqpoint{-0.048611in}{0.000000in}}%
\pgfusepath{stroke,fill}%
}%
\begin{pgfscope}%
\pgfsys@transformshift{0.696435in}{1.988270in}%
\pgfsys@useobject{currentmarker}{}%
\end{pgfscope}%
\end{pgfscope}%
\begin{pgfscope}%
\definecolor{textcolor}{rgb}{0.000000,0.000000,0.000000}%
\pgfsetstrokecolor{textcolor}%
\pgfsetfillcolor{textcolor}%
\pgftext[x=0.289968in, y=1.935508in, left, base]{\color{textcolor}\sffamily\fontsize{10.000000}{12.000000}\selectfont 1.00}%
\end{pgfscope}%
\begin{pgfscope}%
\definecolor{textcolor}{rgb}{0.000000,0.000000,0.000000}%
\pgfsetstrokecolor{textcolor}%
\pgfsetfillcolor{textcolor}%
\pgftext[x=0.234413in,y=1.291603in,,bottom,rotate=90.000000]{\color{textcolor}\sffamily\fontsize{10.000000}{12.000000}\selectfont Share of converged nodes}%
\end{pgfscope}%
\begin{pgfscope}%
\pgfpathrectangle{\pgfqpoint{0.696435in}{0.521603in}}{\pgfqpoint{2.325000in}{1.540000in}}%
\pgfusepath{clip}%
\pgfsetbuttcap%
\pgfsetmiterjoin%
\pgfsetlinewidth{1.003750pt}%
\definecolor{currentstroke}{rgb}{0.121569,0.466667,0.705882}%
\pgfsetstrokecolor{currentstroke}%
\pgfsetdash{}{0pt}%
\pgfpathmoveto{\pgfqpoint{0.875282in}{0.521603in}}%
\pgfpathlineto{\pgfqpoint{0.875282in}{0.523695in}}%
\pgfpathlineto{\pgfqpoint{0.893166in}{0.524584in}}%
\pgfpathlineto{\pgfqpoint{0.893166in}{0.524859in}}%
\pgfpathlineto{\pgfqpoint{0.907474in}{0.525819in}}%
\pgfpathlineto{\pgfqpoint{0.907474in}{0.526191in}}%
\pgfpathlineto{\pgfqpoint{0.918205in}{0.527027in}}%
\pgfpathlineto{\pgfqpoint{0.918205in}{0.527494in}}%
\pgfpathlineto{\pgfqpoint{0.928935in}{0.528530in}}%
\pgfpathlineto{\pgfqpoint{0.928935in}{0.529105in}}%
\pgfpathlineto{\pgfqpoint{0.936089in}{0.529715in}}%
\pgfpathlineto{\pgfqpoint{0.936089in}{0.530362in}}%
\pgfpathlineto{\pgfqpoint{0.943243in}{0.531053in}}%
\pgfpathlineto{\pgfqpoint{0.943243in}{0.531780in}}%
\pgfpathlineto{\pgfqpoint{0.950397in}{0.532558in}}%
\pgfpathlineto{\pgfqpoint{0.950397in}{0.533386in}}%
\pgfpathlineto{\pgfqpoint{0.957551in}{0.534265in}}%
\pgfpathlineto{\pgfqpoint{0.957551in}{0.535192in}}%
\pgfpathlineto{\pgfqpoint{0.964705in}{0.536157in}}%
\pgfpathlineto{\pgfqpoint{0.964705in}{0.537177in}}%
\pgfpathlineto{\pgfqpoint{0.971858in}{0.538267in}}%
\pgfpathlineto{\pgfqpoint{0.971858in}{0.539400in}}%
\pgfpathlineto{\pgfqpoint{0.975435in}{0.539400in}}%
\pgfpathlineto{\pgfqpoint{0.975435in}{0.540598in}}%
\pgfpathlineto{\pgfqpoint{0.979012in}{0.540598in}}%
\pgfpathlineto{\pgfqpoint{0.979012in}{0.541862in}}%
\pgfpathlineto{\pgfqpoint{0.982589in}{0.541862in}}%
\pgfpathlineto{\pgfqpoint{0.982589in}{0.543183in}}%
\pgfpathlineto{\pgfqpoint{0.986166in}{0.543183in}}%
\pgfpathlineto{\pgfqpoint{0.986166in}{0.544578in}}%
\pgfpathlineto{\pgfqpoint{0.989743in}{0.544578in}}%
\pgfpathlineto{\pgfqpoint{0.989743in}{0.546029in}}%
\pgfpathlineto{\pgfqpoint{0.993320in}{0.546029in}}%
\pgfpathlineto{\pgfqpoint{0.993320in}{0.547557in}}%
\pgfpathlineto{\pgfqpoint{0.996897in}{0.547557in}}%
\pgfpathlineto{\pgfqpoint{0.996897in}{0.549145in}}%
\pgfpathlineto{\pgfqpoint{1.000474in}{0.549145in}}%
\pgfpathlineto{\pgfqpoint{1.000474in}{0.550808in}}%
\pgfpathlineto{\pgfqpoint{1.004051in}{0.550808in}}%
\pgfpathlineto{\pgfqpoint{1.004051in}{0.552537in}}%
\pgfpathlineto{\pgfqpoint{1.007628in}{0.552537in}}%
\pgfpathlineto{\pgfqpoint{1.007628in}{0.554343in}}%
\pgfpathlineto{\pgfqpoint{1.011205in}{0.554343in}}%
\pgfpathlineto{\pgfqpoint{1.011205in}{0.556229in}}%
\pgfpathlineto{\pgfqpoint{1.014782in}{0.556229in}}%
\pgfpathlineto{\pgfqpoint{1.014782in}{0.558196in}}%
\pgfpathlineto{\pgfqpoint{1.018358in}{0.558196in}}%
\pgfpathlineto{\pgfqpoint{1.018358in}{0.560242in}}%
\pgfpathlineto{\pgfqpoint{1.021935in}{0.560242in}}%
\pgfpathlineto{\pgfqpoint{1.021935in}{0.562376in}}%
\pgfpathlineto{\pgfqpoint{1.025512in}{0.562376in}}%
\pgfpathlineto{\pgfqpoint{1.025512in}{0.564560in}}%
\pgfpathlineto{\pgfqpoint{1.029089in}{0.564560in}}%
\pgfpathlineto{\pgfqpoint{1.029089in}{0.566841in}}%
\pgfpathlineto{\pgfqpoint{1.032666in}{0.566841in}}%
\pgfpathlineto{\pgfqpoint{1.032666in}{0.569227in}}%
\pgfpathlineto{\pgfqpoint{1.036243in}{0.569227in}}%
\pgfpathlineto{\pgfqpoint{1.036243in}{0.571697in}}%
\pgfpathlineto{\pgfqpoint{1.039820in}{0.571697in}}%
\pgfpathlineto{\pgfqpoint{1.039820in}{0.574254in}}%
\pgfpathlineto{\pgfqpoint{1.043397in}{0.574254in}}%
\pgfpathlineto{\pgfqpoint{1.043397in}{0.576898in}}%
\pgfpathlineto{\pgfqpoint{1.046974in}{0.576898in}}%
\pgfpathlineto{\pgfqpoint{1.046974in}{0.579638in}}%
\pgfpathlineto{\pgfqpoint{1.050551in}{0.579638in}}%
\pgfpathlineto{\pgfqpoint{1.050551in}{0.582476in}}%
\pgfpathlineto{\pgfqpoint{1.054128in}{0.582476in}}%
\pgfpathlineto{\pgfqpoint{1.054128in}{0.585405in}}%
\pgfpathlineto{\pgfqpoint{1.057705in}{0.585405in}}%
\pgfpathlineto{\pgfqpoint{1.057705in}{0.588428in}}%
\pgfpathlineto{\pgfqpoint{1.061282in}{0.588428in}}%
\pgfpathlineto{\pgfqpoint{1.061282in}{0.591537in}}%
\pgfpathlineto{\pgfqpoint{1.064858in}{0.591537in}}%
\pgfpathlineto{\pgfqpoint{1.064858in}{0.594767in}}%
\pgfpathlineto{\pgfqpoint{1.068435in}{0.594767in}}%
\pgfpathlineto{\pgfqpoint{1.068435in}{0.598080in}}%
\pgfpathlineto{\pgfqpoint{1.072012in}{0.598080in}}%
\pgfpathlineto{\pgfqpoint{1.072012in}{0.601506in}}%
\pgfpathlineto{\pgfqpoint{1.075589in}{0.601506in}}%
\pgfpathlineto{\pgfqpoint{1.075589in}{0.605031in}}%
\pgfpathlineto{\pgfqpoint{1.079166in}{0.605031in}}%
\pgfpathlineto{\pgfqpoint{1.079166in}{0.608670in}}%
\pgfpathlineto{\pgfqpoint{1.082743in}{0.608670in}}%
\pgfpathlineto{\pgfqpoint{1.082743in}{0.612403in}}%
\pgfpathlineto{\pgfqpoint{1.086320in}{0.612403in}}%
\pgfpathlineto{\pgfqpoint{1.086320in}{0.616239in}}%
\pgfpathlineto{\pgfqpoint{1.089897in}{0.616239in}}%
\pgfpathlineto{\pgfqpoint{1.089897in}{0.620175in}}%
\pgfpathlineto{\pgfqpoint{1.093474in}{0.620175in}}%
\pgfpathlineto{\pgfqpoint{1.093474in}{0.624236in}}%
\pgfpathlineto{\pgfqpoint{1.097051in}{0.624236in}}%
\pgfpathlineto{\pgfqpoint{1.097051in}{0.628412in}}%
\pgfpathlineto{\pgfqpoint{1.100628in}{0.628412in}}%
\pgfpathlineto{\pgfqpoint{1.100628in}{0.632696in}}%
\pgfpathlineto{\pgfqpoint{1.104205in}{0.632696in}}%
\pgfpathlineto{\pgfqpoint{1.104205in}{0.637096in}}%
\pgfpathlineto{\pgfqpoint{1.107782in}{0.637096in}}%
\pgfpathlineto{\pgfqpoint{1.107782in}{0.641599in}}%
\pgfpathlineto{\pgfqpoint{1.111358in}{0.641599in}}%
\pgfpathlineto{\pgfqpoint{1.111358in}{0.646217in}}%
\pgfpathlineto{\pgfqpoint{1.114935in}{0.646217in}}%
\pgfpathlineto{\pgfqpoint{1.114935in}{0.650946in}}%
\pgfpathlineto{\pgfqpoint{1.118512in}{0.650946in}}%
\pgfpathlineto{\pgfqpoint{1.118512in}{0.655797in}}%
\pgfpathlineto{\pgfqpoint{1.122089in}{0.655797in}}%
\pgfpathlineto{\pgfqpoint{1.122089in}{0.660764in}}%
\pgfpathlineto{\pgfqpoint{1.125666in}{0.660764in}}%
\pgfpathlineto{\pgfqpoint{1.125666in}{0.665870in}}%
\pgfpathlineto{\pgfqpoint{1.129243in}{0.665870in}}%
\pgfpathlineto{\pgfqpoint{1.129243in}{0.671067in}}%
\pgfpathlineto{\pgfqpoint{1.132820in}{0.671067in}}%
\pgfpathlineto{\pgfqpoint{1.132820in}{0.676416in}}%
\pgfpathlineto{\pgfqpoint{1.136397in}{0.676416in}}%
\pgfpathlineto{\pgfqpoint{1.136397in}{0.681870in}}%
\pgfpathlineto{\pgfqpoint{1.139974in}{0.681870in}}%
\pgfpathlineto{\pgfqpoint{1.139974in}{0.687389in}}%
\pgfpathlineto{\pgfqpoint{1.143551in}{0.687389in}}%
\pgfpathlineto{\pgfqpoint{1.143551in}{0.693072in}}%
\pgfpathlineto{\pgfqpoint{1.147128in}{0.693072in}}%
\pgfpathlineto{\pgfqpoint{1.147128in}{0.698849in}}%
\pgfpathlineto{\pgfqpoint{1.150705in}{0.698849in}}%
\pgfpathlineto{\pgfqpoint{1.150705in}{0.704756in}}%
\pgfpathlineto{\pgfqpoint{1.154282in}{0.704756in}}%
\pgfpathlineto{\pgfqpoint{1.154282in}{0.710800in}}%
\pgfpathlineto{\pgfqpoint{1.157858in}{0.710800in}}%
\pgfpathlineto{\pgfqpoint{1.157858in}{0.716937in}}%
\pgfpathlineto{\pgfqpoint{1.161435in}{0.716937in}}%
\pgfpathlineto{\pgfqpoint{1.161435in}{0.723175in}}%
\pgfpathlineto{\pgfqpoint{1.165012in}{0.723175in}}%
\pgfpathlineto{\pgfqpoint{1.165012in}{0.729530in}}%
\pgfpathlineto{\pgfqpoint{1.168589in}{0.729530in}}%
\pgfpathlineto{\pgfqpoint{1.168589in}{0.736005in}}%
\pgfpathlineto{\pgfqpoint{1.172166in}{0.736005in}}%
\pgfpathlineto{\pgfqpoint{1.172166in}{0.742587in}}%
\pgfpathlineto{\pgfqpoint{1.175743in}{0.742587in}}%
\pgfpathlineto{\pgfqpoint{1.175743in}{0.749290in}}%
\pgfpathlineto{\pgfqpoint{1.179320in}{0.749290in}}%
\pgfpathlineto{\pgfqpoint{1.179320in}{0.756091in}}%
\pgfpathlineto{\pgfqpoint{1.182897in}{0.756091in}}%
\pgfpathlineto{\pgfqpoint{1.182897in}{0.763007in}}%
\pgfpathlineto{\pgfqpoint{1.186474in}{0.763007in}}%
\pgfpathlineto{\pgfqpoint{1.186474in}{0.770040in}}%
\pgfpathlineto{\pgfqpoint{1.190051in}{0.770040in}}%
\pgfpathlineto{\pgfqpoint{1.190051in}{0.777145in}}%
\pgfpathlineto{\pgfqpoint{1.193628in}{0.777145in}}%
\pgfpathlineto{\pgfqpoint{1.193628in}{0.784370in}}%
\pgfpathlineto{\pgfqpoint{1.197205in}{0.784370in}}%
\pgfpathlineto{\pgfqpoint{1.197205in}{0.791723in}}%
\pgfpathlineto{\pgfqpoint{1.200782in}{0.791723in}}%
\pgfpathlineto{\pgfqpoint{1.200782in}{0.799173in}}%
\pgfpathlineto{\pgfqpoint{1.204358in}{0.799173in}}%
\pgfpathlineto{\pgfqpoint{1.204358in}{0.806690in}}%
\pgfpathlineto{\pgfqpoint{1.207935in}{0.806690in}}%
\pgfpathlineto{\pgfqpoint{1.207935in}{0.814352in}}%
\pgfpathlineto{\pgfqpoint{1.211512in}{0.814352in}}%
\pgfpathlineto{\pgfqpoint{1.211512in}{0.822097in}}%
\pgfpathlineto{\pgfqpoint{1.215089in}{0.822097in}}%
\pgfpathlineto{\pgfqpoint{1.215089in}{0.829949in}}%
\pgfpathlineto{\pgfqpoint{1.218666in}{0.829949in}}%
\pgfpathlineto{\pgfqpoint{1.218666in}{0.837874in}}%
\pgfpathlineto{\pgfqpoint{1.222243in}{0.837874in}}%
\pgfpathlineto{\pgfqpoint{1.222243in}{0.845903in}}%
\pgfpathlineto{\pgfqpoint{1.225820in}{0.845903in}}%
\pgfpathlineto{\pgfqpoint{1.225820in}{0.854042in}}%
\pgfpathlineto{\pgfqpoint{1.229397in}{0.854042in}}%
\pgfpathlineto{\pgfqpoint{1.229397in}{0.862272in}}%
\pgfpathlineto{\pgfqpoint{1.232974in}{0.862272in}}%
\pgfpathlineto{\pgfqpoint{1.232974in}{0.870570in}}%
\pgfpathlineto{\pgfqpoint{1.236551in}{0.870570in}}%
\pgfpathlineto{\pgfqpoint{1.236551in}{0.878966in}}%
\pgfpathlineto{\pgfqpoint{1.240128in}{0.878966in}}%
\pgfpathlineto{\pgfqpoint{1.240128in}{0.887415in}}%
\pgfpathlineto{\pgfqpoint{1.243705in}{0.887415in}}%
\pgfpathlineto{\pgfqpoint{1.243705in}{0.895952in}}%
\pgfpathlineto{\pgfqpoint{1.247282in}{0.895952in}}%
\pgfpathlineto{\pgfqpoint{1.247282in}{0.904599in}}%
\pgfpathlineto{\pgfqpoint{1.250858in}{0.904599in}}%
\pgfpathlineto{\pgfqpoint{1.250858in}{0.913304in}}%
\pgfpathlineto{\pgfqpoint{1.254435in}{0.913304in}}%
\pgfpathlineto{\pgfqpoint{1.254435in}{0.922103in}}%
\pgfpathlineto{\pgfqpoint{1.258012in}{0.922103in}}%
\pgfpathlineto{\pgfqpoint{1.258012in}{0.930995in}}%
\pgfpathlineto{\pgfqpoint{1.261589in}{0.930995in}}%
\pgfpathlineto{\pgfqpoint{1.261589in}{0.939941in}}%
\pgfpathlineto{\pgfqpoint{1.265166in}{0.939941in}}%
\pgfpathlineto{\pgfqpoint{1.265166in}{0.948949in}}%
\pgfpathlineto{\pgfqpoint{1.268743in}{0.948949in}}%
\pgfpathlineto{\pgfqpoint{1.268743in}{0.958021in}}%
\pgfpathlineto{\pgfqpoint{1.272320in}{0.958021in}}%
\pgfpathlineto{\pgfqpoint{1.272320in}{0.967145in}}%
\pgfpathlineto{\pgfqpoint{1.275897in}{0.967145in}}%
\pgfpathlineto{\pgfqpoint{1.275897in}{0.976323in}}%
\pgfpathlineto{\pgfqpoint{1.279474in}{0.976323in}}%
\pgfpathlineto{\pgfqpoint{1.279474in}{0.985533in}}%
\pgfpathlineto{\pgfqpoint{1.283051in}{0.985533in}}%
\pgfpathlineto{\pgfqpoint{1.283051in}{0.994866in}}%
\pgfpathlineto{\pgfqpoint{1.286628in}{0.994866in}}%
\pgfpathlineto{\pgfqpoint{1.286628in}{1.004215in}}%
\pgfpathlineto{\pgfqpoint{1.290205in}{1.004215in}}%
\pgfpathlineto{\pgfqpoint{1.290205in}{1.013610in}}%
\pgfpathlineto{\pgfqpoint{1.293782in}{1.013610in}}%
\pgfpathlineto{\pgfqpoint{1.293782in}{1.023071in}}%
\pgfpathlineto{\pgfqpoint{1.297358in}{1.023071in}}%
\pgfpathlineto{\pgfqpoint{1.297358in}{1.032563in}}%
\pgfpathlineto{\pgfqpoint{1.300935in}{1.032563in}}%
\pgfpathlineto{\pgfqpoint{1.300935in}{1.042089in}}%
\pgfpathlineto{\pgfqpoint{1.304512in}{1.042089in}}%
\pgfpathlineto{\pgfqpoint{1.304512in}{1.051638in}}%
\pgfpathlineto{\pgfqpoint{1.308089in}{1.051638in}}%
\pgfpathlineto{\pgfqpoint{1.308089in}{1.061241in}}%
\pgfpathlineto{\pgfqpoint{1.311666in}{1.061241in}}%
\pgfpathlineto{\pgfqpoint{1.311666in}{1.070873in}}%
\pgfpathlineto{\pgfqpoint{1.315243in}{1.070873in}}%
\pgfpathlineto{\pgfqpoint{1.315243in}{1.080536in}}%
\pgfpathlineto{\pgfqpoint{1.318820in}{1.080536in}}%
\pgfpathlineto{\pgfqpoint{1.318820in}{1.090250in}}%
\pgfpathlineto{\pgfqpoint{1.322397in}{1.090250in}}%
\pgfpathlineto{\pgfqpoint{1.322397in}{1.099939in}}%
\pgfpathlineto{\pgfqpoint{1.325974in}{1.099939in}}%
\pgfpathlineto{\pgfqpoint{1.325974in}{1.109657in}}%
\pgfpathlineto{\pgfqpoint{1.329551in}{1.109657in}}%
\pgfpathlineto{\pgfqpoint{1.329551in}{1.119386in}}%
\pgfpathlineto{\pgfqpoint{1.333128in}{1.119386in}}%
\pgfpathlineto{\pgfqpoint{1.333128in}{1.129104in}}%
\pgfpathlineto{\pgfqpoint{1.336705in}{1.129104in}}%
\pgfpathlineto{\pgfqpoint{1.336705in}{1.138835in}}%
\pgfpathlineto{\pgfqpoint{1.340282in}{1.138835in}}%
\pgfpathlineto{\pgfqpoint{1.340282in}{1.148597in}}%
\pgfpathlineto{\pgfqpoint{1.343858in}{1.148597in}}%
\pgfpathlineto{\pgfqpoint{1.343858in}{1.158368in}}%
\pgfpathlineto{\pgfqpoint{1.347435in}{1.158368in}}%
\pgfpathlineto{\pgfqpoint{1.347435in}{1.168123in}}%
\pgfpathlineto{\pgfqpoint{1.351012in}{1.168123in}}%
\pgfpathlineto{\pgfqpoint{1.351012in}{1.177871in}}%
\pgfpathlineto{\pgfqpoint{1.354589in}{1.177871in}}%
\pgfpathlineto{\pgfqpoint{1.354589in}{1.187611in}}%
\pgfpathlineto{\pgfqpoint{1.358166in}{1.187611in}}%
\pgfpathlineto{\pgfqpoint{1.358166in}{1.197326in}}%
\pgfpathlineto{\pgfqpoint{1.361743in}{1.197326in}}%
\pgfpathlineto{\pgfqpoint{1.361743in}{1.207026in}}%
\pgfpathlineto{\pgfqpoint{1.365320in}{1.207026in}}%
\pgfpathlineto{\pgfqpoint{1.365320in}{1.216715in}}%
\pgfpathlineto{\pgfqpoint{1.368897in}{1.216715in}}%
\pgfpathlineto{\pgfqpoint{1.368897in}{1.226354in}}%
\pgfpathlineto{\pgfqpoint{1.372474in}{1.226354in}}%
\pgfpathlineto{\pgfqpoint{1.372474in}{1.235974in}}%
\pgfpathlineto{\pgfqpoint{1.376051in}{1.235974in}}%
\pgfpathlineto{\pgfqpoint{1.376051in}{1.245562in}}%
\pgfpathlineto{\pgfqpoint{1.379628in}{1.245562in}}%
\pgfpathlineto{\pgfqpoint{1.379628in}{1.255164in}}%
\pgfpathlineto{\pgfqpoint{1.383205in}{1.255164in}}%
\pgfpathlineto{\pgfqpoint{1.383205in}{1.264718in}}%
\pgfpathlineto{\pgfqpoint{1.386782in}{1.264718in}}%
\pgfpathlineto{\pgfqpoint{1.386782in}{1.274218in}}%
\pgfpathlineto{\pgfqpoint{1.390358in}{1.274218in}}%
\pgfpathlineto{\pgfqpoint{1.390358in}{1.283665in}}%
\pgfpathlineto{\pgfqpoint{1.393935in}{1.283665in}}%
\pgfpathlineto{\pgfqpoint{1.393935in}{1.293082in}}%
\pgfpathlineto{\pgfqpoint{1.397512in}{1.293082in}}%
\pgfpathlineto{\pgfqpoint{1.397512in}{1.302432in}}%
\pgfpathlineto{\pgfqpoint{1.401089in}{1.302432in}}%
\pgfpathlineto{\pgfqpoint{1.401089in}{1.311765in}}%
\pgfpathlineto{\pgfqpoint{1.404666in}{1.311765in}}%
\pgfpathlineto{\pgfqpoint{1.404666in}{1.321029in}}%
\pgfpathlineto{\pgfqpoint{1.408243in}{1.321029in}}%
\pgfpathlineto{\pgfqpoint{1.408243in}{1.330236in}}%
\pgfpathlineto{\pgfqpoint{1.411820in}{1.330236in}}%
\pgfpathlineto{\pgfqpoint{1.411820in}{1.339394in}}%
\pgfpathlineto{\pgfqpoint{1.415397in}{1.339394in}}%
\pgfpathlineto{\pgfqpoint{1.415397in}{1.348508in}}%
\pgfpathlineto{\pgfqpoint{1.418974in}{1.348508in}}%
\pgfpathlineto{\pgfqpoint{1.418974in}{1.357513in}}%
\pgfpathlineto{\pgfqpoint{1.422551in}{1.357513in}}%
\pgfpathlineto{\pgfqpoint{1.422551in}{1.366509in}}%
\pgfpathlineto{\pgfqpoint{1.426128in}{1.366509in}}%
\pgfpathlineto{\pgfqpoint{1.426128in}{1.375437in}}%
\pgfpathlineto{\pgfqpoint{1.429705in}{1.375437in}}%
\pgfpathlineto{\pgfqpoint{1.429705in}{1.384298in}}%
\pgfpathlineto{\pgfqpoint{1.433282in}{1.384298in}}%
\pgfpathlineto{\pgfqpoint{1.433282in}{1.393061in}}%
\pgfpathlineto{\pgfqpoint{1.436858in}{1.393061in}}%
\pgfpathlineto{\pgfqpoint{1.436858in}{1.401733in}}%
\pgfpathlineto{\pgfqpoint{1.440435in}{1.401733in}}%
\pgfpathlineto{\pgfqpoint{1.440435in}{1.410371in}}%
\pgfpathlineto{\pgfqpoint{1.444012in}{1.410371in}}%
\pgfpathlineto{\pgfqpoint{1.444012in}{1.418937in}}%
\pgfpathlineto{\pgfqpoint{1.447589in}{1.418937in}}%
\pgfpathlineto{\pgfqpoint{1.447589in}{1.427422in}}%
\pgfpathlineto{\pgfqpoint{1.451166in}{1.427422in}}%
\pgfpathlineto{\pgfqpoint{1.451166in}{1.435794in}}%
\pgfpathlineto{\pgfqpoint{1.454743in}{1.435794in}}%
\pgfpathlineto{\pgfqpoint{1.454743in}{1.444113in}}%
\pgfpathlineto{\pgfqpoint{1.458320in}{1.444113in}}%
\pgfpathlineto{\pgfqpoint{1.458320in}{1.452347in}}%
\pgfpathlineto{\pgfqpoint{1.461897in}{1.452347in}}%
\pgfpathlineto{\pgfqpoint{1.461897in}{1.460502in}}%
\pgfpathlineto{\pgfqpoint{1.465474in}{1.460502in}}%
\pgfpathlineto{\pgfqpoint{1.465474in}{1.468536in}}%
\pgfpathlineto{\pgfqpoint{1.469051in}{1.468536in}}%
\pgfpathlineto{\pgfqpoint{1.469051in}{1.476522in}}%
\pgfpathlineto{\pgfqpoint{1.472628in}{1.476522in}}%
\pgfpathlineto{\pgfqpoint{1.472628in}{1.484410in}}%
\pgfpathlineto{\pgfqpoint{1.476205in}{1.484410in}}%
\pgfpathlineto{\pgfqpoint{1.476205in}{1.492196in}}%
\pgfpathlineto{\pgfqpoint{1.479782in}{1.492196in}}%
\pgfpathlineto{\pgfqpoint{1.479782in}{1.499903in}}%
\pgfpathlineto{\pgfqpoint{1.483358in}{1.499903in}}%
\pgfpathlineto{\pgfqpoint{1.483358in}{1.507534in}}%
\pgfpathlineto{\pgfqpoint{1.486935in}{1.507534in}}%
\pgfpathlineto{\pgfqpoint{1.486935in}{1.515080in}}%
\pgfpathlineto{\pgfqpoint{1.490512in}{1.515080in}}%
\pgfpathlineto{\pgfqpoint{1.490512in}{1.522497in}}%
\pgfpathlineto{\pgfqpoint{1.494089in}{1.522497in}}%
\pgfpathlineto{\pgfqpoint{1.494089in}{1.529833in}}%
\pgfpathlineto{\pgfqpoint{1.497666in}{1.529833in}}%
\pgfpathlineto{\pgfqpoint{1.497666in}{1.537065in}}%
\pgfpathlineto{\pgfqpoint{1.501243in}{1.537065in}}%
\pgfpathlineto{\pgfqpoint{1.501243in}{1.544195in}}%
\pgfpathlineto{\pgfqpoint{1.504820in}{1.544195in}}%
\pgfpathlineto{\pgfqpoint{1.504820in}{1.551225in}}%
\pgfpathlineto{\pgfqpoint{1.508397in}{1.551225in}}%
\pgfpathlineto{\pgfqpoint{1.508397in}{1.558157in}}%
\pgfpathlineto{\pgfqpoint{1.511974in}{1.558157in}}%
\pgfpathlineto{\pgfqpoint{1.511974in}{1.564993in}}%
\pgfpathlineto{\pgfqpoint{1.515551in}{1.564993in}}%
\pgfpathlineto{\pgfqpoint{1.515551in}{1.571744in}}%
\pgfpathlineto{\pgfqpoint{1.519128in}{1.571744in}}%
\pgfpathlineto{\pgfqpoint{1.519128in}{1.578377in}}%
\pgfpathlineto{\pgfqpoint{1.522705in}{1.578377in}}%
\pgfpathlineto{\pgfqpoint{1.522705in}{1.584914in}}%
\pgfpathlineto{\pgfqpoint{1.526282in}{1.584914in}}%
\pgfpathlineto{\pgfqpoint{1.526282in}{1.591385in}}%
\pgfpathlineto{\pgfqpoint{1.529858in}{1.591385in}}%
\pgfpathlineto{\pgfqpoint{1.529858in}{1.597737in}}%
\pgfpathlineto{\pgfqpoint{1.533435in}{1.597737in}}%
\pgfpathlineto{\pgfqpoint{1.533435in}{1.603988in}}%
\pgfpathlineto{\pgfqpoint{1.537012in}{1.603988in}}%
\pgfpathlineto{\pgfqpoint{1.537012in}{1.610163in}}%
\pgfpathlineto{\pgfqpoint{1.540589in}{1.610163in}}%
\pgfpathlineto{\pgfqpoint{1.540589in}{1.616226in}}%
\pgfpathlineto{\pgfqpoint{1.544166in}{1.616226in}}%
\pgfpathlineto{\pgfqpoint{1.544166in}{1.622195in}}%
\pgfpathlineto{\pgfqpoint{1.547743in}{1.622195in}}%
\pgfpathlineto{\pgfqpoint{1.547743in}{1.628088in}}%
\pgfpathlineto{\pgfqpoint{1.551320in}{1.628088in}}%
\pgfpathlineto{\pgfqpoint{1.551320in}{1.633876in}}%
\pgfpathlineto{\pgfqpoint{1.554897in}{1.633876in}}%
\pgfpathlineto{\pgfqpoint{1.554897in}{1.639564in}}%
\pgfpathlineto{\pgfqpoint{1.558474in}{1.639564in}}%
\pgfpathlineto{\pgfqpoint{1.558474in}{1.645165in}}%
\pgfpathlineto{\pgfqpoint{1.562051in}{1.645165in}}%
\pgfpathlineto{\pgfqpoint{1.562051in}{1.650666in}}%
\pgfpathlineto{\pgfqpoint{1.565628in}{1.650666in}}%
\pgfpathlineto{\pgfqpoint{1.565628in}{1.656074in}}%
\pgfpathlineto{\pgfqpoint{1.569205in}{1.656074in}}%
\pgfpathlineto{\pgfqpoint{1.569205in}{1.661384in}}%
\pgfpathlineto{\pgfqpoint{1.572782in}{1.661384in}}%
\pgfpathlineto{\pgfqpoint{1.572782in}{1.666641in}}%
\pgfpathlineto{\pgfqpoint{1.576358in}{1.666641in}}%
\pgfpathlineto{\pgfqpoint{1.576358in}{1.671766in}}%
\pgfpathlineto{\pgfqpoint{1.579935in}{1.671766in}}%
\pgfpathlineto{\pgfqpoint{1.579935in}{1.676825in}}%
\pgfpathlineto{\pgfqpoint{1.583512in}{1.676825in}}%
\pgfpathlineto{\pgfqpoint{1.583512in}{1.681796in}}%
\pgfpathlineto{\pgfqpoint{1.587089in}{1.681796in}}%
\pgfpathlineto{\pgfqpoint{1.587089in}{1.686653in}}%
\pgfpathlineto{\pgfqpoint{1.590666in}{1.686653in}}%
\pgfpathlineto{\pgfqpoint{1.590666in}{1.691434in}}%
\pgfpathlineto{\pgfqpoint{1.594243in}{1.691434in}}%
\pgfpathlineto{\pgfqpoint{1.594243in}{1.696137in}}%
\pgfpathlineto{\pgfqpoint{1.597820in}{1.696137in}}%
\pgfpathlineto{\pgfqpoint{1.597820in}{1.700779in}}%
\pgfpathlineto{\pgfqpoint{1.601397in}{1.700779in}}%
\pgfpathlineto{\pgfqpoint{1.601397in}{1.705308in}}%
\pgfpathlineto{\pgfqpoint{1.604974in}{1.705308in}}%
\pgfpathlineto{\pgfqpoint{1.604974in}{1.709756in}}%
\pgfpathlineto{\pgfqpoint{1.608551in}{1.709756in}}%
\pgfpathlineto{\pgfqpoint{1.608551in}{1.714130in}}%
\pgfpathlineto{\pgfqpoint{1.612128in}{1.714130in}}%
\pgfpathlineto{\pgfqpoint{1.612128in}{1.718419in}}%
\pgfpathlineto{\pgfqpoint{1.615705in}{1.718419in}}%
\pgfpathlineto{\pgfqpoint{1.615705in}{1.722619in}}%
\pgfpathlineto{\pgfqpoint{1.619282in}{1.722619in}}%
\pgfpathlineto{\pgfqpoint{1.619282in}{1.726748in}}%
\pgfpathlineto{\pgfqpoint{1.622858in}{1.726748in}}%
\pgfpathlineto{\pgfqpoint{1.622858in}{1.730796in}}%
\pgfpathlineto{\pgfqpoint{1.626435in}{1.730796in}}%
\pgfpathlineto{\pgfqpoint{1.626435in}{1.734765in}}%
\pgfpathlineto{\pgfqpoint{1.630012in}{1.734765in}}%
\pgfpathlineto{\pgfqpoint{1.630012in}{1.738669in}}%
\pgfpathlineto{\pgfqpoint{1.633589in}{1.738669in}}%
\pgfpathlineto{\pgfqpoint{1.633589in}{1.742506in}}%
\pgfpathlineto{\pgfqpoint{1.637166in}{1.742506in}}%
\pgfpathlineto{\pgfqpoint{1.637166in}{1.746257in}}%
\pgfpathlineto{\pgfqpoint{1.640743in}{1.746257in}}%
\pgfpathlineto{\pgfqpoint{1.640743in}{1.749937in}}%
\pgfpathlineto{\pgfqpoint{1.644320in}{1.749937in}}%
\pgfpathlineto{\pgfqpoint{1.644320in}{1.753546in}}%
\pgfpathlineto{\pgfqpoint{1.647897in}{1.753546in}}%
\pgfpathlineto{\pgfqpoint{1.647897in}{1.757103in}}%
\pgfpathlineto{\pgfqpoint{1.651474in}{1.757103in}}%
\pgfpathlineto{\pgfqpoint{1.651474in}{1.760596in}}%
\pgfpathlineto{\pgfqpoint{1.655051in}{1.760596in}}%
\pgfpathlineto{\pgfqpoint{1.655051in}{1.764016in}}%
\pgfpathlineto{\pgfqpoint{1.658628in}{1.764016in}}%
\pgfpathlineto{\pgfqpoint{1.658628in}{1.767352in}}%
\pgfpathlineto{\pgfqpoint{1.662205in}{1.767352in}}%
\pgfpathlineto{\pgfqpoint{1.662205in}{1.770647in}}%
\pgfpathlineto{\pgfqpoint{1.665782in}{1.770647in}}%
\pgfpathlineto{\pgfqpoint{1.665782in}{1.773861in}}%
\pgfpathlineto{\pgfqpoint{1.669358in}{1.773861in}}%
\pgfpathlineto{\pgfqpoint{1.669358in}{1.777025in}}%
\pgfpathlineto{\pgfqpoint{1.672935in}{1.777025in}}%
\pgfpathlineto{\pgfqpoint{1.672935in}{1.780127in}}%
\pgfpathlineto{\pgfqpoint{1.676512in}{1.780127in}}%
\pgfpathlineto{\pgfqpoint{1.676512in}{1.783159in}}%
\pgfpathlineto{\pgfqpoint{1.680089in}{1.783159in}}%
\pgfpathlineto{\pgfqpoint{1.680089in}{1.786122in}}%
\pgfpathlineto{\pgfqpoint{1.683666in}{1.786122in}}%
\pgfpathlineto{\pgfqpoint{1.683666in}{1.789062in}}%
\pgfpathlineto{\pgfqpoint{1.687243in}{1.789062in}}%
\pgfpathlineto{\pgfqpoint{1.687243in}{1.791953in}}%
\pgfpathlineto{\pgfqpoint{1.690820in}{1.791953in}}%
\pgfpathlineto{\pgfqpoint{1.690820in}{1.794761in}}%
\pgfpathlineto{\pgfqpoint{1.694397in}{1.794761in}}%
\pgfpathlineto{\pgfqpoint{1.694397in}{1.797540in}}%
\pgfpathlineto{\pgfqpoint{1.697974in}{1.797540in}}%
\pgfpathlineto{\pgfqpoint{1.697974in}{1.800233in}}%
\pgfpathlineto{\pgfqpoint{1.701551in}{1.800233in}}%
\pgfpathlineto{\pgfqpoint{1.701551in}{1.802879in}}%
\pgfpathlineto{\pgfqpoint{1.705128in}{1.802879in}}%
\pgfpathlineto{\pgfqpoint{1.705128in}{1.805470in}}%
\pgfpathlineto{\pgfqpoint{1.708705in}{1.805470in}}%
\pgfpathlineto{\pgfqpoint{1.708705in}{1.808013in}}%
\pgfpathlineto{\pgfqpoint{1.712282in}{1.808013in}}%
\pgfpathlineto{\pgfqpoint{1.712282in}{1.810512in}}%
\pgfpathlineto{\pgfqpoint{1.715858in}{1.810512in}}%
\pgfpathlineto{\pgfqpoint{1.715858in}{1.812959in}}%
\pgfpathlineto{\pgfqpoint{1.719435in}{1.812959in}}%
\pgfpathlineto{\pgfqpoint{1.719435in}{1.815368in}}%
\pgfpathlineto{\pgfqpoint{1.723012in}{1.815368in}}%
\pgfpathlineto{\pgfqpoint{1.723012in}{1.817740in}}%
\pgfpathlineto{\pgfqpoint{1.726589in}{1.817740in}}%
\pgfpathlineto{\pgfqpoint{1.726589in}{1.820057in}}%
\pgfpathlineto{\pgfqpoint{1.730166in}{1.820057in}}%
\pgfpathlineto{\pgfqpoint{1.730166in}{1.822333in}}%
\pgfpathlineto{\pgfqpoint{1.733743in}{1.822333in}}%
\pgfpathlineto{\pgfqpoint{1.733743in}{1.824574in}}%
\pgfpathlineto{\pgfqpoint{1.737320in}{1.824574in}}%
\pgfpathlineto{\pgfqpoint{1.737320in}{1.826771in}}%
\pgfpathlineto{\pgfqpoint{1.740897in}{1.826771in}}%
\pgfpathlineto{\pgfqpoint{1.740897in}{1.828917in}}%
\pgfpathlineto{\pgfqpoint{1.744474in}{1.828917in}}%
\pgfpathlineto{\pgfqpoint{1.744474in}{1.831020in}}%
\pgfpathlineto{\pgfqpoint{1.748051in}{1.831020in}}%
\pgfpathlineto{\pgfqpoint{1.748051in}{1.833088in}}%
\pgfpathlineto{\pgfqpoint{1.751628in}{1.833088in}}%
\pgfpathlineto{\pgfqpoint{1.751628in}{1.835127in}}%
\pgfpathlineto{\pgfqpoint{1.755205in}{1.835127in}}%
\pgfpathlineto{\pgfqpoint{1.755205in}{1.837113in}}%
\pgfpathlineto{\pgfqpoint{1.758782in}{1.837113in}}%
\pgfpathlineto{\pgfqpoint{1.758782in}{1.839062in}}%
\pgfpathlineto{\pgfqpoint{1.762358in}{1.839062in}}%
\pgfpathlineto{\pgfqpoint{1.762358in}{1.840976in}}%
\pgfpathlineto{\pgfqpoint{1.765935in}{1.840976in}}%
\pgfpathlineto{\pgfqpoint{1.765935in}{1.842853in}}%
\pgfpathlineto{\pgfqpoint{1.769512in}{1.842853in}}%
\pgfpathlineto{\pgfqpoint{1.769512in}{1.844706in}}%
\pgfpathlineto{\pgfqpoint{1.773089in}{1.844706in}}%
\pgfpathlineto{\pgfqpoint{1.773089in}{1.846523in}}%
\pgfpathlineto{\pgfqpoint{1.776666in}{1.846523in}}%
\pgfpathlineto{\pgfqpoint{1.776666in}{1.848305in}}%
\pgfpathlineto{\pgfqpoint{1.780243in}{1.848305in}}%
\pgfpathlineto{\pgfqpoint{1.780243in}{1.850061in}}%
\pgfpathlineto{\pgfqpoint{1.783820in}{1.850061in}}%
\pgfpathlineto{\pgfqpoint{1.783820in}{1.851778in}}%
\pgfpathlineto{\pgfqpoint{1.787397in}{1.851778in}}%
\pgfpathlineto{\pgfqpoint{1.787397in}{1.853469in}}%
\pgfpathlineto{\pgfqpoint{1.790974in}{1.853469in}}%
\pgfpathlineto{\pgfqpoint{1.790974in}{1.855127in}}%
\pgfpathlineto{\pgfqpoint{1.794551in}{1.855127in}}%
\pgfpathlineto{\pgfqpoint{1.794551in}{1.856763in}}%
\pgfpathlineto{\pgfqpoint{1.798128in}{1.856763in}}%
\pgfpathlineto{\pgfqpoint{1.798128in}{1.858378in}}%
\pgfpathlineto{\pgfqpoint{1.801705in}{1.858378in}}%
\pgfpathlineto{\pgfqpoint{1.801705in}{1.859946in}}%
\pgfpathlineto{\pgfqpoint{1.805282in}{1.859946in}}%
\pgfpathlineto{\pgfqpoint{1.805282in}{1.861509in}}%
\pgfpathlineto{\pgfqpoint{1.808858in}{1.861509in}}%
\pgfpathlineto{\pgfqpoint{1.808858in}{1.863029in}}%
\pgfpathlineto{\pgfqpoint{1.812435in}{1.863029in}}%
\pgfpathlineto{\pgfqpoint{1.812435in}{1.864527in}}%
\pgfpathlineto{\pgfqpoint{1.816012in}{1.864527in}}%
\pgfpathlineto{\pgfqpoint{1.816012in}{1.866000in}}%
\pgfpathlineto{\pgfqpoint{1.819589in}{1.866000in}}%
\pgfpathlineto{\pgfqpoint{1.819589in}{1.867440in}}%
\pgfpathlineto{\pgfqpoint{1.823166in}{1.867440in}}%
\pgfpathlineto{\pgfqpoint{1.823166in}{1.868852in}}%
\pgfpathlineto{\pgfqpoint{1.826743in}{1.868852in}}%
\pgfpathlineto{\pgfqpoint{1.826743in}{1.870249in}}%
\pgfpathlineto{\pgfqpoint{1.830320in}{1.870249in}}%
\pgfpathlineto{\pgfqpoint{1.830320in}{1.871617in}}%
\pgfpathlineto{\pgfqpoint{1.833897in}{1.871617in}}%
\pgfpathlineto{\pgfqpoint{1.833897in}{1.872956in}}%
\pgfpathlineto{\pgfqpoint{1.837474in}{1.872956in}}%
\pgfpathlineto{\pgfqpoint{1.837474in}{1.874283in}}%
\pgfpathlineto{\pgfqpoint{1.841051in}{1.874283in}}%
\pgfpathlineto{\pgfqpoint{1.841051in}{1.875591in}}%
\pgfpathlineto{\pgfqpoint{1.844628in}{1.875591in}}%
\pgfpathlineto{\pgfqpoint{1.844628in}{1.876871in}}%
\pgfpathlineto{\pgfqpoint{1.848205in}{1.876871in}}%
\pgfpathlineto{\pgfqpoint{1.848205in}{1.878122in}}%
\pgfpathlineto{\pgfqpoint{1.851782in}{1.878122in}}%
\pgfpathlineto{\pgfqpoint{1.851782in}{1.879365in}}%
\pgfpathlineto{\pgfqpoint{1.855358in}{1.879365in}}%
\pgfpathlineto{\pgfqpoint{1.855358in}{1.880576in}}%
\pgfpathlineto{\pgfqpoint{1.858935in}{1.880576in}}%
\pgfpathlineto{\pgfqpoint{1.858935in}{1.881787in}}%
\pgfpathlineto{\pgfqpoint{1.862512in}{1.881787in}}%
\pgfpathlineto{\pgfqpoint{1.862512in}{1.882963in}}%
\pgfpathlineto{\pgfqpoint{1.866089in}{1.882963in}}%
\pgfpathlineto{\pgfqpoint{1.866089in}{1.884126in}}%
\pgfpathlineto{\pgfqpoint{1.869666in}{1.884126in}}%
\pgfpathlineto{\pgfqpoint{1.869666in}{1.885276in}}%
\pgfpathlineto{\pgfqpoint{1.873243in}{1.885276in}}%
\pgfpathlineto{\pgfqpoint{1.873243in}{1.886406in}}%
\pgfpathlineto{\pgfqpoint{1.876820in}{1.886406in}}%
\pgfpathlineto{\pgfqpoint{1.876820in}{1.887525in}}%
\pgfpathlineto{\pgfqpoint{1.883974in}{1.888622in}}%
\pgfpathlineto{\pgfqpoint{1.883974in}{1.889701in}}%
\pgfpathlineto{\pgfqpoint{1.891128in}{1.890763in}}%
\pgfpathlineto{\pgfqpoint{1.891128in}{1.891824in}}%
\pgfpathlineto{\pgfqpoint{1.898281in}{1.892854in}}%
\pgfpathlineto{\pgfqpoint{1.898281in}{1.893867in}}%
\pgfpathlineto{\pgfqpoint{1.905435in}{1.894862in}}%
\pgfpathlineto{\pgfqpoint{1.905435in}{1.895851in}}%
\pgfpathlineto{\pgfqpoint{1.912589in}{1.896843in}}%
\pgfpathlineto{\pgfqpoint{1.912589in}{1.897804in}}%
\pgfpathlineto{\pgfqpoint{1.919743in}{1.898756in}}%
\pgfpathlineto{\pgfqpoint{1.919743in}{1.899684in}}%
\pgfpathlineto{\pgfqpoint{1.926897in}{1.900602in}}%
\pgfpathlineto{\pgfqpoint{1.926897in}{1.901518in}}%
\pgfpathlineto{\pgfqpoint{1.934051in}{1.902414in}}%
\pgfpathlineto{\pgfqpoint{1.934051in}{1.903288in}}%
\pgfpathlineto{\pgfqpoint{1.941205in}{1.904159in}}%
\pgfpathlineto{\pgfqpoint{1.941205in}{1.905012in}}%
\pgfpathlineto{\pgfqpoint{1.948358in}{1.905862in}}%
\pgfpathlineto{\pgfqpoint{1.948358in}{1.906698in}}%
\pgfpathlineto{\pgfqpoint{1.955512in}{1.907520in}}%
\pgfpathlineto{\pgfqpoint{1.955512in}{1.908331in}}%
\pgfpathlineto{\pgfqpoint{1.962666in}{1.909133in}}%
\pgfpathlineto{\pgfqpoint{1.962666in}{1.909925in}}%
\pgfpathlineto{\pgfqpoint{1.969820in}{1.910703in}}%
\pgfpathlineto{\pgfqpoint{1.969820in}{1.911475in}}%
\pgfpathlineto{\pgfqpoint{1.976974in}{1.912239in}}%
\pgfpathlineto{\pgfqpoint{1.976974in}{1.912981in}}%
\pgfpathlineto{\pgfqpoint{1.984128in}{1.913721in}}%
\pgfpathlineto{\pgfqpoint{1.984128in}{1.914447in}}%
\pgfpathlineto{\pgfqpoint{1.991281in}{1.915165in}}%
\pgfpathlineto{\pgfqpoint{1.991281in}{1.915884in}}%
\pgfpathlineto{\pgfqpoint{1.998435in}{1.916584in}}%
\pgfpathlineto{\pgfqpoint{1.998435in}{1.917276in}}%
\pgfpathlineto{\pgfqpoint{2.005589in}{1.917952in}}%
\pgfpathlineto{\pgfqpoint{2.005589in}{1.918626in}}%
\pgfpathlineto{\pgfqpoint{2.012743in}{1.919290in}}%
\pgfpathlineto{\pgfqpoint{2.012743in}{1.919946in}}%
\pgfpathlineto{\pgfqpoint{2.019897in}{1.920592in}}%
\pgfpathlineto{\pgfqpoint{2.019897in}{1.921231in}}%
\pgfpathlineto{\pgfqpoint{2.027051in}{1.921866in}}%
\pgfpathlineto{\pgfqpoint{2.027051in}{1.922487in}}%
\pgfpathlineto{\pgfqpoint{2.034205in}{1.923104in}}%
\pgfpathlineto{\pgfqpoint{2.034205in}{1.923712in}}%
\pgfpathlineto{\pgfqpoint{2.041358in}{1.924320in}}%
\pgfpathlineto{\pgfqpoint{2.041358in}{1.924916in}}%
\pgfpathlineto{\pgfqpoint{2.048512in}{1.925497in}}%
\pgfpathlineto{\pgfqpoint{2.048512in}{1.926076in}}%
\pgfpathlineto{\pgfqpoint{2.055666in}{1.926651in}}%
\pgfpathlineto{\pgfqpoint{2.055666in}{1.927224in}}%
\pgfpathlineto{\pgfqpoint{2.066397in}{1.928324in}}%
\pgfpathlineto{\pgfqpoint{2.066397in}{1.928863in}}%
\pgfpathlineto{\pgfqpoint{2.077128in}{1.929932in}}%
\pgfpathlineto{\pgfqpoint{2.077128in}{1.930454in}}%
\pgfpathlineto{\pgfqpoint{2.087858in}{1.931469in}}%
\pgfpathlineto{\pgfqpoint{2.087858in}{1.931972in}}%
\pgfpathlineto{\pgfqpoint{2.098589in}{1.932961in}}%
\pgfpathlineto{\pgfqpoint{2.098589in}{1.933450in}}%
\pgfpathlineto{\pgfqpoint{2.109320in}{1.934407in}}%
\pgfpathlineto{\pgfqpoint{2.109320in}{1.934875in}}%
\pgfpathlineto{\pgfqpoint{2.120051in}{1.935791in}}%
\pgfpathlineto{\pgfqpoint{2.120051in}{1.936244in}}%
\pgfpathlineto{\pgfqpoint{2.130781in}{1.937121in}}%
\pgfpathlineto{\pgfqpoint{2.130781in}{1.937551in}}%
\pgfpathlineto{\pgfqpoint{2.141512in}{1.938411in}}%
\pgfpathlineto{\pgfqpoint{2.141512in}{1.938833in}}%
\pgfpathlineto{\pgfqpoint{2.152243in}{1.939656in}}%
\pgfpathlineto{\pgfqpoint{2.152243in}{1.940068in}}%
\pgfpathlineto{\pgfqpoint{2.162974in}{1.940866in}}%
\pgfpathlineto{\pgfqpoint{2.162974in}{1.941258in}}%
\pgfpathlineto{\pgfqpoint{2.173705in}{1.942025in}}%
\pgfpathlineto{\pgfqpoint{2.173705in}{1.942401in}}%
\pgfpathlineto{\pgfqpoint{2.188012in}{1.943509in}}%
\pgfpathlineto{\pgfqpoint{2.188012in}{1.943869in}}%
\pgfpathlineto{\pgfqpoint{2.202320in}{1.944917in}}%
\pgfpathlineto{\pgfqpoint{2.202320in}{1.945259in}}%
\pgfpathlineto{\pgfqpoint{2.216628in}{1.946269in}}%
\pgfpathlineto{\pgfqpoint{2.216628in}{1.946596in}}%
\pgfpathlineto{\pgfqpoint{2.230935in}{1.947559in}}%
\pgfpathlineto{\pgfqpoint{2.230935in}{1.947873in}}%
\pgfpathlineto{\pgfqpoint{2.245243in}{1.948800in}}%
\pgfpathlineto{\pgfqpoint{2.245243in}{1.949109in}}%
\pgfpathlineto{\pgfqpoint{2.259551in}{1.949978in}}%
\pgfpathlineto{\pgfqpoint{2.259551in}{1.950261in}}%
\pgfpathlineto{\pgfqpoint{2.273858in}{1.951101in}}%
\pgfpathlineto{\pgfqpoint{2.273858in}{1.951376in}}%
\pgfpathlineto{\pgfqpoint{2.291743in}{1.952429in}}%
\pgfpathlineto{\pgfqpoint{2.291743in}{1.952685in}}%
\pgfpathlineto{\pgfqpoint{2.309628in}{1.953681in}}%
\pgfpathlineto{\pgfqpoint{2.309628in}{1.953928in}}%
\pgfpathlineto{\pgfqpoint{2.327512in}{1.954887in}}%
\pgfpathlineto{\pgfqpoint{2.327512in}{1.955117in}}%
\pgfpathlineto{\pgfqpoint{2.345397in}{1.956012in}}%
\pgfpathlineto{\pgfqpoint{2.345397in}{1.956231in}}%
\pgfpathlineto{\pgfqpoint{2.366858in}{1.957277in}}%
\pgfpathlineto{\pgfqpoint{2.366858in}{1.957480in}}%
\pgfpathlineto{\pgfqpoint{2.388320in}{1.958469in}}%
\pgfpathlineto{\pgfqpoint{2.388320in}{1.958658in}}%
\pgfpathlineto{\pgfqpoint{2.413358in}{1.959765in}}%
\pgfpathlineto{\pgfqpoint{2.413358in}{1.959943in}}%
\pgfpathlineto{\pgfqpoint{2.438397in}{1.960976in}}%
\pgfpathlineto{\pgfqpoint{2.438397in}{1.961141in}}%
\pgfpathlineto{\pgfqpoint{2.467012in}{1.962250in}}%
\pgfpathlineto{\pgfqpoint{2.467012in}{1.962405in}}%
\pgfpathlineto{\pgfqpoint{2.495628in}{1.963433in}}%
\pgfpathlineto{\pgfqpoint{2.495628in}{1.963574in}}%
\pgfpathlineto{\pgfqpoint{2.527820in}{1.964649in}}%
\pgfpathlineto{\pgfqpoint{2.527820in}{1.964780in}}%
\pgfpathlineto{\pgfqpoint{2.560012in}{1.965783in}}%
\pgfpathlineto{\pgfqpoint{2.560012in}{1.965901in}}%
\pgfpathlineto{\pgfqpoint{2.595781in}{1.966942in}}%
\pgfpathlineto{\pgfqpoint{2.595781in}{1.967051in}}%
\pgfpathlineto{\pgfqpoint{2.635128in}{1.968112in}}%
\pgfpathlineto{\pgfqpoint{2.635128in}{1.968213in}}%
\pgfpathlineto{\pgfqpoint{2.678051in}{1.969282in}}%
\pgfpathlineto{\pgfqpoint{2.678051in}{1.969375in}}%
\pgfpathlineto{\pgfqpoint{2.724551in}{1.970440in}}%
\pgfpathlineto{\pgfqpoint{2.724551in}{1.970525in}}%
\pgfpathlineto{\pgfqpoint{2.774628in}{1.971575in}}%
\pgfpathlineto{\pgfqpoint{2.774628in}{1.971650in}}%
\pgfpathlineto{\pgfqpoint{2.828281in}{1.972693in}}%
\pgfpathlineto{\pgfqpoint{2.828281in}{1.972763in}}%
\pgfpathlineto{\pgfqpoint{2.889089in}{1.973845in}}%
\pgfpathlineto{\pgfqpoint{2.889089in}{1.973910in}}%
\pgfpathlineto{\pgfqpoint{2.957051in}{1.975018in}}%
\pgfpathlineto{\pgfqpoint{2.957051in}{1.975075in}}%
\pgfpathlineto{\pgfqpoint{3.031435in}{1.976182in}}%
\pgfpathlineto{\pgfqpoint{3.031435in}{1.976182in}}%
\pgfusepath{stroke}%
\end{pgfscope}%
\begin{pgfscope}%
\pgfsetrectcap%
\pgfsetmiterjoin%
\pgfsetlinewidth{0.803000pt}%
\definecolor{currentstroke}{rgb}{0.000000,0.000000,0.000000}%
\pgfsetstrokecolor{currentstroke}%
\pgfsetdash{}{0pt}%
\pgfpathmoveto{\pgfqpoint{0.696435in}{0.521603in}}%
\pgfpathlineto{\pgfqpoint{0.696435in}{2.061603in}}%
\pgfusepath{stroke}%
\end{pgfscope}%
\begin{pgfscope}%
\pgfsetrectcap%
\pgfsetmiterjoin%
\pgfsetlinewidth{0.803000pt}%
\definecolor{currentstroke}{rgb}{0.000000,0.000000,0.000000}%
\pgfsetstrokecolor{currentstroke}%
\pgfsetdash{}{0pt}%
\pgfpathmoveto{\pgfqpoint{3.021435in}{0.521603in}}%
\pgfpathlineto{\pgfqpoint{3.021435in}{2.061603in}}%
\pgfusepath{stroke}%
\end{pgfscope}%
\begin{pgfscope}%
\pgfsetrectcap%
\pgfsetmiterjoin%
\pgfsetlinewidth{0.803000pt}%
\definecolor{currentstroke}{rgb}{0.000000,0.000000,0.000000}%
\pgfsetstrokecolor{currentstroke}%
\pgfsetdash{}{0pt}%
\pgfpathmoveto{\pgfqpoint{0.696435in}{0.521603in}}%
\pgfpathlineto{\pgfqpoint{3.021435in}{0.521603in}}%
\pgfusepath{stroke}%
\end{pgfscope}%
\begin{pgfscope}%
\pgfsetrectcap%
\pgfsetmiterjoin%
\pgfsetlinewidth{0.803000pt}%
\definecolor{currentstroke}{rgb}{0.000000,0.000000,0.000000}%
\pgfsetstrokecolor{currentstroke}%
\pgfsetdash{}{0pt}%
\pgfpathmoveto{\pgfqpoint{0.696435in}{2.061603in}}%
\pgfpathlineto{\pgfqpoint{3.021435in}{2.061603in}}%
\pgfusepath{stroke}%
\end{pgfscope}%
\end{pgfpicture}%
\makeatother%
\endgroup%

%% file: figs/sim/waited_buckets_0_plot.pgf
\begingroup%
\makeatletter%
\begin{pgfpicture}%
\pgfpathrectangle{\pgfpointorigin}{\pgfqpoint{3.121435in}{2.161603in}}%
\pgfusepath{use as bounding box, clip}%
\begin{pgfscope}%
\pgfsetbuttcap%
\pgfsetmiterjoin%
\definecolor{currentfill}{rgb}{1.000000,1.000000,1.000000}%
\pgfsetfillcolor{currentfill}%
\pgfsetlinewidth{0.000000pt}%
\definecolor{currentstroke}{rgb}{1.000000,1.000000,1.000000}%
\pgfsetstrokecolor{currentstroke}%
\pgfsetdash{}{0pt}%
\pgfpathmoveto{\pgfqpoint{0.000000in}{0.000000in}}%
\pgfpathlineto{\pgfqpoint{3.121435in}{0.000000in}}%
\pgfpathlineto{\pgfqpoint{3.121435in}{2.161603in}}%
\pgfpathlineto{\pgfqpoint{0.000000in}{2.161603in}}%
\pgfpathclose%
\pgfusepath{fill}%
\end{pgfscope}%
\begin{pgfscope}%
\pgfsetbuttcap%
\pgfsetmiterjoin%
\definecolor{currentfill}{rgb}{1.000000,1.000000,1.000000}%
\pgfsetfillcolor{currentfill}%
\pgfsetlinewidth{0.000000pt}%
\definecolor{currentstroke}{rgb}{0.000000,0.000000,0.000000}%
\pgfsetstrokecolor{currentstroke}%
\pgfsetstrokeopacity{0.000000}%
\pgfsetdash{}{0pt}%
\pgfpathmoveto{\pgfqpoint{0.696435in}{0.521603in}}%
\pgfpathlineto{\pgfqpoint{3.021435in}{0.521603in}}%
\pgfpathlineto{\pgfqpoint{3.021435in}{2.061603in}}%
\pgfpathlineto{\pgfqpoint{0.696435in}{2.061603in}}%
\pgfpathclose%
\pgfusepath{fill}%
\end{pgfscope}%
\begin{pgfscope}%
\pgfsetbuttcap%
\pgfsetroundjoin%
\definecolor{currentfill}{rgb}{0.000000,0.000000,0.000000}%
\pgfsetfillcolor{currentfill}%
\pgfsetlinewidth{0.803000pt}%
\definecolor{currentstroke}{rgb}{0.000000,0.000000,0.000000}%
\pgfsetstrokecolor{currentstroke}%
\pgfsetdash{}{0pt}%
\pgfsys@defobject{currentmarker}{\pgfqpoint{0.000000in}{-0.048611in}}{\pgfqpoint{0.000000in}{0.000000in}}{%
\pgfpathmoveto{\pgfqpoint{0.000000in}{0.000000in}}%
\pgfpathlineto{\pgfqpoint{0.000000in}{-0.048611in}}%
\pgfusepath{stroke,fill}%
}%
\begin{pgfscope}%
\pgfsys@transformshift{0.802117in}{0.521603in}%
\pgfsys@useobject{currentmarker}{}%
\end{pgfscope}%
\end{pgfscope}%
\begin{pgfscope}%
\definecolor{textcolor}{rgb}{0.000000,0.000000,0.000000}%
\pgfsetstrokecolor{textcolor}%
\pgfsetfillcolor{textcolor}%
\pgftext[x=0.802117in,y=0.424381in,,top]{\color{textcolor}\sffamily\fontsize{10.000000}{12.000000}\selectfont 0}%
\end{pgfscope}%
\begin{pgfscope}%
\pgfsetbuttcap%
\pgfsetroundjoin%
\definecolor{currentfill}{rgb}{0.000000,0.000000,0.000000}%
\pgfsetfillcolor{currentfill}%
\pgfsetlinewidth{0.803000pt}%
\definecolor{currentstroke}{rgb}{0.000000,0.000000,0.000000}%
\pgfsetstrokecolor{currentstroke}%
\pgfsetdash{}{0pt}%
\pgfsys@defobject{currentmarker}{\pgfqpoint{0.000000in}{-0.048611in}}{\pgfqpoint{0.000000in}{0.000000in}}{%
\pgfpathmoveto{\pgfqpoint{0.000000in}{0.000000in}}%
\pgfpathlineto{\pgfqpoint{0.000000in}{-0.048611in}}%
\pgfusepath{stroke,fill}%
}%
\begin{pgfscope}%
\pgfsys@transformshift{1.406013in}{0.521603in}%
\pgfsys@useobject{currentmarker}{}%
\end{pgfscope}%
\end{pgfscope}%
\begin{pgfscope}%
\definecolor{textcolor}{rgb}{0.000000,0.000000,0.000000}%
\pgfsetstrokecolor{textcolor}%
\pgfsetfillcolor{textcolor}%
\pgftext[x=1.406013in,y=0.424381in,,top]{\color{textcolor}\sffamily\fontsize{10.000000}{12.000000}\selectfont 50}%
\end{pgfscope}%
\begin{pgfscope}%
\pgfsetbuttcap%
\pgfsetroundjoin%
\definecolor{currentfill}{rgb}{0.000000,0.000000,0.000000}%
\pgfsetfillcolor{currentfill}%
\pgfsetlinewidth{0.803000pt}%
\definecolor{currentstroke}{rgb}{0.000000,0.000000,0.000000}%
\pgfsetstrokecolor{currentstroke}%
\pgfsetdash{}{0pt}%
\pgfsys@defobject{currentmarker}{\pgfqpoint{0.000000in}{-0.048611in}}{\pgfqpoint{0.000000in}{0.000000in}}{%
\pgfpathmoveto{\pgfqpoint{0.000000in}{0.000000in}}%
\pgfpathlineto{\pgfqpoint{0.000000in}{-0.048611in}}%
\pgfusepath{stroke,fill}%
}%
\begin{pgfscope}%
\pgfsys@transformshift{2.009909in}{0.521603in}%
\pgfsys@useobject{currentmarker}{}%
\end{pgfscope}%
\end{pgfscope}%
\begin{pgfscope}%
\definecolor{textcolor}{rgb}{0.000000,0.000000,0.000000}%
\pgfsetstrokecolor{textcolor}%
\pgfsetfillcolor{textcolor}%
\pgftext[x=2.009909in,y=0.424381in,,top]{\color{textcolor}\sffamily\fontsize{10.000000}{12.000000}\selectfont 100}%
\end{pgfscope}%
\begin{pgfscope}%
\pgfsetbuttcap%
\pgfsetroundjoin%
\definecolor{currentfill}{rgb}{0.000000,0.000000,0.000000}%
\pgfsetfillcolor{currentfill}%
\pgfsetlinewidth{0.803000pt}%
\definecolor{currentstroke}{rgb}{0.000000,0.000000,0.000000}%
\pgfsetstrokecolor{currentstroke}%
\pgfsetdash{}{0pt}%
\pgfsys@defobject{currentmarker}{\pgfqpoint{0.000000in}{-0.048611in}}{\pgfqpoint{0.000000in}{0.000000in}}{%
\pgfpathmoveto{\pgfqpoint{0.000000in}{0.000000in}}%
\pgfpathlineto{\pgfqpoint{0.000000in}{-0.048611in}}%
\pgfusepath{stroke,fill}%
}%
\begin{pgfscope}%
\pgfsys@transformshift{2.613806in}{0.521603in}%
\pgfsys@useobject{currentmarker}{}%
\end{pgfscope}%
\end{pgfscope}%
\begin{pgfscope}%
\definecolor{textcolor}{rgb}{0.000000,0.000000,0.000000}%
\pgfsetstrokecolor{textcolor}%
\pgfsetfillcolor{textcolor}%
\pgftext[x=2.613806in,y=0.424381in,,top]{\color{textcolor}\sffamily\fontsize{10.000000}{12.000000}\selectfont 150}%
\end{pgfscope}%
\begin{pgfscope}%
\definecolor{textcolor}{rgb}{0.000000,0.000000,0.000000}%
\pgfsetstrokecolor{textcolor}%
\pgfsetfillcolor{textcolor}%
\pgftext[x=1.858935in,y=0.234413in,,top]{\color{textcolor}\sffamily\fontsize{10.000000}{12.000000}\selectfont Wait time in seconds}%
\end{pgfscope}%
\begin{pgfscope}%
\pgfsetbuttcap%
\pgfsetroundjoin%
\definecolor{currentfill}{rgb}{0.000000,0.000000,0.000000}%
\pgfsetfillcolor{currentfill}%
\pgfsetlinewidth{0.803000pt}%
\definecolor{currentstroke}{rgb}{0.000000,0.000000,0.000000}%
\pgfsetstrokecolor{currentstroke}%
\pgfsetdash{}{0pt}%
\pgfsys@defobject{currentmarker}{\pgfqpoint{-0.048611in}{0.000000in}}{\pgfqpoint{-0.000000in}{0.000000in}}{%
\pgfpathmoveto{\pgfqpoint{-0.000000in}{0.000000in}}%
\pgfpathlineto{\pgfqpoint{-0.048611in}{0.000000in}}%
\pgfusepath{stroke,fill}%
}%
\begin{pgfscope}%
\pgfsys@transformshift{0.696435in}{0.521603in}%
\pgfsys@useobject{currentmarker}{}%
\end{pgfscope}%
\end{pgfscope}%
\begin{pgfscope}%
\definecolor{textcolor}{rgb}{0.000000,0.000000,0.000000}%
\pgfsetstrokecolor{textcolor}%
\pgfsetfillcolor{textcolor}%
\pgftext[x=0.289968in, y=0.468842in, left, base]{\color{textcolor}\sffamily\fontsize{10.000000}{12.000000}\selectfont 0.00}%
\end{pgfscope}%
\begin{pgfscope}%
\pgfsetbuttcap%
\pgfsetroundjoin%
\definecolor{currentfill}{rgb}{0.000000,0.000000,0.000000}%
\pgfsetfillcolor{currentfill}%
\pgfsetlinewidth{0.803000pt}%
\definecolor{currentstroke}{rgb}{0.000000,0.000000,0.000000}%
\pgfsetstrokecolor{currentstroke}%
\pgfsetdash{}{0pt}%
\pgfsys@defobject{currentmarker}{\pgfqpoint{-0.048611in}{0.000000in}}{\pgfqpoint{-0.000000in}{0.000000in}}{%
\pgfpathmoveto{\pgfqpoint{-0.000000in}{0.000000in}}%
\pgfpathlineto{\pgfqpoint{-0.048611in}{0.000000in}}%
\pgfusepath{stroke,fill}%
}%
\begin{pgfscope}%
\pgfsys@transformshift{0.696435in}{0.888270in}%
\pgfsys@useobject{currentmarker}{}%
\end{pgfscope}%
\end{pgfscope}%
\begin{pgfscope}%
\definecolor{textcolor}{rgb}{0.000000,0.000000,0.000000}%
\pgfsetstrokecolor{textcolor}%
\pgfsetfillcolor{textcolor}%
\pgftext[x=0.289968in, y=0.835508in, left, base]{\color{textcolor}\sffamily\fontsize{10.000000}{12.000000}\selectfont 0.25}%
\end{pgfscope}%
\begin{pgfscope}%
\pgfsetbuttcap%
\pgfsetroundjoin%
\definecolor{currentfill}{rgb}{0.000000,0.000000,0.000000}%
\pgfsetfillcolor{currentfill}%
\pgfsetlinewidth{0.803000pt}%
\definecolor{currentstroke}{rgb}{0.000000,0.000000,0.000000}%
\pgfsetstrokecolor{currentstroke}%
\pgfsetdash{}{0pt}%
\pgfsys@defobject{currentmarker}{\pgfqpoint{-0.048611in}{0.000000in}}{\pgfqpoint{-0.000000in}{0.000000in}}{%
\pgfpathmoveto{\pgfqpoint{-0.000000in}{0.000000in}}%
\pgfpathlineto{\pgfqpoint{-0.048611in}{0.000000in}}%
\pgfusepath{stroke,fill}%
}%
\begin{pgfscope}%
\pgfsys@transformshift{0.696435in}{1.254937in}%
\pgfsys@useobject{currentmarker}{}%
\end{pgfscope}%
\end{pgfscope}%
\begin{pgfscope}%
\definecolor{textcolor}{rgb}{0.000000,0.000000,0.000000}%
\pgfsetstrokecolor{textcolor}%
\pgfsetfillcolor{textcolor}%
\pgftext[x=0.289968in, y=1.202175in, left, base]{\color{textcolor}\sffamily\fontsize{10.000000}{12.000000}\selectfont 0.50}%
\end{pgfscope}%
\begin{pgfscope}%
\pgfsetbuttcap%
\pgfsetroundjoin%
\definecolor{currentfill}{rgb}{0.000000,0.000000,0.000000}%
\pgfsetfillcolor{currentfill}%
\pgfsetlinewidth{0.803000pt}%
\definecolor{currentstroke}{rgb}{0.000000,0.000000,0.000000}%
\pgfsetstrokecolor{currentstroke}%
\pgfsetdash{}{0pt}%
\pgfsys@defobject{currentmarker}{\pgfqpoint{-0.048611in}{0.000000in}}{\pgfqpoint{-0.000000in}{0.000000in}}{%
\pgfpathmoveto{\pgfqpoint{-0.000000in}{0.000000in}}%
\pgfpathlineto{\pgfqpoint{-0.048611in}{0.000000in}}%
\pgfusepath{stroke,fill}%
}%
\begin{pgfscope}%
\pgfsys@transformshift{0.696435in}{1.621603in}%
\pgfsys@useobject{currentmarker}{}%
\end{pgfscope}%
\end{pgfscope}%
\begin{pgfscope}%
\definecolor{textcolor}{rgb}{0.000000,0.000000,0.000000}%
\pgfsetstrokecolor{textcolor}%
\pgfsetfillcolor{textcolor}%
\pgftext[x=0.289968in, y=1.568842in, left, base]{\color{textcolor}\sffamily\fontsize{10.000000}{12.000000}\selectfont 0.75}%
\end{pgfscope}%
\begin{pgfscope}%
\pgfsetbuttcap%
\pgfsetroundjoin%
\definecolor{currentfill}{rgb}{0.000000,0.000000,0.000000}%
\pgfsetfillcolor{currentfill}%
\pgfsetlinewidth{0.803000pt}%
\definecolor{currentstroke}{rgb}{0.000000,0.000000,0.000000}%
\pgfsetstrokecolor{currentstroke}%
\pgfsetdash{}{0pt}%
\pgfsys@defobject{currentmarker}{\pgfqpoint{-0.048611in}{0.000000in}}{\pgfqpoint{-0.000000in}{0.000000in}}{%
\pgfpathmoveto{\pgfqpoint{-0.000000in}{0.000000in}}%
\pgfpathlineto{\pgfqpoint{-0.048611in}{0.000000in}}%
\pgfusepath{stroke,fill}%
}%
\begin{pgfscope}%
\pgfsys@transformshift{0.696435in}{1.988270in}%
\pgfsys@useobject{currentmarker}{}%
\end{pgfscope}%
\end{pgfscope}%
\begin{pgfscope}%
\definecolor{textcolor}{rgb}{0.000000,0.000000,0.000000}%
\pgfsetstrokecolor{textcolor}%
\pgfsetfillcolor{textcolor}%
\pgftext[x=0.289968in, y=1.935508in, left, base]{\color{textcolor}\sffamily\fontsize{10.000000}{12.000000}\selectfont 1.00}%
\end{pgfscope}%
\begin{pgfscope}%
\definecolor{textcolor}{rgb}{0.000000,0.000000,0.000000}%
\pgfsetstrokecolor{textcolor}%
\pgfsetfillcolor{textcolor}%
\pgftext[x=0.234413in,y=1.291603in,,bottom,rotate=90.000000]{\color{textcolor}\sffamily\fontsize{10.000000}{12.000000}\selectfont Share of messages}%
\end{pgfscope}%
\begin{pgfscope}%
\pgfpathrectangle{\pgfqpoint{0.696435in}{0.521603in}}{\pgfqpoint{2.325000in}{1.540000in}}%
\pgfusepath{clip}%
\pgfsetbuttcap%
\pgfsetmiterjoin%
\pgfsetlinewidth{1.003750pt}%
\definecolor{currentstroke}{rgb}{0.121569,0.466667,0.705882}%
\pgfsetstrokecolor{currentstroke}%
\pgfsetdash{}{0pt}%
\pgfpathmoveto{\pgfqpoint{0.802117in}{0.521603in}}%
\pgfpathlineto{\pgfqpoint{0.802117in}{0.522537in}}%
\pgfpathlineto{\pgfqpoint{0.823254in}{0.522537in}}%
\pgfpathlineto{\pgfqpoint{0.823254in}{0.523451in}}%
\pgfpathlineto{\pgfqpoint{0.844390in}{0.523451in}}%
\pgfpathlineto{\pgfqpoint{0.844390in}{0.524683in}}%
\pgfpathlineto{\pgfqpoint{0.865526in}{0.524683in}}%
\pgfpathlineto{\pgfqpoint{0.865526in}{0.526800in}}%
\pgfpathlineto{\pgfqpoint{0.886663in}{0.526800in}}%
\pgfpathlineto{\pgfqpoint{0.886663in}{0.529101in}}%
\pgfpathlineto{\pgfqpoint{0.907799in}{0.529101in}}%
\pgfpathlineto{\pgfqpoint{0.907799in}{0.531666in}}%
\pgfpathlineto{\pgfqpoint{0.928935in}{0.531666in}}%
\pgfpathlineto{\pgfqpoint{0.928935in}{0.535508in}}%
\pgfpathlineto{\pgfqpoint{0.950072in}{0.535508in}}%
\pgfpathlineto{\pgfqpoint{0.950072in}{0.539176in}}%
\pgfpathlineto{\pgfqpoint{0.971208in}{0.539176in}}%
\pgfpathlineto{\pgfqpoint{0.971208in}{0.543830in}}%
\pgfpathlineto{\pgfqpoint{0.992345in}{0.543830in}}%
\pgfpathlineto{\pgfqpoint{0.992345in}{0.548979in}}%
\pgfpathlineto{\pgfqpoint{1.013481in}{0.548979in}}%
\pgfpathlineto{\pgfqpoint{1.013481in}{0.554484in}}%
\pgfpathlineto{\pgfqpoint{1.034617in}{0.554484in}}%
\pgfpathlineto{\pgfqpoint{1.034617in}{0.560942in}}%
\pgfpathlineto{\pgfqpoint{1.055754in}{0.560942in}}%
\pgfpathlineto{\pgfqpoint{1.055754in}{0.568201in}}%
\pgfpathlineto{\pgfqpoint{1.076890in}{0.568201in}}%
\pgfpathlineto{\pgfqpoint{1.076890in}{0.575330in}}%
\pgfpathlineto{\pgfqpoint{1.098026in}{0.575330in}}%
\pgfpathlineto{\pgfqpoint{1.098026in}{0.584005in}}%
\pgfpathlineto{\pgfqpoint{1.119163in}{0.584005in}}%
\pgfpathlineto{\pgfqpoint{1.119163in}{0.592757in}}%
\pgfpathlineto{\pgfqpoint{1.140299in}{0.592757in}}%
\pgfpathlineto{\pgfqpoint{1.140299in}{0.602073in}}%
\pgfpathlineto{\pgfqpoint{1.161435in}{0.602073in}}%
\pgfpathlineto{\pgfqpoint{1.161435in}{0.612152in}}%
\pgfpathlineto{\pgfqpoint{1.182572in}{0.612152in}}%
\pgfpathlineto{\pgfqpoint{1.182572in}{0.623014in}}%
\pgfpathlineto{\pgfqpoint{1.203708in}{0.623014in}}%
\pgfpathlineto{\pgfqpoint{1.203708in}{0.633842in}}%
\pgfpathlineto{\pgfqpoint{1.224845in}{0.633842in}}%
\pgfpathlineto{\pgfqpoint{1.224845in}{0.646602in}}%
\pgfpathlineto{\pgfqpoint{1.245981in}{0.646602in}}%
\pgfpathlineto{\pgfqpoint{1.245981in}{0.659029in}}%
\pgfpathlineto{\pgfqpoint{1.267117in}{0.659029in}}%
\pgfpathlineto{\pgfqpoint{1.267117in}{0.672876in}}%
\pgfpathlineto{\pgfqpoint{1.288254in}{0.672876in}}%
\pgfpathlineto{\pgfqpoint{1.288254in}{0.687176in}}%
\pgfpathlineto{\pgfqpoint{1.309390in}{0.687176in}}%
\pgfpathlineto{\pgfqpoint{1.309390in}{0.702786in}}%
\pgfpathlineto{\pgfqpoint{1.330526in}{0.702786in}}%
\pgfpathlineto{\pgfqpoint{1.330526in}{0.718049in}}%
\pgfpathlineto{\pgfqpoint{1.351663in}{0.718049in}}%
\pgfpathlineto{\pgfqpoint{1.351663in}{0.736003in}}%
\pgfpathlineto{\pgfqpoint{1.372799in}{0.736003in}}%
\pgfpathlineto{\pgfqpoint{1.372799in}{0.753024in}}%
\pgfpathlineto{\pgfqpoint{1.393935in}{0.753024in}}%
\pgfpathlineto{\pgfqpoint{1.393935in}{0.771557in}}%
\pgfpathlineto{\pgfqpoint{1.415072in}{0.771557in}}%
\pgfpathlineto{\pgfqpoint{1.415072in}{0.791008in}}%
\pgfpathlineto{\pgfqpoint{1.436208in}{0.791008in}}%
\pgfpathlineto{\pgfqpoint{1.436208in}{0.811464in}}%
\pgfpathlineto{\pgfqpoint{1.457345in}{0.811464in}}%
\pgfpathlineto{\pgfqpoint{1.457345in}{0.831688in}}%
\pgfpathlineto{\pgfqpoint{1.478481in}{0.831688in}}%
\pgfpathlineto{\pgfqpoint{1.478481in}{0.854671in}}%
\pgfpathlineto{\pgfqpoint{1.499617in}{0.854671in}}%
\pgfpathlineto{\pgfqpoint{1.499617in}{0.876705in}}%
\pgfpathlineto{\pgfqpoint{1.520754in}{0.876705in}}%
\pgfpathlineto{\pgfqpoint{1.520754in}{0.901698in}}%
\pgfpathlineto{\pgfqpoint{1.541890in}{0.901698in}}%
\pgfpathlineto{\pgfqpoint{1.541890in}{0.926164in}}%
\pgfpathlineto{\pgfqpoint{1.563026in}{0.926164in}}%
\pgfpathlineto{\pgfqpoint{1.563026in}{0.952455in}}%
\pgfpathlineto{\pgfqpoint{1.584163in}{0.952455in}}%
\pgfpathlineto{\pgfqpoint{1.584163in}{0.979356in}}%
\pgfpathlineto{\pgfqpoint{1.605299in}{0.979356in}}%
\pgfpathlineto{\pgfqpoint{1.605299in}{1.008419in}}%
\pgfpathlineto{\pgfqpoint{1.626435in}{1.008419in}}%
\pgfpathlineto{\pgfqpoint{1.626435in}{1.035376in}}%
\pgfpathlineto{\pgfqpoint{1.647572in}{1.035376in}}%
\pgfpathlineto{\pgfqpoint{1.647572in}{1.067309in}}%
\pgfpathlineto{\pgfqpoint{1.668708in}{1.067309in}}%
\pgfpathlineto{\pgfqpoint{1.668708in}{1.098032in}}%
\pgfpathlineto{\pgfqpoint{1.689845in}{1.098032in}}%
\pgfpathlineto{\pgfqpoint{1.689845in}{1.130456in}}%
\pgfpathlineto{\pgfqpoint{1.710981in}{1.130456in}}%
\pgfpathlineto{\pgfqpoint{1.710981in}{1.164760in}}%
\pgfpathlineto{\pgfqpoint{1.732117in}{1.164760in}}%
\pgfpathlineto{\pgfqpoint{1.732117in}{1.200787in}}%
\pgfpathlineto{\pgfqpoint{1.753254in}{1.200787in}}%
\pgfpathlineto{\pgfqpoint{1.753254in}{1.235381in}}%
\pgfpathlineto{\pgfqpoint{1.774390in}{1.235381in}}%
\pgfpathlineto{\pgfqpoint{1.774390in}{1.272325in}}%
\pgfpathlineto{\pgfqpoint{1.795526in}{1.272325in}}%
\pgfpathlineto{\pgfqpoint{1.795526in}{1.307811in}}%
\pgfpathlineto{\pgfqpoint{1.816663in}{1.307811in}}%
\pgfpathlineto{\pgfqpoint{1.816663in}{1.345182in}}%
\pgfpathlineto{\pgfqpoint{1.837799in}{1.345182in}}%
\pgfpathlineto{\pgfqpoint{1.837799in}{1.378598in}}%
\pgfpathlineto{\pgfqpoint{1.858935in}{1.378598in}}%
\pgfpathlineto{\pgfqpoint{1.858935in}{1.414398in}}%
\pgfpathlineto{\pgfqpoint{1.880072in}{1.414398in}}%
\pgfpathlineto{\pgfqpoint{1.880072in}{1.446179in}}%
\pgfpathlineto{\pgfqpoint{1.901208in}{1.446179in}}%
\pgfpathlineto{\pgfqpoint{1.901208in}{1.476337in}}%
\pgfpathlineto{\pgfqpoint{1.922345in}{1.476337in}}%
\pgfpathlineto{\pgfqpoint{1.922345in}{1.506265in}}%
\pgfpathlineto{\pgfqpoint{1.943481in}{1.506265in}}%
\pgfpathlineto{\pgfqpoint{1.943481in}{1.535106in}}%
\pgfpathlineto{\pgfqpoint{1.964617in}{1.535106in}}%
\pgfpathlineto{\pgfqpoint{1.964617in}{1.561202in}}%
\pgfpathlineto{\pgfqpoint{1.985754in}{1.561202in}}%
\pgfpathlineto{\pgfqpoint{1.985754in}{1.589771in}}%
\pgfpathlineto{\pgfqpoint{2.006890in}{1.589771in}}%
\pgfpathlineto{\pgfqpoint{2.006890in}{1.614049in}}%
\pgfpathlineto{\pgfqpoint{2.028026in}{1.614049in}}%
\pgfpathlineto{\pgfqpoint{2.028026in}{1.639386in}}%
\pgfpathlineto{\pgfqpoint{2.049163in}{1.639386in}}%
\pgfpathlineto{\pgfqpoint{2.049163in}{1.663586in}}%
\pgfpathlineto{\pgfqpoint{2.070299in}{1.663586in}}%
\pgfpathlineto{\pgfqpoint{2.070299in}{1.686422in}}%
\pgfpathlineto{\pgfqpoint{2.091435in}{1.686422in}}%
\pgfpathlineto{\pgfqpoint{2.091435in}{1.707844in}}%
\pgfpathlineto{\pgfqpoint{2.112572in}{1.707844in}}%
\pgfpathlineto{\pgfqpoint{2.112572in}{1.730371in}}%
\pgfpathlineto{\pgfqpoint{2.133708in}{1.730371in}}%
\pgfpathlineto{\pgfqpoint{2.133708in}{1.749366in}}%
\pgfpathlineto{\pgfqpoint{2.154845in}{1.749366in}}%
\pgfpathlineto{\pgfqpoint{2.154845in}{1.769470in}}%
\pgfpathlineto{\pgfqpoint{2.175981in}{1.769470in}}%
\pgfpathlineto{\pgfqpoint{2.175981in}{1.787930in}}%
\pgfpathlineto{\pgfqpoint{2.197117in}{1.787930in}}%
\pgfpathlineto{\pgfqpoint{2.197117in}{1.805570in}}%
\pgfpathlineto{\pgfqpoint{2.218254in}{1.805570in}}%
\pgfpathlineto{\pgfqpoint{2.218254in}{1.822068in}}%
\pgfpathlineto{\pgfqpoint{2.239390in}{1.822068in}}%
\pgfpathlineto{\pgfqpoint{2.239390in}{1.838302in}}%
\pgfpathlineto{\pgfqpoint{2.260526in}{1.838302in}}%
\pgfpathlineto{\pgfqpoint{2.260526in}{1.851964in}}%
\pgfpathlineto{\pgfqpoint{2.281663in}{1.851964in}}%
\pgfpathlineto{\pgfqpoint{2.281663in}{1.866842in}}%
\pgfpathlineto{\pgfqpoint{2.302799in}{1.866842in}}%
\pgfpathlineto{\pgfqpoint{2.302799in}{1.878746in}}%
\pgfpathlineto{\pgfqpoint{2.323935in}{1.878746in}}%
\pgfpathlineto{\pgfqpoint{2.323935in}{1.890515in}}%
\pgfpathlineto{\pgfqpoint{2.345072in}{1.890515in}}%
\pgfpathlineto{\pgfqpoint{2.345072in}{1.901810in}}%
\pgfpathlineto{\pgfqpoint{2.366208in}{1.901810in}}%
\pgfpathlineto{\pgfqpoint{2.366208in}{1.911579in}}%
\pgfpathlineto{\pgfqpoint{2.387345in}{1.911579in}}%
\pgfpathlineto{\pgfqpoint{2.387345in}{1.920361in}}%
\pgfpathlineto{\pgfqpoint{2.408481in}{1.920361in}}%
\pgfpathlineto{\pgfqpoint{2.408481in}{1.929674in}}%
\pgfpathlineto{\pgfqpoint{2.429617in}{1.929674in}}%
\pgfpathlineto{\pgfqpoint{2.429617in}{1.936565in}}%
\pgfpathlineto{\pgfqpoint{2.450754in}{1.936565in}}%
\pgfpathlineto{\pgfqpoint{2.450754in}{1.943721in}}%
\pgfpathlineto{\pgfqpoint{2.471890in}{1.943721in}}%
\pgfpathlineto{\pgfqpoint{2.471890in}{1.950494in}}%
\pgfpathlineto{\pgfqpoint{2.493026in}{1.950494in}}%
\pgfpathlineto{\pgfqpoint{2.493026in}{1.955630in}}%
\pgfpathlineto{\pgfqpoint{2.514163in}{1.955630in}}%
\pgfpathlineto{\pgfqpoint{2.514163in}{1.960620in}}%
\pgfpathlineto{\pgfqpoint{2.535299in}{1.960620in}}%
\pgfpathlineto{\pgfqpoint{2.535299in}{1.965595in}}%
\pgfpathlineto{\pgfqpoint{2.556435in}{1.965595in}}%
\pgfpathlineto{\pgfqpoint{2.556435in}{1.968928in}}%
\pgfpathlineto{\pgfqpoint{2.577572in}{1.968928in}}%
\pgfpathlineto{\pgfqpoint{2.577572in}{1.972576in}}%
\pgfpathlineto{\pgfqpoint{2.598708in}{1.972576in}}%
\pgfpathlineto{\pgfqpoint{2.598708in}{1.975653in}}%
\pgfpathlineto{\pgfqpoint{2.619845in}{1.975653in}}%
\pgfpathlineto{\pgfqpoint{2.619845in}{1.977930in}}%
\pgfpathlineto{\pgfqpoint{2.640981in}{1.977930in}}%
\pgfpathlineto{\pgfqpoint{2.640981in}{1.980203in}}%
\pgfpathlineto{\pgfqpoint{2.662117in}{1.980203in}}%
\pgfpathlineto{\pgfqpoint{2.662117in}{1.982182in}}%
\pgfpathlineto{\pgfqpoint{2.683254in}{1.982182in}}%
\pgfpathlineto{\pgfqpoint{2.683254in}{1.983470in}}%
\pgfpathlineto{\pgfqpoint{2.704390in}{1.983470in}}%
\pgfpathlineto{\pgfqpoint{2.704390in}{1.984901in}}%
\pgfpathlineto{\pgfqpoint{2.725526in}{1.984901in}}%
\pgfpathlineto{\pgfqpoint{2.725526in}{1.985831in}}%
\pgfpathlineto{\pgfqpoint{2.746663in}{1.985831in}}%
\pgfpathlineto{\pgfqpoint{2.746663in}{1.986536in}}%
\pgfpathlineto{\pgfqpoint{2.767799in}{1.986536in}}%
\pgfpathlineto{\pgfqpoint{2.767799in}{1.987231in}}%
\pgfpathlineto{\pgfqpoint{2.788935in}{1.987231in}}%
\pgfpathlineto{\pgfqpoint{2.788935in}{1.987604in}}%
\pgfpathlineto{\pgfqpoint{2.810072in}{1.987604in}}%
\pgfpathlineto{\pgfqpoint{2.810072in}{1.987849in}}%
\pgfpathlineto{\pgfqpoint{2.831208in}{1.987849in}}%
\pgfpathlineto{\pgfqpoint{2.831208in}{1.988131in}}%
\pgfpathlineto{\pgfqpoint{2.894617in}{1.988225in}}%
\pgfpathlineto{\pgfqpoint{2.894617in}{1.988270in}}%
\pgfpathlineto{\pgfqpoint{2.915754in}{1.988270in}}%
\pgfpathlineto{\pgfqpoint{2.915754in}{0.521603in}}%
\pgfpathlineto{\pgfqpoint{2.915754in}{0.521603in}}%
\pgfusepath{stroke}%
\end{pgfscope}%
\begin{pgfscope}%
\pgfpathrectangle{\pgfqpoint{0.696435in}{0.521603in}}{\pgfqpoint{2.325000in}{1.540000in}}%
\pgfusepath{clip}%
\pgfsetbuttcap%
\pgfsetmiterjoin%
\pgfsetlinewidth{1.003750pt}%
\definecolor{currentstroke}{rgb}{1.000000,0.498039,0.054902}%
\pgfsetstrokecolor{currentstroke}%
\pgfsetdash{}{0pt}%
\pgfpathmoveto{\pgfqpoint{0.802117in}{0.521603in}}%
\pgfpathlineto{\pgfqpoint{0.802117in}{0.523260in}}%
\pgfpathlineto{\pgfqpoint{0.822626in}{0.523260in}}%
\pgfpathlineto{\pgfqpoint{0.822626in}{0.524959in}}%
\pgfpathlineto{\pgfqpoint{0.843134in}{0.524959in}}%
\pgfpathlineto{\pgfqpoint{0.843134in}{0.526798in}}%
\pgfpathlineto{\pgfqpoint{0.863642in}{0.526798in}}%
\pgfpathlineto{\pgfqpoint{0.863642in}{0.530746in}}%
\pgfpathlineto{\pgfqpoint{0.884150in}{0.530746in}}%
\pgfpathlineto{\pgfqpoint{0.884150in}{0.534692in}}%
\pgfpathlineto{\pgfqpoint{0.904659in}{0.534692in}}%
\pgfpathlineto{\pgfqpoint{0.904659in}{0.539263in}}%
\pgfpathlineto{\pgfqpoint{0.925167in}{0.539263in}}%
\pgfpathlineto{\pgfqpoint{0.925167in}{0.546248in}}%
\pgfpathlineto{\pgfqpoint{0.945675in}{0.546248in}}%
\pgfpathlineto{\pgfqpoint{0.945675in}{0.553549in}}%
\pgfpathlineto{\pgfqpoint{0.966184in}{0.553549in}}%
\pgfpathlineto{\pgfqpoint{0.966184in}{0.561650in}}%
\pgfpathlineto{\pgfqpoint{0.986692in}{0.561650in}}%
\pgfpathlineto{\pgfqpoint{0.986692in}{0.572843in}}%
\pgfpathlineto{\pgfqpoint{1.007200in}{0.572843in}}%
\pgfpathlineto{\pgfqpoint{1.007200in}{0.584021in}}%
\pgfpathlineto{\pgfqpoint{1.027709in}{0.584021in}}%
\pgfpathlineto{\pgfqpoint{1.027709in}{0.596689in}}%
\pgfpathlineto{\pgfqpoint{1.048217in}{0.596689in}}%
\pgfpathlineto{\pgfqpoint{1.048217in}{0.612588in}}%
\pgfpathlineto{\pgfqpoint{1.068725in}{0.612588in}}%
\pgfpathlineto{\pgfqpoint{1.068725in}{0.628126in}}%
\pgfpathlineto{\pgfqpoint{1.089234in}{0.628126in}}%
\pgfpathlineto{\pgfqpoint{1.089234in}{0.645607in}}%
\pgfpathlineto{\pgfqpoint{1.109742in}{0.645607in}}%
\pgfpathlineto{\pgfqpoint{1.109742in}{0.666052in}}%
\pgfpathlineto{\pgfqpoint{1.130250in}{0.666052in}}%
\pgfpathlineto{\pgfqpoint{1.130250in}{0.686105in}}%
\pgfpathlineto{\pgfqpoint{1.150759in}{0.686105in}}%
\pgfpathlineto{\pgfqpoint{1.150759in}{0.707364in}}%
\pgfpathlineto{\pgfqpoint{1.171267in}{0.707364in}}%
\pgfpathlineto{\pgfqpoint{1.171267in}{0.730888in}}%
\pgfpathlineto{\pgfqpoint{1.191775in}{0.730888in}}%
\pgfpathlineto{\pgfqpoint{1.191775in}{0.753890in}}%
\pgfpathlineto{\pgfqpoint{1.212283in}{0.753890in}}%
\pgfpathlineto{\pgfqpoint{1.212283in}{0.777892in}}%
\pgfpathlineto{\pgfqpoint{1.232792in}{0.777892in}}%
\pgfpathlineto{\pgfqpoint{1.232792in}{0.802680in}}%
\pgfpathlineto{\pgfqpoint{1.253300in}{0.802680in}}%
\pgfpathlineto{\pgfqpoint{1.253300in}{0.829078in}}%
\pgfpathlineto{\pgfqpoint{1.273808in}{0.829078in}}%
\pgfpathlineto{\pgfqpoint{1.273808in}{0.854919in}}%
\pgfpathlineto{\pgfqpoint{1.294317in}{0.854919in}}%
\pgfpathlineto{\pgfqpoint{1.294317in}{0.881002in}}%
\pgfpathlineto{\pgfqpoint{1.314825in}{0.881002in}}%
\pgfpathlineto{\pgfqpoint{1.314825in}{0.909232in}}%
\pgfpathlineto{\pgfqpoint{1.335333in}{0.909232in}}%
\pgfpathlineto{\pgfqpoint{1.335333in}{0.936751in}}%
\pgfpathlineto{\pgfqpoint{1.355842in}{0.936751in}}%
\pgfpathlineto{\pgfqpoint{1.355842in}{0.965243in}}%
\pgfpathlineto{\pgfqpoint{1.376350in}{0.965243in}}%
\pgfpathlineto{\pgfqpoint{1.376350in}{0.994116in}}%
\pgfpathlineto{\pgfqpoint{1.396858in}{0.994116in}}%
\pgfpathlineto{\pgfqpoint{1.396858in}{1.022098in}}%
\pgfpathlineto{\pgfqpoint{1.417367in}{1.022098in}}%
\pgfpathlineto{\pgfqpoint{1.417367in}{1.052046in}}%
\pgfpathlineto{\pgfqpoint{1.437875in}{1.052046in}}%
\pgfpathlineto{\pgfqpoint{1.437875in}{1.081802in}}%
\pgfpathlineto{\pgfqpoint{1.458383in}{1.081802in}}%
\pgfpathlineto{\pgfqpoint{1.458383in}{1.111124in}}%
\pgfpathlineto{\pgfqpoint{1.478892in}{1.111124in}}%
\pgfpathlineto{\pgfqpoint{1.478892in}{1.141071in}}%
\pgfpathlineto{\pgfqpoint{1.499400in}{1.141071in}}%
\pgfpathlineto{\pgfqpoint{1.499400in}{1.171784in}}%
\pgfpathlineto{\pgfqpoint{1.519908in}{1.171784in}}%
\pgfpathlineto{\pgfqpoint{1.519908in}{1.203173in}}%
\pgfpathlineto{\pgfqpoint{1.540416in}{1.203173in}}%
\pgfpathlineto{\pgfqpoint{1.540416in}{1.233841in}}%
\pgfpathlineto{\pgfqpoint{1.560925in}{1.233841in}}%
\pgfpathlineto{\pgfqpoint{1.560925in}{1.265880in}}%
\pgfpathlineto{\pgfqpoint{1.581433in}{1.265880in}}%
\pgfpathlineto{\pgfqpoint{1.581433in}{1.297332in}}%
\pgfpathlineto{\pgfqpoint{1.601941in}{1.297332in}}%
\pgfpathlineto{\pgfqpoint{1.601941in}{1.327994in}}%
\pgfpathlineto{\pgfqpoint{1.622450in}{1.327994in}}%
\pgfpathlineto{\pgfqpoint{1.622450in}{1.359422in}}%
\pgfpathlineto{\pgfqpoint{1.642958in}{1.359422in}}%
\pgfpathlineto{\pgfqpoint{1.642958in}{1.389978in}}%
\pgfpathlineto{\pgfqpoint{1.663466in}{1.389978in}}%
\pgfpathlineto{\pgfqpoint{1.663466in}{1.420432in}}%
\pgfpathlineto{\pgfqpoint{1.683975in}{1.420432in}}%
\pgfpathlineto{\pgfqpoint{1.683975in}{1.450188in}}%
\pgfpathlineto{\pgfqpoint{1.704483in}{1.450188in}}%
\pgfpathlineto{\pgfqpoint{1.704483in}{1.479790in}}%
\pgfpathlineto{\pgfqpoint{1.724991in}{1.479790in}}%
\pgfpathlineto{\pgfqpoint{1.724991in}{1.509713in}}%
\pgfpathlineto{\pgfqpoint{1.745500in}{1.509713in}}%
\pgfpathlineto{\pgfqpoint{1.745500in}{1.539804in}}%
\pgfpathlineto{\pgfqpoint{1.766008in}{1.539804in}}%
\pgfpathlineto{\pgfqpoint{1.766008in}{1.569987in}}%
\pgfpathlineto{\pgfqpoint{1.786516in}{1.569987in}}%
\pgfpathlineto{\pgfqpoint{1.786516in}{1.600522in}}%
\pgfpathlineto{\pgfqpoint{1.807025in}{1.600522in}}%
\pgfpathlineto{\pgfqpoint{1.807025in}{1.629263in}}%
\pgfpathlineto{\pgfqpoint{1.827533in}{1.629263in}}%
\pgfpathlineto{\pgfqpoint{1.827533in}{1.662616in}}%
\pgfpathlineto{\pgfqpoint{1.848041in}{1.662616in}}%
\pgfpathlineto{\pgfqpoint{1.848041in}{1.695745in}}%
\pgfpathlineto{\pgfqpoint{1.868549in}{1.695745in}}%
\pgfpathlineto{\pgfqpoint{1.868549in}{1.729270in}}%
\pgfpathlineto{\pgfqpoint{1.889058in}{1.729270in}}%
\pgfpathlineto{\pgfqpoint{1.889058in}{1.755413in}}%
\pgfpathlineto{\pgfqpoint{1.909566in}{1.755413in}}%
\pgfpathlineto{\pgfqpoint{1.909566in}{1.780764in}}%
\pgfpathlineto{\pgfqpoint{1.930074in}{1.780764in}}%
\pgfpathlineto{\pgfqpoint{1.930074in}{1.806268in}}%
\pgfpathlineto{\pgfqpoint{1.950583in}{1.806268in}}%
\pgfpathlineto{\pgfqpoint{1.950583in}{1.826223in}}%
\pgfpathlineto{\pgfqpoint{1.971091in}{1.826223in}}%
\pgfpathlineto{\pgfqpoint{1.971091in}{1.845513in}}%
\pgfpathlineto{\pgfqpoint{1.991599in}{1.845513in}}%
\pgfpathlineto{\pgfqpoint{1.991599in}{1.864632in}}%
\pgfpathlineto{\pgfqpoint{2.012108in}{1.864632in}}%
\pgfpathlineto{\pgfqpoint{2.012108in}{1.879198in}}%
\pgfpathlineto{\pgfqpoint{2.032616in}{1.879198in}}%
\pgfpathlineto{\pgfqpoint{2.032616in}{1.893761in}}%
\pgfpathlineto{\pgfqpoint{2.053124in}{1.893761in}}%
\pgfpathlineto{\pgfqpoint{2.053124in}{1.907461in}}%
\pgfpathlineto{\pgfqpoint{2.073633in}{1.907461in}}%
\pgfpathlineto{\pgfqpoint{2.073633in}{1.917651in}}%
\pgfpathlineto{\pgfqpoint{2.094141in}{1.917651in}}%
\pgfpathlineto{\pgfqpoint{2.094141in}{1.927796in}}%
\pgfpathlineto{\pgfqpoint{2.114649in}{1.927796in}}%
\pgfpathlineto{\pgfqpoint{2.114649in}{1.937188in}}%
\pgfpathlineto{\pgfqpoint{2.135157in}{1.937188in}}%
\pgfpathlineto{\pgfqpoint{2.135157in}{1.944062in}}%
\pgfpathlineto{\pgfqpoint{2.155666in}{1.944062in}}%
\pgfpathlineto{\pgfqpoint{2.155666in}{1.950787in}}%
\pgfpathlineto{\pgfqpoint{2.176174in}{1.950787in}}%
\pgfpathlineto{\pgfqpoint{2.176174in}{1.957055in}}%
\pgfpathlineto{\pgfqpoint{2.196682in}{1.957055in}}%
\pgfpathlineto{\pgfqpoint{2.196682in}{1.961605in}}%
\pgfpathlineto{\pgfqpoint{2.217191in}{1.961605in}}%
\pgfpathlineto{\pgfqpoint{2.217191in}{1.966040in}}%
\pgfpathlineto{\pgfqpoint{2.237699in}{1.966040in}}%
\pgfpathlineto{\pgfqpoint{2.237699in}{1.969925in}}%
\pgfpathlineto{\pgfqpoint{2.258207in}{1.969925in}}%
\pgfpathlineto{\pgfqpoint{2.258207in}{1.972767in}}%
\pgfpathlineto{\pgfqpoint{2.278716in}{1.972767in}}%
\pgfpathlineto{\pgfqpoint{2.278716in}{1.975649in}}%
\pgfpathlineto{\pgfqpoint{2.299224in}{1.975649in}}%
\pgfpathlineto{\pgfqpoint{2.299224in}{1.978008in}}%
\pgfpathlineto{\pgfqpoint{2.319732in}{1.978008in}}%
\pgfpathlineto{\pgfqpoint{2.319732in}{1.979785in}}%
\pgfpathlineto{\pgfqpoint{2.340241in}{1.979785in}}%
\pgfpathlineto{\pgfqpoint{2.340241in}{1.981593in}}%
\pgfpathlineto{\pgfqpoint{2.360749in}{1.981593in}}%
\pgfpathlineto{\pgfqpoint{2.360749in}{1.982997in}}%
\pgfpathlineto{\pgfqpoint{2.381257in}{1.982997in}}%
\pgfpathlineto{\pgfqpoint{2.381257in}{1.984037in}}%
\pgfpathlineto{\pgfqpoint{2.401766in}{1.984037in}}%
\pgfpathlineto{\pgfqpoint{2.401766in}{1.985092in}}%
\pgfpathlineto{\pgfqpoint{2.422274in}{1.985092in}}%
\pgfpathlineto{\pgfqpoint{2.422274in}{1.985844in}}%
\pgfpathlineto{\pgfqpoint{2.442782in}{1.985844in}}%
\pgfpathlineto{\pgfqpoint{2.442782in}{1.986392in}}%
\pgfpathlineto{\pgfqpoint{2.463290in}{1.986392in}}%
\pgfpathlineto{\pgfqpoint{2.463290in}{1.986936in}}%
\pgfpathlineto{\pgfqpoint{2.483799in}{1.986936in}}%
\pgfpathlineto{\pgfqpoint{2.483799in}{1.987292in}}%
\pgfpathlineto{\pgfqpoint{2.504307in}{1.987292in}}%
\pgfpathlineto{\pgfqpoint{2.504307in}{1.987551in}}%
\pgfpathlineto{\pgfqpoint{2.524815in}{1.987551in}}%
\pgfpathlineto{\pgfqpoint{2.524815in}{1.987808in}}%
\pgfpathlineto{\pgfqpoint{2.545324in}{1.987808in}}%
\pgfpathlineto{\pgfqpoint{2.545324in}{1.987952in}}%
\pgfpathlineto{\pgfqpoint{2.586340in}{1.988045in}}%
\pgfpathlineto{\pgfqpoint{2.586340in}{1.988139in}}%
\pgfpathlineto{\pgfqpoint{2.668374in}{1.988243in}}%
\pgfpathlineto{\pgfqpoint{2.668374in}{1.988253in}}%
\pgfpathlineto{\pgfqpoint{2.852948in}{1.988270in}}%
\pgfpathlineto{\pgfqpoint{2.852948in}{0.521603in}}%
\pgfpathlineto{\pgfqpoint{2.852948in}{0.521603in}}%
\pgfusepath{stroke}%
\end{pgfscope}%
\begin{pgfscope}%
\pgfsetrectcap%
\pgfsetmiterjoin%
\pgfsetlinewidth{0.803000pt}%
\definecolor{currentstroke}{rgb}{0.000000,0.000000,0.000000}%
\pgfsetstrokecolor{currentstroke}%
\pgfsetdash{}{0pt}%
\pgfpathmoveto{\pgfqpoint{0.696435in}{0.521603in}}%
\pgfpathlineto{\pgfqpoint{0.696435in}{2.061603in}}%
\pgfusepath{stroke}%
\end{pgfscope}%
\begin{pgfscope}%
\pgfsetrectcap%
\pgfsetmiterjoin%
\pgfsetlinewidth{0.803000pt}%
\definecolor{currentstroke}{rgb}{0.000000,0.000000,0.000000}%
\pgfsetstrokecolor{currentstroke}%
\pgfsetdash{}{0pt}%
\pgfpathmoveto{\pgfqpoint{3.021435in}{0.521603in}}%
\pgfpathlineto{\pgfqpoint{3.021435in}{2.061603in}}%
\pgfusepath{stroke}%
\end{pgfscope}%
\begin{pgfscope}%
\pgfsetrectcap%
\pgfsetmiterjoin%
\pgfsetlinewidth{0.803000pt}%
\definecolor{currentstroke}{rgb}{0.000000,0.000000,0.000000}%
\pgfsetstrokecolor{currentstroke}%
\pgfsetdash{}{0pt}%
\pgfpathmoveto{\pgfqpoint{0.696435in}{0.521603in}}%
\pgfpathlineto{\pgfqpoint{3.021435in}{0.521603in}}%
\pgfusepath{stroke}%
\end{pgfscope}%
\begin{pgfscope}%
\pgfsetrectcap%
\pgfsetmiterjoin%
\pgfsetlinewidth{0.803000pt}%
\definecolor{currentstroke}{rgb}{0.000000,0.000000,0.000000}%
\pgfsetstrokecolor{currentstroke}%
\pgfsetdash{}{0pt}%
\pgfpathmoveto{\pgfqpoint{0.696435in}{2.061603in}}%
\pgfpathlineto{\pgfqpoint{3.021435in}{2.061603in}}%
\pgfusepath{stroke}%
\end{pgfscope}%
\begin{pgfscope}%
\pgfsetbuttcap%
\pgfsetmiterjoin%
\definecolor{currentfill}{rgb}{1.000000,1.000000,1.000000}%
\pgfsetfillcolor{currentfill}%
\pgfsetfillopacity{0.800000}%
\pgfsetlinewidth{1.003750pt}%
\definecolor{currentstroke}{rgb}{0.800000,0.800000,0.800000}%
\pgfsetstrokecolor{currentstroke}%
\pgfsetstrokeopacity{0.800000}%
\pgfsetdash{}{0pt}%
\pgfpathmoveto{\pgfqpoint{1.793691in}{0.569798in}}%
\pgfpathlineto{\pgfqpoint{2.953963in}{0.569798in}}%
\pgfpathquadraticcurveto{\pgfqpoint{2.973241in}{0.569798in}}{\pgfqpoint{2.973241in}{0.589076in}}%
\pgfpathlineto{\pgfqpoint{2.973241in}{0.862391in}}%
\pgfpathquadraticcurveto{\pgfqpoint{2.973241in}{0.881668in}}{\pgfqpoint{2.953963in}{0.881668in}}%
\pgfpathlineto{\pgfqpoint{1.793691in}{0.881668in}}%
\pgfpathquadraticcurveto{\pgfqpoint{1.774413in}{0.881668in}}{\pgfqpoint{1.774413in}{0.862391in}}%
\pgfpathlineto{\pgfqpoint{1.774413in}{0.589076in}}%
\pgfpathquadraticcurveto{\pgfqpoint{1.774413in}{0.569798in}}{\pgfqpoint{1.793691in}{0.569798in}}%
\pgfpathclose%
\pgfusepath{stroke,fill}%
\end{pgfscope}%
\begin{pgfscope}%
\pgfsetbuttcap%
\pgfsetmiterjoin%
\pgfsetlinewidth{1.003750pt}%
\definecolor{currentstroke}{rgb}{0.121569,0.466667,0.705882}%
\pgfsetstrokecolor{currentstroke}%
\pgfsetdash{}{0pt}%
\pgfpathmoveto{\pgfqpoint{1.812969in}{0.769880in}}%
\pgfpathlineto{\pgfqpoint{2.005747in}{0.769880in}}%
\pgfpathlineto{\pgfqpoint{2.005747in}{0.837352in}}%
\pgfpathlineto{\pgfqpoint{1.812969in}{0.837352in}}%
\pgfpathclose%
\pgfusepath{stroke}%
\end{pgfscope}%
\begin{pgfscope}%
\definecolor{textcolor}{rgb}{0.000000,0.000000,0.000000}%
\pgfsetstrokecolor{textcolor}%
\pgfsetfillcolor{textcolor}%
\pgftext[x=2.082858in,y=0.769880in,left,base]{\color{textcolor}\sffamily\fontsize{6.940000}{8.328000}\selectfont lnd}%
\end{pgfscope}%
\begin{pgfscope}%
\pgfsetbuttcap%
\pgfsetmiterjoin%
\pgfsetlinewidth{1.003750pt}%
\definecolor{currentstroke}{rgb}{1.000000,0.498039,0.054902}%
\pgfsetstrokecolor{currentstroke}%
\pgfsetdash{}{0pt}%
\pgfpathmoveto{\pgfqpoint{1.812969in}{0.628403in}}%
\pgfpathlineto{\pgfqpoint{2.005747in}{0.628403in}}%
\pgfpathlineto{\pgfqpoint{2.005747in}{0.695875in}}%
\pgfpathlineto{\pgfqpoint{1.812969in}{0.695875in}}%
\pgfpathclose%
\pgfusepath{stroke}%
\end{pgfscope}%
\begin{pgfscope}%
\definecolor{textcolor}{rgb}{0.000000,0.000000,0.000000}%
\pgfsetstrokecolor{textcolor}%
\pgfsetfillcolor{textcolor}%
\pgftext[x=2.082858in,y=0.628403in,left,base]{\color{textcolor}\sffamily\fontsize{6.940000}{8.328000}\selectfont lnd-no-keepalives}%
\end{pgfscope}%
\end{pgfpicture}%
\makeatother%
\endgroup%

%% file: figs/eval/bandwidth_conns_0_plot.pgf
\begingroup%
\makeatletter%
\begin{pgfpicture}%
\pgfpathrectangle{\pgfpointorigin}{\pgfqpoint{3.201170in}{2.161603in}}%
\pgfusepath{use as bounding box, clip}%
\begin{pgfscope}%
\pgfsetbuttcap%
\pgfsetmiterjoin%
\definecolor{currentfill}{rgb}{1.000000,1.000000,1.000000}%
\pgfsetfillcolor{currentfill}%
\pgfsetlinewidth{0.000000pt}%
\definecolor{currentstroke}{rgb}{1.000000,1.000000,1.000000}%
\pgfsetstrokecolor{currentstroke}%
\pgfsetdash{}{0pt}%
\pgfpathmoveto{\pgfqpoint{0.000000in}{0.000000in}}%
\pgfpathlineto{\pgfqpoint{3.201170in}{0.000000in}}%
\pgfpathlineto{\pgfqpoint{3.201170in}{2.161603in}}%
\pgfpathlineto{\pgfqpoint{0.000000in}{2.161603in}}%
\pgfpathclose%
\pgfusepath{fill}%
\end{pgfscope}%
\begin{pgfscope}%
\pgfsetbuttcap%
\pgfsetmiterjoin%
\definecolor{currentfill}{rgb}{1.000000,1.000000,1.000000}%
\pgfsetfillcolor{currentfill}%
\pgfsetlinewidth{0.000000pt}%
\definecolor{currentstroke}{rgb}{0.000000,0.000000,0.000000}%
\pgfsetstrokecolor{currentstroke}%
\pgfsetstrokeopacity{0.000000}%
\pgfsetdash{}{0pt}%
\pgfpathmoveto{\pgfqpoint{0.652287in}{0.521603in}}%
\pgfpathlineto{\pgfqpoint{2.977287in}{0.521603in}}%
\pgfpathlineto{\pgfqpoint{2.977287in}{2.061603in}}%
\pgfpathlineto{\pgfqpoint{0.652287in}{2.061603in}}%
\pgfpathclose%
\pgfusepath{fill}%
\end{pgfscope}%
\begin{pgfscope}%
\pgfpathrectangle{\pgfqpoint{0.652287in}{0.521603in}}{\pgfqpoint{2.325000in}{1.540000in}}%
\pgfusepath{clip}%
\pgfsetbuttcap%
\pgfsetmiterjoin%
\definecolor{currentfill}{rgb}{0.121569,0.466667,0.705882}%
\pgfsetfillcolor{currentfill}%
\pgfsetlinewidth{0.000000pt}%
\definecolor{currentstroke}{rgb}{0.000000,0.000000,0.000000}%
\pgfsetstrokecolor{currentstroke}%
\pgfsetstrokeopacity{0.000000}%
\pgfsetdash{}{0pt}%
\pgfpathmoveto{\pgfqpoint{0.757968in}{0.521603in}}%
\pgfpathlineto{\pgfqpoint{0.980456in}{0.521603in}}%
\pgfpathlineto{\pgfqpoint{0.980456in}{0.732130in}}%
\pgfpathlineto{\pgfqpoint{0.757968in}{0.732130in}}%
\pgfpathclose%
\pgfusepath{fill}%
\end{pgfscope}%
\begin{pgfscope}%
\pgfpathrectangle{\pgfqpoint{0.652287in}{0.521603in}}{\pgfqpoint{2.325000in}{1.540000in}}%
\pgfusepath{clip}%
\pgfsetbuttcap%
\pgfsetmiterjoin%
\definecolor{currentfill}{rgb}{0.121569,0.466667,0.705882}%
\pgfsetfillcolor{currentfill}%
\pgfsetlinewidth{0.000000pt}%
\definecolor{currentstroke}{rgb}{0.000000,0.000000,0.000000}%
\pgfsetstrokecolor{currentstroke}%
\pgfsetstrokeopacity{0.000000}%
\pgfsetdash{}{0pt}%
\pgfpathmoveto{\pgfqpoint{1.314188in}{0.521603in}}%
\pgfpathlineto{\pgfqpoint{1.536676in}{0.521603in}}%
\pgfpathlineto{\pgfqpoint{1.536676in}{0.914835in}}%
\pgfpathlineto{\pgfqpoint{1.314188in}{0.914835in}}%
\pgfpathclose%
\pgfusepath{fill}%
\end{pgfscope}%
\begin{pgfscope}%
\pgfpathrectangle{\pgfqpoint{0.652287in}{0.521603in}}{\pgfqpoint{2.325000in}{1.540000in}}%
\pgfusepath{clip}%
\pgfsetbuttcap%
\pgfsetmiterjoin%
\definecolor{currentfill}{rgb}{0.121569,0.466667,0.705882}%
\pgfsetfillcolor{currentfill}%
\pgfsetlinewidth{0.000000pt}%
\definecolor{currentstroke}{rgb}{0.000000,0.000000,0.000000}%
\pgfsetstrokecolor{currentstroke}%
\pgfsetstrokeopacity{0.000000}%
\pgfsetdash{}{0pt}%
\pgfpathmoveto{\pgfqpoint{1.870409in}{0.521603in}}%
\pgfpathlineto{\pgfqpoint{2.092897in}{0.521603in}}%
\pgfpathlineto{\pgfqpoint{2.092897in}{1.272854in}}%
\pgfpathlineto{\pgfqpoint{1.870409in}{1.272854in}}%
\pgfpathclose%
\pgfusepath{fill}%
\end{pgfscope}%
\begin{pgfscope}%
\pgfpathrectangle{\pgfqpoint{0.652287in}{0.521603in}}{\pgfqpoint{2.325000in}{1.540000in}}%
\pgfusepath{clip}%
\pgfsetbuttcap%
\pgfsetmiterjoin%
\definecolor{currentfill}{rgb}{0.121569,0.466667,0.705882}%
\pgfsetfillcolor{currentfill}%
\pgfsetlinewidth{0.000000pt}%
\definecolor{currentstroke}{rgb}{0.000000,0.000000,0.000000}%
\pgfsetstrokecolor{currentstroke}%
\pgfsetstrokeopacity{0.000000}%
\pgfsetdash{}{0pt}%
\pgfpathmoveto{\pgfqpoint{2.426629in}{0.521603in}}%
\pgfpathlineto{\pgfqpoint{2.649117in}{0.521603in}}%
\pgfpathlineto{\pgfqpoint{2.649117in}{1.988270in}}%
\pgfpathlineto{\pgfqpoint{2.426629in}{1.988270in}}%
\pgfpathclose%
\pgfusepath{fill}%
\end{pgfscope}%
\begin{pgfscope}%
\pgfpathrectangle{\pgfqpoint{0.652287in}{0.521603in}}{\pgfqpoint{2.325000in}{1.540000in}}%
\pgfusepath{clip}%
\pgfsetbuttcap%
\pgfsetmiterjoin%
\definecolor{currentfill}{rgb}{1.000000,0.498039,0.054902}%
\pgfsetfillcolor{currentfill}%
\pgfsetlinewidth{0.000000pt}%
\definecolor{currentstroke}{rgb}{0.000000,0.000000,0.000000}%
\pgfsetstrokecolor{currentstroke}%
\pgfsetstrokeopacity{0.000000}%
\pgfsetdash{}{0pt}%
\pgfpathmoveto{\pgfqpoint{0.980456in}{0.521603in}}%
\pgfpathlineto{\pgfqpoint{1.202944in}{0.521603in}}%
\pgfpathlineto{\pgfqpoint{1.202944in}{0.601122in}}%
\pgfpathlineto{\pgfqpoint{0.980456in}{0.601122in}}%
\pgfpathclose%
\pgfusepath{fill}%
\end{pgfscope}%
\begin{pgfscope}%
\pgfpathrectangle{\pgfqpoint{0.652287in}{0.521603in}}{\pgfqpoint{2.325000in}{1.540000in}}%
\pgfusepath{clip}%
\pgfsetbuttcap%
\pgfsetmiterjoin%
\definecolor{currentfill}{rgb}{1.000000,0.498039,0.054902}%
\pgfsetfillcolor{currentfill}%
\pgfsetlinewidth{0.000000pt}%
\definecolor{currentstroke}{rgb}{0.000000,0.000000,0.000000}%
\pgfsetstrokecolor{currentstroke}%
\pgfsetstrokeopacity{0.000000}%
\pgfsetdash{}{0pt}%
\pgfpathmoveto{\pgfqpoint{1.536676in}{0.521603in}}%
\pgfpathlineto{\pgfqpoint{1.759165in}{0.521603in}}%
\pgfpathlineto{\pgfqpoint{1.759165in}{0.603987in}}%
\pgfpathlineto{\pgfqpoint{1.536676in}{0.603987in}}%
\pgfpathclose%
\pgfusepath{fill}%
\end{pgfscope}%
\begin{pgfscope}%
\pgfpathrectangle{\pgfqpoint{0.652287in}{0.521603in}}{\pgfqpoint{2.325000in}{1.540000in}}%
\pgfusepath{clip}%
\pgfsetbuttcap%
\pgfsetmiterjoin%
\definecolor{currentfill}{rgb}{1.000000,0.498039,0.054902}%
\pgfsetfillcolor{currentfill}%
\pgfsetlinewidth{0.000000pt}%
\definecolor{currentstroke}{rgb}{0.000000,0.000000,0.000000}%
\pgfsetstrokecolor{currentstroke}%
\pgfsetstrokeopacity{0.000000}%
\pgfsetdash{}{0pt}%
\pgfpathmoveto{\pgfqpoint{2.092897in}{0.521603in}}%
\pgfpathlineto{\pgfqpoint{2.315385in}{0.521603in}}%
\pgfpathlineto{\pgfqpoint{2.315385in}{0.610672in}}%
\pgfpathlineto{\pgfqpoint{2.092897in}{0.610672in}}%
\pgfpathclose%
\pgfusepath{fill}%
\end{pgfscope}%
\begin{pgfscope}%
\pgfpathrectangle{\pgfqpoint{0.652287in}{0.521603in}}{\pgfqpoint{2.325000in}{1.540000in}}%
\pgfusepath{clip}%
\pgfsetbuttcap%
\pgfsetmiterjoin%
\definecolor{currentfill}{rgb}{1.000000,0.498039,0.054902}%
\pgfsetfillcolor{currentfill}%
\pgfsetlinewidth{0.000000pt}%
\definecolor{currentstroke}{rgb}{0.000000,0.000000,0.000000}%
\pgfsetstrokecolor{currentstroke}%
\pgfsetstrokeopacity{0.000000}%
\pgfsetdash{}{0pt}%
\pgfpathmoveto{\pgfqpoint{2.649117in}{0.521603in}}%
\pgfpathlineto{\pgfqpoint{2.871605in}{0.521603in}}%
\pgfpathlineto{\pgfqpoint{2.871605in}{0.610714in}}%
\pgfpathlineto{\pgfqpoint{2.649117in}{0.610714in}}%
\pgfpathclose%
\pgfusepath{fill}%
\end{pgfscope}%
\begin{pgfscope}%
\pgfsetbuttcap%
\pgfsetroundjoin%
\definecolor{currentfill}{rgb}{0.000000,0.000000,0.000000}%
\pgfsetfillcolor{currentfill}%
\pgfsetlinewidth{0.803000pt}%
\definecolor{currentstroke}{rgb}{0.000000,0.000000,0.000000}%
\pgfsetstrokecolor{currentstroke}%
\pgfsetdash{}{0pt}%
\pgfsys@defobject{currentmarker}{\pgfqpoint{0.000000in}{-0.048611in}}{\pgfqpoint{0.000000in}{0.000000in}}{%
\pgfpathmoveto{\pgfqpoint{0.000000in}{0.000000in}}%
\pgfpathlineto{\pgfqpoint{0.000000in}{-0.048611in}}%
\pgfusepath{stroke,fill}%
}%
\begin{pgfscope}%
\pgfsys@transformshift{0.980456in}{0.521603in}%
\pgfsys@useobject{currentmarker}{}%
\end{pgfscope}%
\end{pgfscope}%
\begin{pgfscope}%
\definecolor{textcolor}{rgb}{0.000000,0.000000,0.000000}%
\pgfsetstrokecolor{textcolor}%
\pgfsetfillcolor{textcolor}%
\pgftext[x=0.980456in,y=0.424381in,,top]{\color{textcolor}\sffamily\fontsize{10.000000}{12.000000}\selectfont 4}%
\end{pgfscope}%
\begin{pgfscope}%
\pgfsetbuttcap%
\pgfsetroundjoin%
\definecolor{currentfill}{rgb}{0.000000,0.000000,0.000000}%
\pgfsetfillcolor{currentfill}%
\pgfsetlinewidth{0.803000pt}%
\definecolor{currentstroke}{rgb}{0.000000,0.000000,0.000000}%
\pgfsetstrokecolor{currentstroke}%
\pgfsetdash{}{0pt}%
\pgfsys@defobject{currentmarker}{\pgfqpoint{0.000000in}{-0.048611in}}{\pgfqpoint{0.000000in}{0.000000in}}{%
\pgfpathmoveto{\pgfqpoint{0.000000in}{0.000000in}}%
\pgfpathlineto{\pgfqpoint{0.000000in}{-0.048611in}}%
\pgfusepath{stroke,fill}%
}%
\begin{pgfscope}%
\pgfsys@transformshift{1.536676in}{0.521603in}%
\pgfsys@useobject{currentmarker}{}%
\end{pgfscope}%
\end{pgfscope}%
\begin{pgfscope}%
\definecolor{textcolor}{rgb}{0.000000,0.000000,0.000000}%
\pgfsetstrokecolor{textcolor}%
\pgfsetfillcolor{textcolor}%
\pgftext[x=1.536676in,y=0.424381in,,top]{\color{textcolor}\sffamily\fontsize{10.000000}{12.000000}\selectfont 8}%
\end{pgfscope}%
\begin{pgfscope}%
\pgfsetbuttcap%
\pgfsetroundjoin%
\definecolor{currentfill}{rgb}{0.000000,0.000000,0.000000}%
\pgfsetfillcolor{currentfill}%
\pgfsetlinewidth{0.803000pt}%
\definecolor{currentstroke}{rgb}{0.000000,0.000000,0.000000}%
\pgfsetstrokecolor{currentstroke}%
\pgfsetdash{}{0pt}%
\pgfsys@defobject{currentmarker}{\pgfqpoint{0.000000in}{-0.048611in}}{\pgfqpoint{0.000000in}{0.000000in}}{%
\pgfpathmoveto{\pgfqpoint{0.000000in}{0.000000in}}%
\pgfpathlineto{\pgfqpoint{0.000000in}{-0.048611in}}%
\pgfusepath{stroke,fill}%
}%
\begin{pgfscope}%
\pgfsys@transformshift{2.092897in}{0.521603in}%
\pgfsys@useobject{currentmarker}{}%
\end{pgfscope}%
\end{pgfscope}%
\begin{pgfscope}%
\definecolor{textcolor}{rgb}{0.000000,0.000000,0.000000}%
\pgfsetstrokecolor{textcolor}%
\pgfsetfillcolor{textcolor}%
\pgftext[x=2.092897in,y=0.424381in,,top]{\color{textcolor}\sffamily\fontsize{10.000000}{12.000000}\selectfont 16}%
\end{pgfscope}%
\begin{pgfscope}%
\pgfsetbuttcap%
\pgfsetroundjoin%
\definecolor{currentfill}{rgb}{0.000000,0.000000,0.000000}%
\pgfsetfillcolor{currentfill}%
\pgfsetlinewidth{0.803000pt}%
\definecolor{currentstroke}{rgb}{0.000000,0.000000,0.000000}%
\pgfsetstrokecolor{currentstroke}%
\pgfsetdash{}{0pt}%
\pgfsys@defobject{currentmarker}{\pgfqpoint{0.000000in}{-0.048611in}}{\pgfqpoint{0.000000in}{0.000000in}}{%
\pgfpathmoveto{\pgfqpoint{0.000000in}{0.000000in}}%
\pgfpathlineto{\pgfqpoint{0.000000in}{-0.048611in}}%
\pgfusepath{stroke,fill}%
}%
\begin{pgfscope}%
\pgfsys@transformshift{2.649117in}{0.521603in}%
\pgfsys@useobject{currentmarker}{}%
\end{pgfscope}%
\end{pgfscope}%
\begin{pgfscope}%
\definecolor{textcolor}{rgb}{0.000000,0.000000,0.000000}%
\pgfsetstrokecolor{textcolor}%
\pgfsetfillcolor{textcolor}%
\pgftext[x=2.649117in,y=0.424381in,,top]{\color{textcolor}\sffamily\fontsize{10.000000}{12.000000}\selectfont 32}%
\end{pgfscope}%
\begin{pgfscope}%
\definecolor{textcolor}{rgb}{0.000000,0.000000,0.000000}%
\pgfsetstrokecolor{textcolor}%
\pgfsetfillcolor{textcolor}%
\pgftext[x=1.814787in,y=0.234413in,,top]{\color{textcolor}\sffamily\fontsize{10.000000}{12.000000}\selectfont Number of active gossip connections}%
\end{pgfscope}%
\begin{pgfscope}%
\pgfsetbuttcap%
\pgfsetroundjoin%
\definecolor{currentfill}{rgb}{0.000000,0.000000,0.000000}%
\pgfsetfillcolor{currentfill}%
\pgfsetlinewidth{0.803000pt}%
\definecolor{currentstroke}{rgb}{0.000000,0.000000,0.000000}%
\pgfsetstrokecolor{currentstroke}%
\pgfsetdash{}{0pt}%
\pgfsys@defobject{currentmarker}{\pgfqpoint{-0.048611in}{0.000000in}}{\pgfqpoint{-0.000000in}{0.000000in}}{%
\pgfpathmoveto{\pgfqpoint{-0.000000in}{0.000000in}}%
\pgfpathlineto{\pgfqpoint{-0.048611in}{0.000000in}}%
\pgfusepath{stroke,fill}%
}%
\begin{pgfscope}%
\pgfsys@transformshift{0.652287in}{0.521603in}%
\pgfsys@useobject{currentmarker}{}%
\end{pgfscope}%
\end{pgfscope}%
\begin{pgfscope}%
\definecolor{textcolor}{rgb}{0.000000,0.000000,0.000000}%
\pgfsetstrokecolor{textcolor}%
\pgfsetfillcolor{textcolor}%
\pgftext[x=0.466699in, y=0.468842in, left, base]{\color{textcolor}\sffamily\fontsize{10.000000}{12.000000}\selectfont 0}%
\end{pgfscope}%
\begin{pgfscope}%
\pgfsetbuttcap%
\pgfsetroundjoin%
\definecolor{currentfill}{rgb}{0.000000,0.000000,0.000000}%
\pgfsetfillcolor{currentfill}%
\pgfsetlinewidth{0.803000pt}%
\definecolor{currentstroke}{rgb}{0.000000,0.000000,0.000000}%
\pgfsetstrokecolor{currentstroke}%
\pgfsetdash{}{0pt}%
\pgfsys@defobject{currentmarker}{\pgfqpoint{-0.048611in}{0.000000in}}{\pgfqpoint{-0.000000in}{0.000000in}}{%
\pgfpathmoveto{\pgfqpoint{-0.000000in}{0.000000in}}%
\pgfpathlineto{\pgfqpoint{-0.048611in}{0.000000in}}%
\pgfusepath{stroke,fill}%
}%
\begin{pgfscope}%
\pgfsys@transformshift{0.652287in}{0.936843in}%
\pgfsys@useobject{currentmarker}{}%
\end{pgfscope}%
\end{pgfscope}%
\begin{pgfscope}%
\definecolor{textcolor}{rgb}{0.000000,0.000000,0.000000}%
\pgfsetstrokecolor{textcolor}%
\pgfsetfillcolor{textcolor}%
\pgftext[x=0.289968in, y=0.884081in, left, base]{\color{textcolor}\sffamily\fontsize{10.000000}{12.000000}\selectfont 100}%
\end{pgfscope}%
\begin{pgfscope}%
\pgfsetbuttcap%
\pgfsetroundjoin%
\definecolor{currentfill}{rgb}{0.000000,0.000000,0.000000}%
\pgfsetfillcolor{currentfill}%
\pgfsetlinewidth{0.803000pt}%
\definecolor{currentstroke}{rgb}{0.000000,0.000000,0.000000}%
\pgfsetstrokecolor{currentstroke}%
\pgfsetdash{}{0pt}%
\pgfsys@defobject{currentmarker}{\pgfqpoint{-0.048611in}{0.000000in}}{\pgfqpoint{-0.000000in}{0.000000in}}{%
\pgfpathmoveto{\pgfqpoint{-0.000000in}{0.000000in}}%
\pgfpathlineto{\pgfqpoint{-0.048611in}{0.000000in}}%
\pgfusepath{stroke,fill}%
}%
\begin{pgfscope}%
\pgfsys@transformshift{0.652287in}{1.352082in}%
\pgfsys@useobject{currentmarker}{}%
\end{pgfscope}%
\end{pgfscope}%
\begin{pgfscope}%
\definecolor{textcolor}{rgb}{0.000000,0.000000,0.000000}%
\pgfsetstrokecolor{textcolor}%
\pgfsetfillcolor{textcolor}%
\pgftext[x=0.289968in, y=1.299320in, left, base]{\color{textcolor}\sffamily\fontsize{10.000000}{12.000000}\selectfont 200}%
\end{pgfscope}%
\begin{pgfscope}%
\pgfsetbuttcap%
\pgfsetroundjoin%
\definecolor{currentfill}{rgb}{0.000000,0.000000,0.000000}%
\pgfsetfillcolor{currentfill}%
\pgfsetlinewidth{0.803000pt}%
\definecolor{currentstroke}{rgb}{0.000000,0.000000,0.000000}%
\pgfsetstrokecolor{currentstroke}%
\pgfsetdash{}{0pt}%
\pgfsys@defobject{currentmarker}{\pgfqpoint{-0.048611in}{0.000000in}}{\pgfqpoint{-0.000000in}{0.000000in}}{%
\pgfpathmoveto{\pgfqpoint{-0.000000in}{0.000000in}}%
\pgfpathlineto{\pgfqpoint{-0.048611in}{0.000000in}}%
\pgfusepath{stroke,fill}%
}%
\begin{pgfscope}%
\pgfsys@transformshift{0.652287in}{1.767321in}%
\pgfsys@useobject{currentmarker}{}%
\end{pgfscope}%
\end{pgfscope}%
\begin{pgfscope}%
\definecolor{textcolor}{rgb}{0.000000,0.000000,0.000000}%
\pgfsetstrokecolor{textcolor}%
\pgfsetfillcolor{textcolor}%
\pgftext[x=0.289968in, y=1.714560in, left, base]{\color{textcolor}\sffamily\fontsize{10.000000}{12.000000}\selectfont 300}%
\end{pgfscope}%
\begin{pgfscope}%
\definecolor{textcolor}{rgb}{0.000000,0.000000,0.000000}%
\pgfsetstrokecolor{textcolor}%
\pgfsetfillcolor{textcolor}%
\pgftext[x=0.234413in,y=1.291603in,,bottom,rotate=90.000000]{\color{textcolor}\sffamily\fontsize{10.000000}{12.000000}\selectfont Bandwidth in GB}%
\end{pgfscope}%
\begin{pgfscope}%
\pgfsetrectcap%
\pgfsetmiterjoin%
\pgfsetlinewidth{0.803000pt}%
\definecolor{currentstroke}{rgb}{0.000000,0.000000,0.000000}%
\pgfsetstrokecolor{currentstroke}%
\pgfsetdash{}{0pt}%
\pgfpathmoveto{\pgfqpoint{0.652287in}{0.521603in}}%
\pgfpathlineto{\pgfqpoint{0.652287in}{2.061603in}}%
\pgfusepath{stroke}%
\end{pgfscope}%
\begin{pgfscope}%
\pgfsetrectcap%
\pgfsetmiterjoin%
\pgfsetlinewidth{0.803000pt}%
\definecolor{currentstroke}{rgb}{0.000000,0.000000,0.000000}%
\pgfsetstrokecolor{currentstroke}%
\pgfsetdash{}{0pt}%
\pgfpathmoveto{\pgfqpoint{2.977287in}{0.521603in}}%
\pgfpathlineto{\pgfqpoint{2.977287in}{2.061603in}}%
\pgfusepath{stroke}%
\end{pgfscope}%
\begin{pgfscope}%
\pgfsetrectcap%
\pgfsetmiterjoin%
\pgfsetlinewidth{0.803000pt}%
\definecolor{currentstroke}{rgb}{0.000000,0.000000,0.000000}%
\pgfsetstrokecolor{currentstroke}%
\pgfsetdash{}{0pt}%
\pgfpathmoveto{\pgfqpoint{0.652287in}{0.521603in}}%
\pgfpathlineto{\pgfqpoint{2.977287in}{0.521603in}}%
\pgfusepath{stroke}%
\end{pgfscope}%
\begin{pgfscope}%
\pgfsetrectcap%
\pgfsetmiterjoin%
\pgfsetlinewidth{0.803000pt}%
\definecolor{currentstroke}{rgb}{0.000000,0.000000,0.000000}%
\pgfsetstrokecolor{currentstroke}%
\pgfsetdash{}{0pt}%
\pgfpathmoveto{\pgfqpoint{0.652287in}{2.061603in}}%
\pgfpathlineto{\pgfqpoint{2.977287in}{2.061603in}}%
\pgfusepath{stroke}%
\end{pgfscope}%
\begin{pgfscope}%
\pgfsetbuttcap%
\pgfsetmiterjoin%
\definecolor{currentfill}{rgb}{1.000000,1.000000,1.000000}%
\pgfsetfillcolor{currentfill}%
\pgfsetfillopacity{0.800000}%
\pgfsetlinewidth{1.003750pt}%
\definecolor{currentstroke}{rgb}{0.800000,0.800000,0.800000}%
\pgfsetstrokecolor{currentstroke}%
\pgfsetstrokeopacity{0.800000}%
\pgfsetdash{}{0pt}%
\pgfpathmoveto{\pgfqpoint{0.719759in}{1.701538in}}%
\pgfpathlineto{\pgfqpoint{1.550530in}{1.701538in}}%
\pgfpathquadraticcurveto{\pgfqpoint{1.569807in}{1.701538in}}{\pgfqpoint{1.569807in}{1.720816in}}%
\pgfpathlineto{\pgfqpoint{1.569807in}{1.994131in}}%
\pgfpathquadraticcurveto{\pgfqpoint{1.569807in}{2.013409in}}{\pgfqpoint{1.550530in}{2.013409in}}%
\pgfpathlineto{\pgfqpoint{0.719759in}{2.013409in}}%
\pgfpathquadraticcurveto{\pgfqpoint{0.700481in}{2.013409in}}{\pgfqpoint{0.700481in}{1.994131in}}%
\pgfpathlineto{\pgfqpoint{0.700481in}{1.720816in}}%
\pgfpathquadraticcurveto{\pgfqpoint{0.700481in}{1.701538in}}{\pgfqpoint{0.719759in}{1.701538in}}%
\pgfpathclose%
\pgfusepath{stroke,fill}%
\end{pgfscope}%
\begin{pgfscope}%
\pgfsetbuttcap%
\pgfsetmiterjoin%
\definecolor{currentfill}{rgb}{0.121569,0.466667,0.705882}%
\pgfsetfillcolor{currentfill}%
\pgfsetlinewidth{0.000000pt}%
\definecolor{currentstroke}{rgb}{0.000000,0.000000,0.000000}%
\pgfsetstrokecolor{currentstroke}%
\pgfsetstrokeopacity{0.000000}%
\pgfsetdash{}{0pt}%
\pgfpathmoveto{\pgfqpoint{0.739037in}{1.901620in}}%
\pgfpathlineto{\pgfqpoint{0.931814in}{1.901620in}}%
\pgfpathlineto{\pgfqpoint{0.931814in}{1.969093in}}%
\pgfpathlineto{\pgfqpoint{0.739037in}{1.969093in}}%
\pgfpathclose%
\pgfusepath{fill}%
\end{pgfscope}%
\begin{pgfscope}%
\definecolor{textcolor}{rgb}{0.000000,0.000000,0.000000}%
\pgfsetstrokecolor{textcolor}%
\pgfsetfillcolor{textcolor}%
\pgftext[x=1.008925in,y=1.901620in,left,base]{\color{textcolor}\sffamily\fontsize{6.940000}{8.328000}\selectfont flooding}%
\end{pgfscope}%
\begin{pgfscope}%
\pgfsetbuttcap%
\pgfsetmiterjoin%
\definecolor{currentfill}{rgb}{1.000000,0.498039,0.054902}%
\pgfsetfillcolor{currentfill}%
\pgfsetlinewidth{0.000000pt}%
\definecolor{currentstroke}{rgb}{0.000000,0.000000,0.000000}%
\pgfsetstrokecolor{currentstroke}%
\pgfsetstrokeopacity{0.000000}%
\pgfsetdash{}{0pt}%
\pgfpathmoveto{\pgfqpoint{0.739037in}{1.760143in}}%
\pgfpathlineto{\pgfqpoint{0.931814in}{1.760143in}}%
\pgfpathlineto{\pgfqpoint{0.931814in}{1.827616in}}%
\pgfpathlineto{\pgfqpoint{0.739037in}{1.827616in}}%
\pgfpathclose%
\pgfusepath{fill}%
\end{pgfscope}%
\begin{pgfscope}%
\definecolor{textcolor}{rgb}{0.000000,0.000000,0.000000}%
\pgfsetstrokecolor{textcolor}%
\pgfsetfillcolor{textcolor}%
\pgftext[x=1.008925in,y=1.760143in,left,base]{\color{textcolor}\sffamily\fontsize{6.940000}{8.328000}\selectfont minisketch}%
\end{pgfscope}%
\end{pgfpicture}%
\makeatother%
\endgroup%

%% file: figs/eval/convdelay_payments_0_plot.pgf
\begingroup%
\makeatletter%
\begin{pgfpicture}%
\pgfpathrectangle{\pgfpointorigin}{\pgfqpoint{3.077287in}{2.161603in}}%
\pgfusepath{use as bounding box, clip}%
\begin{pgfscope}%
\pgfsetbuttcap%
\pgfsetmiterjoin%
\definecolor{currentfill}{rgb}{1.000000,1.000000,1.000000}%
\pgfsetfillcolor{currentfill}%
\pgfsetlinewidth{0.000000pt}%
\definecolor{currentstroke}{rgb}{1.000000,1.000000,1.000000}%
\pgfsetstrokecolor{currentstroke}%
\pgfsetdash{}{0pt}%
\pgfpathmoveto{\pgfqpoint{0.000000in}{0.000000in}}%
\pgfpathlineto{\pgfqpoint{3.077287in}{0.000000in}}%
\pgfpathlineto{\pgfqpoint{3.077287in}{2.161603in}}%
\pgfpathlineto{\pgfqpoint{0.000000in}{2.161603in}}%
\pgfpathclose%
\pgfusepath{fill}%
\end{pgfscope}%
\begin{pgfscope}%
\pgfsetbuttcap%
\pgfsetmiterjoin%
\definecolor{currentfill}{rgb}{1.000000,1.000000,1.000000}%
\pgfsetfillcolor{currentfill}%
\pgfsetlinewidth{0.000000pt}%
\definecolor{currentstroke}{rgb}{0.000000,0.000000,0.000000}%
\pgfsetstrokecolor{currentstroke}%
\pgfsetstrokeopacity{0.000000}%
\pgfsetdash{}{0pt}%
\pgfpathmoveto{\pgfqpoint{0.652287in}{0.521603in}}%
\pgfpathlineto{\pgfqpoint{2.977287in}{0.521603in}}%
\pgfpathlineto{\pgfqpoint{2.977287in}{2.061603in}}%
\pgfpathlineto{\pgfqpoint{0.652287in}{2.061603in}}%
\pgfpathclose%
\pgfusepath{fill}%
\end{pgfscope}%
\begin{pgfscope}%
\pgfpathrectangle{\pgfqpoint{0.652287in}{0.521603in}}{\pgfqpoint{2.325000in}{1.540000in}}%
\pgfusepath{clip}%
\pgfsetbuttcap%
\pgfsetroundjoin%
\definecolor{currentfill}{rgb}{0.000000,0.000000,0.000000}%
\pgfsetfillcolor{currentfill}%
\pgfsetlinewidth{1.505625pt}%
\definecolor{currentstroke}{rgb}{0.000000,0.000000,0.000000}%
\pgfsetstrokecolor{currentstroke}%
\pgfsetdash{}{0pt}%
\pgfsys@defobject{currentmarker}{\pgfqpoint{-0.041667in}{-0.041667in}}{\pgfqpoint{0.041667in}{0.041667in}}{%
\pgfpathmoveto{\pgfqpoint{-0.041667in}{-0.041667in}}%
\pgfpathlineto{\pgfqpoint{0.041667in}{0.041667in}}%
\pgfpathmoveto{\pgfqpoint{-0.041667in}{0.041667in}}%
\pgfpathlineto{\pgfqpoint{0.041667in}{-0.041667in}}%
\pgfusepath{stroke,fill}%
}%
\begin{pgfscope}%
\pgfsys@transformshift{2.871605in}{1.991603in}%
\pgfsys@useobject{currentmarker}{}%
\end{pgfscope}%
\begin{pgfscope}%
\pgfsys@transformshift{2.870400in}{1.968309in}%
\pgfsys@useobject{currentmarker}{}%
\end{pgfscope}%
\begin{pgfscope}%
\pgfsys@transformshift{1.175230in}{0.987610in}%
\pgfsys@useobject{currentmarker}{}%
\end{pgfscope}%
\begin{pgfscope}%
\pgfsys@transformshift{1.183163in}{0.964315in}%
\pgfsys@useobject{currentmarker}{}%
\end{pgfscope}%
\begin{pgfscope}%
\pgfsys@transformshift{0.759173in}{0.600921in}%
\pgfsys@useobject{currentmarker}{}%
\end{pgfscope}%
\begin{pgfscope}%
\pgfsys@transformshift{0.765859in}{0.596262in}%
\pgfsys@useobject{currentmarker}{}%
\end{pgfscope}%
\begin{pgfscope}%
\pgfsys@transformshift{0.761706in}{0.591603in}%
\pgfsys@useobject{currentmarker}{}%
\end{pgfscope}%
\begin{pgfscope}%
\pgfsys@transformshift{0.759380in}{0.593933in}%
\pgfsys@useobject{currentmarker}{}%
\end{pgfscope}%
\begin{pgfscope}%
\pgfsys@transformshift{0.757968in}{0.598592in}%
\pgfsys@useobject{currentmarker}{}%
\end{pgfscope}%
\begin{pgfscope}%
\pgfsys@transformshift{0.834510in}{0.666146in}%
\pgfsys@useobject{currentmarker}{}%
\end{pgfscope}%
\begin{pgfscope}%
\pgfsys@transformshift{0.838622in}{0.689440in}%
\pgfsys@useobject{currentmarker}{}%
\end{pgfscope}%
\begin{pgfscope}%
\pgfsys@transformshift{0.840532in}{0.689440in}%
\pgfsys@useobject{currentmarker}{}%
\end{pgfscope}%
\begin{pgfscope}%
\pgfsys@transformshift{0.839867in}{0.659157in}%
\pgfsys@useobject{currentmarker}{}%
\end{pgfscope}%
\begin{pgfscope}%
\pgfsys@transformshift{2.053029in}{1.402252in}%
\pgfsys@useobject{currentmarker}{}%
\end{pgfscope}%
\begin{pgfscope}%
\pgfsys@transformshift{1.861530in}{1.325380in}%
\pgfsys@useobject{currentmarker}{}%
\end{pgfscope}%
\begin{pgfscope}%
\pgfsys@transformshift{2.056351in}{1.507078in}%
\pgfsys@useobject{currentmarker}{}%
\end{pgfscope}%
\begin{pgfscope}%
\pgfsys@transformshift{1.867302in}{1.227543in}%
\pgfsys@useobject{currentmarker}{}%
\end{pgfscope}%
\end{pgfscope}%
\begin{pgfscope}%
\pgfsetbuttcap%
\pgfsetroundjoin%
\definecolor{currentfill}{rgb}{0.000000,0.000000,0.000000}%
\pgfsetfillcolor{currentfill}%
\pgfsetlinewidth{0.803000pt}%
\definecolor{currentstroke}{rgb}{0.000000,0.000000,0.000000}%
\pgfsetstrokecolor{currentstroke}%
\pgfsetdash{}{0pt}%
\pgfsys@defobject{currentmarker}{\pgfqpoint{0.000000in}{-0.048611in}}{\pgfqpoint{0.000000in}{0.000000in}}{%
\pgfpathmoveto{\pgfqpoint{0.000000in}{0.000000in}}%
\pgfpathlineto{\pgfqpoint{0.000000in}{-0.048611in}}%
\pgfusepath{stroke,fill}%
}%
\begin{pgfscope}%
\pgfsys@transformshift{0.754563in}{0.521603in}%
\pgfsys@useobject{currentmarker}{}%
\end{pgfscope}%
\end{pgfscope}%
\begin{pgfscope}%
\definecolor{textcolor}{rgb}{0.000000,0.000000,0.000000}%
\pgfsetstrokecolor{textcolor}%
\pgfsetfillcolor{textcolor}%
\pgftext[x=0.754563in,y=0.424381in,,top]{\color{textcolor}\sffamily\fontsize{10.000000}{12.000000}\selectfont 0}%
\end{pgfscope}%
\begin{pgfscope}%
\pgfsetbuttcap%
\pgfsetroundjoin%
\definecolor{currentfill}{rgb}{0.000000,0.000000,0.000000}%
\pgfsetfillcolor{currentfill}%
\pgfsetlinewidth{0.803000pt}%
\definecolor{currentstroke}{rgb}{0.000000,0.000000,0.000000}%
\pgfsetstrokecolor{currentstroke}%
\pgfsetdash{}{0pt}%
\pgfsys@defobject{currentmarker}{\pgfqpoint{0.000000in}{-0.048611in}}{\pgfqpoint{0.000000in}{0.000000in}}{%
\pgfpathmoveto{\pgfqpoint{0.000000in}{0.000000in}}%
\pgfpathlineto{\pgfqpoint{0.000000in}{-0.048611in}}%
\pgfusepath{stroke,fill}%
}%
\begin{pgfscope}%
\pgfsys@transformshift{1.585182in}{0.521603in}%
\pgfsys@useobject{currentmarker}{}%
\end{pgfscope}%
\end{pgfscope}%
\begin{pgfscope}%
\definecolor{textcolor}{rgb}{0.000000,0.000000,0.000000}%
\pgfsetstrokecolor{textcolor}%
\pgfsetfillcolor{textcolor}%
\pgftext[x=1.585182in,y=0.424381in,,top]{\color{textcolor}\sffamily\fontsize{10.000000}{12.000000}\selectfont 200}%
\end{pgfscope}%
\begin{pgfscope}%
\pgfsetbuttcap%
\pgfsetroundjoin%
\definecolor{currentfill}{rgb}{0.000000,0.000000,0.000000}%
\pgfsetfillcolor{currentfill}%
\pgfsetlinewidth{0.803000pt}%
\definecolor{currentstroke}{rgb}{0.000000,0.000000,0.000000}%
\pgfsetstrokecolor{currentstroke}%
\pgfsetdash{}{0pt}%
\pgfsys@defobject{currentmarker}{\pgfqpoint{0.000000in}{-0.048611in}}{\pgfqpoint{0.000000in}{0.000000in}}{%
\pgfpathmoveto{\pgfqpoint{0.000000in}{0.000000in}}%
\pgfpathlineto{\pgfqpoint{0.000000in}{-0.048611in}}%
\pgfusepath{stroke,fill}%
}%
\begin{pgfscope}%
\pgfsys@transformshift{2.415802in}{0.521603in}%
\pgfsys@useobject{currentmarker}{}%
\end{pgfscope}%
\end{pgfscope}%
\begin{pgfscope}%
\definecolor{textcolor}{rgb}{0.000000,0.000000,0.000000}%
\pgfsetstrokecolor{textcolor}%
\pgfsetfillcolor{textcolor}%
\pgftext[x=2.415802in,y=0.424381in,,top]{\color{textcolor}\sffamily\fontsize{10.000000}{12.000000}\selectfont 400}%
\end{pgfscope}%
\begin{pgfscope}%
\definecolor{textcolor}{rgb}{0.000000,0.000000,0.000000}%
\pgfsetstrokecolor{textcolor}%
\pgfsetfillcolor{textcolor}%
\pgftext[x=1.814787in,y=0.234413in,,top]{\color{textcolor}\sffamily\fontsize{10.000000}{12.000000}\selectfont Convergence delay in seconds}%
\end{pgfscope}%
\begin{pgfscope}%
\pgfsetbuttcap%
\pgfsetroundjoin%
\definecolor{currentfill}{rgb}{0.000000,0.000000,0.000000}%
\pgfsetfillcolor{currentfill}%
\pgfsetlinewidth{0.803000pt}%
\definecolor{currentstroke}{rgb}{0.000000,0.000000,0.000000}%
\pgfsetstrokecolor{currentstroke}%
\pgfsetdash{}{0pt}%
\pgfsys@defobject{currentmarker}{\pgfqpoint{-0.048611in}{0.000000in}}{\pgfqpoint{-0.000000in}{0.000000in}}{%
\pgfpathmoveto{\pgfqpoint{-0.000000in}{0.000000in}}%
\pgfpathlineto{\pgfqpoint{-0.048611in}{0.000000in}}%
\pgfusepath{stroke,fill}%
}%
\begin{pgfscope}%
\pgfsys@transformshift{0.652287in}{0.589274in}%
\pgfsys@useobject{currentmarker}{}%
\end{pgfscope}%
\end{pgfscope}%
\begin{pgfscope}%
\definecolor{textcolor}{rgb}{0.000000,0.000000,0.000000}%
\pgfsetstrokecolor{textcolor}%
\pgfsetfillcolor{textcolor}%
\pgftext[x=0.466699in, y=0.536512in, left, base]{\color{textcolor}\sffamily\fontsize{10.000000}{12.000000}\selectfont 0}%
\end{pgfscope}%
\begin{pgfscope}%
\pgfsetbuttcap%
\pgfsetroundjoin%
\definecolor{currentfill}{rgb}{0.000000,0.000000,0.000000}%
\pgfsetfillcolor{currentfill}%
\pgfsetlinewidth{0.803000pt}%
\definecolor{currentstroke}{rgb}{0.000000,0.000000,0.000000}%
\pgfsetstrokecolor{currentstroke}%
\pgfsetdash{}{0pt}%
\pgfsys@defobject{currentmarker}{\pgfqpoint{-0.048611in}{0.000000in}}{\pgfqpoint{-0.000000in}{0.000000in}}{%
\pgfpathmoveto{\pgfqpoint{-0.000000in}{0.000000in}}%
\pgfpathlineto{\pgfqpoint{-0.048611in}{0.000000in}}%
\pgfusepath{stroke,fill}%
}%
\begin{pgfscope}%
\pgfsys@transformshift{0.652287in}{1.055164in}%
\pgfsys@useobject{currentmarker}{}%
\end{pgfscope}%
\end{pgfscope}%
\begin{pgfscope}%
\definecolor{textcolor}{rgb}{0.000000,0.000000,0.000000}%
\pgfsetstrokecolor{textcolor}%
\pgfsetfillcolor{textcolor}%
\pgftext[x=0.289968in, y=1.002403in, left, base]{\color{textcolor}\sffamily\fontsize{10.000000}{12.000000}\selectfont 200}%
\end{pgfscope}%
\begin{pgfscope}%
\pgfsetbuttcap%
\pgfsetroundjoin%
\definecolor{currentfill}{rgb}{0.000000,0.000000,0.000000}%
\pgfsetfillcolor{currentfill}%
\pgfsetlinewidth{0.803000pt}%
\definecolor{currentstroke}{rgb}{0.000000,0.000000,0.000000}%
\pgfsetstrokecolor{currentstroke}%
\pgfsetdash{}{0pt}%
\pgfsys@defobject{currentmarker}{\pgfqpoint{-0.048611in}{0.000000in}}{\pgfqpoint{-0.000000in}{0.000000in}}{%
\pgfpathmoveto{\pgfqpoint{-0.000000in}{0.000000in}}%
\pgfpathlineto{\pgfqpoint{-0.048611in}{0.000000in}}%
\pgfusepath{stroke,fill}%
}%
\begin{pgfscope}%
\pgfsys@transformshift{0.652287in}{1.521054in}%
\pgfsys@useobject{currentmarker}{}%
\end{pgfscope}%
\end{pgfscope}%
\begin{pgfscope}%
\definecolor{textcolor}{rgb}{0.000000,0.000000,0.000000}%
\pgfsetstrokecolor{textcolor}%
\pgfsetfillcolor{textcolor}%
\pgftext[x=0.289968in, y=1.468293in, left, base]{\color{textcolor}\sffamily\fontsize{10.000000}{12.000000}\selectfont 400}%
\end{pgfscope}%
\begin{pgfscope}%
\pgfsetbuttcap%
\pgfsetroundjoin%
\definecolor{currentfill}{rgb}{0.000000,0.000000,0.000000}%
\pgfsetfillcolor{currentfill}%
\pgfsetlinewidth{0.803000pt}%
\definecolor{currentstroke}{rgb}{0.000000,0.000000,0.000000}%
\pgfsetstrokecolor{currentstroke}%
\pgfsetdash{}{0pt}%
\pgfsys@defobject{currentmarker}{\pgfqpoint{-0.048611in}{0.000000in}}{\pgfqpoint{-0.000000in}{0.000000in}}{%
\pgfpathmoveto{\pgfqpoint{-0.000000in}{0.000000in}}%
\pgfpathlineto{\pgfqpoint{-0.048611in}{0.000000in}}%
\pgfusepath{stroke,fill}%
}%
\begin{pgfscope}%
\pgfsys@transformshift{0.652287in}{1.986944in}%
\pgfsys@useobject{currentmarker}{}%
\end{pgfscope}%
\end{pgfscope}%
\begin{pgfscope}%
\definecolor{textcolor}{rgb}{0.000000,0.000000,0.000000}%
\pgfsetstrokecolor{textcolor}%
\pgfsetfillcolor{textcolor}%
\pgftext[x=0.289968in, y=1.934183in, left, base]{\color{textcolor}\sffamily\fontsize{10.000000}{12.000000}\selectfont 600}%
\end{pgfscope}%
\begin{pgfscope}%
\definecolor{textcolor}{rgb}{0.000000,0.000000,0.000000}%
\pgfsetstrokecolor{textcolor}%
\pgfsetfillcolor{textcolor}%
\pgftext[x=0.234413in,y=1.291603in,,bottom,rotate=90.000000]{\color{textcolor}\sffamily\fontsize{10.000000}{12.000000}\selectfont Payment attempts}%
\end{pgfscope}%
\begin{pgfscope}%
\pgfpathrectangle{\pgfqpoint{0.652287in}{0.521603in}}{\pgfqpoint{2.325000in}{1.540000in}}%
\pgfusepath{clip}%
\pgfsetrectcap%
\pgfsetroundjoin%
\pgfsetlinewidth{1.003750pt}%
\definecolor{currentstroke}{rgb}{0.000000,0.000000,0.000000}%
\pgfsetstrokecolor{currentstroke}%
\pgfsetdash{}{0pt}%
\pgfpathmoveto{\pgfqpoint{2.871605in}{1.974022in}}%
\pgfpathlineto{\pgfqpoint{2.870400in}{1.973250in}}%
\pgfpathlineto{\pgfqpoint{1.175230in}{0.885881in}}%
\pgfpathlineto{\pgfqpoint{1.183163in}{0.890970in}}%
\pgfpathlineto{\pgfqpoint{0.759173in}{0.619001in}}%
\pgfpathlineto{\pgfqpoint{0.765859in}{0.623290in}}%
\pgfpathlineto{\pgfqpoint{0.761706in}{0.620626in}}%
\pgfpathlineto{\pgfqpoint{0.759380in}{0.619134in}}%
\pgfpathlineto{\pgfqpoint{0.757968in}{0.618228in}}%
\pgfpathlineto{\pgfqpoint{0.834510in}{0.667326in}}%
\pgfpathlineto{\pgfqpoint{0.838622in}{0.669963in}}%
\pgfpathlineto{\pgfqpoint{0.840532in}{0.671189in}}%
\pgfpathlineto{\pgfqpoint{0.839867in}{0.670763in}}%
\pgfpathlineto{\pgfqpoint{2.053029in}{1.448946in}}%
\pgfpathlineto{\pgfqpoint{1.861530in}{1.326109in}}%
\pgfpathlineto{\pgfqpoint{2.056351in}{1.451077in}}%
\pgfpathlineto{\pgfqpoint{1.867302in}{1.329812in}}%
\pgfusepath{stroke}%
\end{pgfscope}%
\begin{pgfscope}%
\pgfsetrectcap%
\pgfsetmiterjoin%
\pgfsetlinewidth{0.803000pt}%
\definecolor{currentstroke}{rgb}{0.000000,0.000000,0.000000}%
\pgfsetstrokecolor{currentstroke}%
\pgfsetdash{}{0pt}%
\pgfpathmoveto{\pgfqpoint{0.652287in}{0.521603in}}%
\pgfpathlineto{\pgfqpoint{0.652287in}{2.061603in}}%
\pgfusepath{stroke}%
\end{pgfscope}%
\begin{pgfscope}%
\pgfsetrectcap%
\pgfsetmiterjoin%
\pgfsetlinewidth{0.803000pt}%
\definecolor{currentstroke}{rgb}{0.000000,0.000000,0.000000}%
\pgfsetstrokecolor{currentstroke}%
\pgfsetdash{}{0pt}%
\pgfpathmoveto{\pgfqpoint{2.977287in}{0.521603in}}%
\pgfpathlineto{\pgfqpoint{2.977287in}{2.061603in}}%
\pgfusepath{stroke}%
\end{pgfscope}%
\begin{pgfscope}%
\pgfsetrectcap%
\pgfsetmiterjoin%
\pgfsetlinewidth{0.803000pt}%
\definecolor{currentstroke}{rgb}{0.000000,0.000000,0.000000}%
\pgfsetstrokecolor{currentstroke}%
\pgfsetdash{}{0pt}%
\pgfpathmoveto{\pgfqpoint{0.652287in}{0.521603in}}%
\pgfpathlineto{\pgfqpoint{2.977287in}{0.521603in}}%
\pgfusepath{stroke}%
\end{pgfscope}%
\begin{pgfscope}%
\pgfsetrectcap%
\pgfsetmiterjoin%
\pgfsetlinewidth{0.803000pt}%
\definecolor{currentstroke}{rgb}{0.000000,0.000000,0.000000}%
\pgfsetstrokecolor{currentstroke}%
\pgfsetdash{}{0pt}%
\pgfpathmoveto{\pgfqpoint{0.652287in}{2.061603in}}%
\pgfpathlineto{\pgfqpoint{2.977287in}{2.061603in}}%
\pgfusepath{stroke}%
\end{pgfscope}%
\end{pgfpicture}%
\makeatother%
\endgroup%